\documentclass[apj,twocolumn]{openjournal}
\usepackage{amsmath}

\usepackage[dvipsnames]{xcolor}
\usepackage[breaklinks,colorlinks,citecolor=blue,urlcolor=blue]{hyperref}

\usepackage{orcidlink}

\renewcommand{\arraystretch}{1.2}  
\usepackage{listings}
\usepackage{color}
\definecolor{dkgreen}{rgb}{0,0.6,0}
\definecolor{gray}{rgb}{0.5,0.5,0.5}
\definecolor{mauve}{rgb}{0.58,0,0.82}
\definecolor{golden}{rgb}{0.86,0.65,0.01}
\usepackage{color}
\usepackage{soul}
\usepackage{amsmath}
\usepackage{xspace}
\usepackage{graphicx}
\usepackage{enumitem}
\usepackage{amssymb}
\usepackage{xifthen}
\usepackage{hyperref}
\usepackage[normalem]{ulem}
\usepackage{multirow}
\usepackage{lineno}

\newcommand{\code}[1]{\texttt{#1}\xspace}

\newcommand{\gaia}{\textit{Gaia}\xspace}
\newcommand{\Gaia}{\gaia}
\newcommand{\unit}[1]{\ensuremath{\mathrm{\,#1}}\xspace}
\newcommand{\feh}{\unit{[Fe/H]}}
\newcommand{\teff}{\ensuremath{T_\mathrm{eff}}\xspace}
\newcommand{\logg}{\ensuremath{\log\,g}\xspace}

\newcommand{\kms}{\unit{km\,s^{-1}}}

\newcommand{\msun}{\unit{M_\odot}}

\newcommand{\mstar}{\unit{M_\star}}

\begin{document}

\title{Chemical Abundances in the Leiptr Stellar Stream: a Disrupted Ultra-faint Dwarf Galaxy?}

\author{Kaia R. Atzberger\,\orcidlink{0000-0001-9649-8103}$^{1,2}$}
\author{Sam A. Usman\,\orcidlink{0000-0003-0918-7185}$^{3,4}$}
\author{Alexander P. Ji\,\orcidlink{0000-0002-4863-8842}$^{3,4}$}
\author{Lara R. Cullinane\,\orcidlink{0000-0001-8536-0547}$^{5}$}
\author{Denis Erkal\,\orcidlink{0000-0002-8448-5505}$^{6}$}
\author{Terese T. Hansen\,\orcidlink{0000-0001-6154-8983}$^{7}$}
\author{Geraint F. Lewis\,\orcidlink{0000-0003-3081-9319}$^{8}$}
\author{Ting S. Li\,\orcidlink{0000-0002-9110-6163}$^{9}$}
\author{Guilherme Limberg\,\orcidlink{0000-0002-9269-8287}$^{10,3,4}$}
\author{Alice Luna\,\orcidlink{0009-0009-9570-0715}$^{3,4}$}
\author{Sarah L. Martell\,\orcidlink{0000-0002-3430-4163}$^{11,12}$}
\author{Madeleine McKenzie\,\orcidlink{0000-0002-1715-1257}$^{13,12}$}
\author{Andrew B. Pace\,\orcidlink{0000-0002-6021-8760}$^{2}$}
\author{Daniel B. Zucker\,\orcidlink{0000-0003-1124-8477}$^{14,15,12}$}

\affiliation{$^1$Department of Astronomy, The Ohio State University, 140 W 18th Avenue, Columbus, OH, 43210, USA}
\affiliation{$^2$Department of Astronomy, University of Virginia, 530 McCormick Road, Charlottesville, VA 22904, USA}
\affiliation{$^3$Department of Astronomy \& Astrophysics, University of Chicago, 5640 S Ellis Avenue, Chicago, IL 60637, USA}
\affiliation{$^4$Kavli Institute for Cosmological Physics, University of Chicago, Chicago, IL 60637, USA}
\affiliation{$^5$Leibniz-Institut für Astrophysik (AIP), An der Sternwarte 16, D-14482 Potsdam, Germany}
\affiliation{$^6$School of Mathematics and Physics, University of Surrey, Guildford, GU2 7XH, UK}
\affiliation{$^7$Department of Astronomy, Stockholm University, AlbaNova University Center, SE-106 91 Stockholm, Sweden}
\affiliation{$^8$Sydney Institute for Astronomy, School of Physics, A28, The University of Sydney, NSW 2006, Australia}
\affiliation{$^9$Department of Astronomy and Astrophysics, University of Toronto, 50 St. George Street, Toronto ON, M5S 3H4, Canada}
\affiliation{$^{10}$IAG, Departamento de Astronomia, Universidade de São Paulo, SP 05508-090 São Paulo, Brazil}
\affiliation{$^{11}$School of Physics, University of New South Wales, Sydney, NSW 2052, Australia}
\affiliation{$^{12}$Centre of Excellence for Astrophysics in Three Dimensions (ASTRO 3D), Australia}
\affiliation{$^{13}$Research School of Astronomy \& Astrophysics, Australian National University, Canberra, ACT 2611, Australia}
\affiliation{$^{14}$School of Mathematical and Physical Sciences, Macquarie University, Sydney, NSW 2109, Australia}
\affiliation{$^{15}$Macquarie University Astrophysics and Space Technologies Research Centre, Sydney, NSW 2109, Australia}

\email{katzberger@email.virginia.edu}

\begin{abstract}
Chemical abundances of stellar streams can be used to determine the nature of a stream's progenitor.
Here we study the progenitor of the recently discovered Leiptr stellar stream, which was previously suggested to be a tidally disrupted halo globular cluster.
We obtain high-resolution spectra of five red giant branch stars selected from the \Gaia DR2 \texttt{STREAMFINDER} catalog with Magellan/MIKE. 
One star is a clear non-member.
The remaining four stars display chemical abundances consistent with those of a low-mass dwarf galaxy:
they have a low mean metallicity, $\langle{\rm[Fe/H]}\rangle = -2.2$; they do not all have identical metallicities; and they display low [$\alpha$/Fe] $\sim 0$ and [Sr/Fe] and [Ba/Fe] $\sim -1$.
This pattern of low $\alpha$ and neutron-capture element abundances is only found in intact dwarf galaxies with stellar mass $\lesssim 10^5 \msun$.
Although more data are needed to be certain, Leiptr's chemistry is consistent with being the lowest-mass dwarf galaxy stream without a known intact progenitor, possibly in the mass range of ultra-faint dwarf galaxies.
Leiptr thus preserves a record of one of the lowest-mass early accretion events into the Milky Way.
\end{abstract}

\maketitle
\section{Introduction}
\label{intro}

Over the course of their lifetimes, stellar systems can be torn apart by the gravitational potential of massive galaxies.
Stellar streams form during the intermediate stage of tidal disruption where they preserve their coherence in phase-space before mixing into a stellar halo \citep{Helmi99}.
Current models of galaxy formation predict that there should be a large population of globular cluster and dwarf galaxy stellar streams in the Milky Way’s stellar halo as a result of past accretion events \citep{Helmi20}.
These systems provide observational evidence to evaluate theoretical models of hierarchical structure formation \citep[e.g.,][]{Johnston98,Helmi99,Freeman02,Bullock05} and quantify the amount of dark matter substructure \citep[e.g.,][]{Ibata02,Johnston02,Carlberg09,Varghese11,Erkal16}.
Large photometric surveys such as the Sloan Digital Sky Survey (\citealt{York00}) and the Dark Energy Survey (\citealt{DES16}) in combination with proper motions from \Gaia data releases \citep{GAIA16,Lindegren21,GAIA22} have revealed over 140 streams \citep{Grillmair16,Shipp18,Mateu18,Mateu2023,Ibata24,Bonaca24}.

It is often unclear whether a given stream's progenitor is a dwarf galaxy, a globular cluster, or possibly a superposition of both \citep[e.g.,][]{Carlberg2018,Malhan2021}.
In some cases, streams are clearly identified with intact progenitors, such as the Sagittarius dwarf galaxy \citep{Ibata2001,Majewski2003}, the Antlia II and Crater II dwarf spheroidals \citep{Ji2021,Coppi2024}, and several globular clusters \citep[e.g.,][]{Simpson2020,Hansen2020,Gull2021}.
The majority of streams do not have associated intact progenitors. 
Instead, a stream's morphology and kinematics can provide a strong prior on the progenitor \citep[e.g.,][]{Erkal2016,Shipp18,Ibata24}. 
However, a definitive classification requires studying a stream's chemical abundances \citep[e.g.,][]{Li19,Li22,Ji20,Martin2022}.

The majority of known streams are thought to have globular cluster progenitors, due to their thin widths and in many cases small or unresolved metallicity dispersions ${\lesssim}0.05$ dex \citep{Li22, Ibata24}.
These include the Phoenix \citep{Wan20,Casey2021} and C-19 streams \citep{Martin2022,Yuan2022} which are below the globular cluster metallicity ``floor'';
the dynamically perturbed GD-1 \citep{Grillmair2006,Price-Whelan2018} and ATLAS-Aliqa Uma streams \citep{Shipp18,Li2021}; disrupted globular clusters embedded within the Indus, Jhelum, and Wukong/LMS-1 dwarf galaxy streams \citep{Hansen2021,Awad2024,Limberg2024};
and the 300S stream, which was originally thought to be a dwarf galaxy stream \citep{Simon2011,Frebel2013,Fu2018} but has since been chemically confirmed to be an accreted globular cluster stream \citep{Li22,Usman24}.

Less common by number are the dwarf galaxy streams.
Many of the first discovered streams, such as the Helmi stream \citep{Helmi1999,Roederer2010,Limberg2021,Matsuno2022b} and Sagittarius stream \citep{Ibata2001,Majewski2003}, were easily identified as dwarf galaxies as these are well-populated streams from relatively massive ($\mstar \gtrsim 10^8 \msun$) dwarf galaxy progenitors.
More recently, several lower-mass dwarf galaxy streams have also been identified. 
Based on their chemical patterns, the Indus and Jhelum streams have progenitor mass $\mstar \sim 10^{6-7} \msun$ \citep{Hansen2021}; the Orphan-Chenab stream \citep{Belokurov2007,Casey2013,Shipp18,Koposov2019,Erkal19} has progenitor mass $\mstar \sim 10^{6.5} \msun$ \citep{Koposov2019,Hawkins23}; the Wukong/LMS-1 stream has progenitor mass $\mstar \sim 10^7 \msun$ \citep{Naidu20,Yuan2020,Limberg2024};
the Cetus/Palca stream has progenitor mass $\mstar \sim 10^6 \msun$ \citep{Newberg2009,Yuan2022cetus,Sitnova2024};
and the Elqui stream has progenitor mass $\mstar \sim 10^6 \msun$ \citep{Shipp18, Ji20, Li22}.
This does not include more disrupted and phase-mixed dwarf galaxies like Gaia-Sausage/Enceladus \citep[GSE,][]{Belokurov18,Haywood18,Helmi18,Carrillo22}, Sequoia \citep{Myeong19,Matsuno2019}, Thamnos \citep{Koppelman2019}, and others \citep{Naidu20,Helmi20,Deason2024}.

Current predictions for globular cluster streams are that the Milky Way has ${\sim}1000$ globular cluster stellar streams that have yet to be fully phase-mixed.
These are predicted by mock stellar streams from the orbital histories of progenitors generated by hierarchical models of globular cluster formation \citep{Pearson2024}.
There are also expected to be many hundreds of dwarf galaxies that have been tidally disrupted around the Milky Way.
A number of strategies have been developed for this over time, such as semi-analytic methods for determining phase-space populations of tidal debris, simulations of satellite galaxy disruption, and clustering stellar dynamics in 4D phase space \citep[e.g.,][]{Johnston98,Helmi99,Santistevan2020,Brauer2022,Gandhi2024}.
However, it is harder to identify the dwarf galaxy streams.
Current stream searches, especially those relying on \textit{Gaia} parallaxes \citep[e.g.,][]{Ibata19,Ibata24,Dodd2023}, mostly search at nearby galactocentric radii, where globular cluster streams are likely more numerous than dwarf galaxy streams \citep{Johnston2008,Gomez2013,Pearson2024,Bonaca24}.
Dwarf galaxy stars have larger velocity dispersions and metallicity spreads, so at fixed stellar mass they are harder to find compared to globular clusters that are more spatially compact and at single metallicities.
Finally, the vast majority of the tidally disrupted dwarf galaxies are very low-mass, with $\mstar \lesssim 10^5$ \msun, i.e. ultra-faint dwarf galaxies \citep[e.g.,][]{Weisz2017, Simon19, Brauer2019, Santistevan2020, Gandhi2024}, so they should be even harder to identify.
Obtaining a full census of the Milky Way's accretion history thus requires extending stream discoveries down to this regime of ultra-faint dwarf galaxies.
Since these ultra-faint dwarf galaxies are also the most dark matter-dominated galaxies that live in the lowest-mass star-forming dark matter halos, understanding the number of these streams in the Milky Way could contribute to understanding reionization and the small scale structure of dark matter \citep[e.g.,][]{Bullock2017,Jethwa2018,Simon19,Nadler2021}.

In recent years, much progress has been made in finding streams from the lowest-mass dwarf galaxies.
The first clear stream around an intact ultra-faint dwarf galaxy was found around the Tucana~III system \citep{DrlicaWagner15,Shipp18}, which has been spectroscopically confirmed \citep{Hansen2017,Li2018}. More recently, a tidal stream likely emanating from the Hydrus I dwarf galaxy was found \citep{Ibata24}.
Several individual stars with chemistry similar to ultra-faint dwarf galaxy stars have also been found throughout the Milky Way halo  \citep[e.g.,][]{Casey2017, Roederer2017}.
Motivated by the discovery of the highly $r$-process (rapid neutron-capture process) enhanced ultra-faint dwarf Reticulum~II \citep{Ji2016b,Roederer2016}, several stellar dynamical groups within the Milky Way that are highly enhanced in $r$-process elements have also been suggested to originate from ultra-faint dwarf galaxies \citep[e.g.,][]{Roederer18,Lee19,Limberg2021UFD,Gudin2021}, but the probability that the stars in these groups all come from individual ultra-faint dwarf galaxies is statistically unlikely \citep{Brauer2022}.
Thus far, there is no progenitor-free bona-fide stellar stream that can be cleanly associated with an ultra-faint dwarf galaxy.
The only claimed candidate, Specter \citep{Chandra2022}, has a high mean metallicity that suggests we are only seeing part of the stream from a much higher-mass progenitor dwarf galaxy.

\begin{deluxetable*}{ccccccccccc}
\tablecolumns{11}
\tablecaption{\label{tab:leiptrstars}Leiptr Stars}
\tablehead{Name & \Gaia DR3 Source ID & ExpTime & SNR & SNR & RV & RA & Dec & $G$ & $BP$ & $RP$
\\  &   & (s) & 4500~\AA\ & 6500~\AA& (\kms) & ($^\circ$)& ($^\circ$)& (mag) & (mag) & (mag)}
\startdata
Leiptr-186 & 2909738294618626048 & 300 & 44 & 80 & 191.1 & 89.11202 & $-28.18913$ & 12.33 & 12.90 & 11.61 \\
Leiptr-208 & 2963688139034857472 & 1500 & 35 & 53 & 144.2 & 84.19212 & $-23.48547$ & 14.84 & 15.26 & 14.25 \\
Leiptr-252\tablenotemark{$*$} & 3181161746481442432 & 5400 & 35 & 53 & $-4.2$ & 71.34197 & $-11.61942$ & 16.21 & 16.63 & 15.61 \\
Leiptr-321 & 5531991922485962624 & 2400 & 29 & 55 & 372.0 & 115.32295 & $-44.58400$ & 15.21 & 15.81 & 14.44 \\
Leiptr-342 & 5561352147121401216 & 300 & 38 & 77 & 337.4 & 108.16171 & $-40.85943$ & 12.43 & 13.14 & 11.61 \\
\enddata
\tablecomments{SNR 4500~\AA\ and 6500~\AA\ is the signal-to-noise per pixel at 4500~\AA\ and 6500~\AA. 
RV is the heliocentric radial velocity with 1 km s$^{-1}$ uncertainty.}
\tablenotetext{$^*$}{Non-member}
\end{deluxetable*}

Here we investigate the progenitor of the Leiptr stellar stream, which was previously thought to be a tidally disrupted globular cluster.
Leiptr was discovered among a group of inner galaxy streams in the \Gaia DR2 catalog by \citet{Ibata19} using the \texttt{STREAMFINDER} algorithm.
They identified candidates from streams with distinct and coherent sky positions, proper motions, and parallaxes that were clearly distinguishable from the Galactic disk or bulge.
They determined Leiptr was 48$^{\circ}$ long and located toward the Galactic anticenter with a heliocentric distance of $7.9\pm 0.4$ kpc, pericenter distance of $12.8\pm 0.6$ kpc, apocenter distance of $85.2\pm 26.0$ kpc, and strong retrograde motion of $L_z$ = $4689\pm 314$ km s$^{-1}$ kpc where $L_z$ is the vertical component of angular momentum in a Galactocentric Cartesian frame.
\citet{Malhan2022} updated these orbital parameters, estimating the pericenter to be 12.3$\pm$ 0.1 kpc and the apocenter to be 45.1 $\pm$ 0.2 kpc, with a resulting eccentricity of 0.57.
Leiptr has the highest $\lvert L_z \rvert$ of stellar streams in their sample, next to Gaia-12.
\citet{Bonaca21} identified Leiptr, Gj\"{o}ll, GD-1, Phlegethon, Ylgr, and Wambelong as potential members of a highly aligned ``plane of streams'' that may share a common origin.
They suggested the retrograde streams are most likely globular clusters from the Sequoia/Arjuna dwarf galaxy \citep{Myeong19,Naidu20} based on photometry and clustering in orbital poles; Gj\"{o}ll in particular has been shown to be connected to the NGC 3201 globular cluster \citep{Hansen2020,Riley20}.
\citet{Bonaca24} independently determined that Leiptr has a mean heliocentric distance of 7.4 kpc, mean Galactic radius of 13.8 kpc, angular length of 73$^{\circ}$, angular width of 0.5$^{\circ}$ (74 pc), and a stellar mass of $10^{3.8}$ \msun.
However, there have been no previous chemical studies of Leiptr.

We present the chemical abundances of four member stars in the Leiptr stream and one non-member using high-resolution spectroscopy. 
We find that Leiptr is likely the first spectroscopically confirmed ultra-faint dwarf galaxy stream without an intact progenitor.
Section~\ref{sec:Observations} explains the observations.
Section \ref{sec:Abundance Analysis} describes the abundance analysis.
Section \ref{sec:Results} compares the chemical abundances of Leiptr to those of halo, dwarf galaxy, and globular cluster stars.
We discuss what these results suggest about Leiptr's chemical signature and progenitor in Section \ref{sec:Discussion} and conclude in Section \ref{sec:Conclusions}.

\section{Observations}\label{sec:Observations}

\begin{figure}[t]
    \centering
    \includegraphics[width=\linewidth]{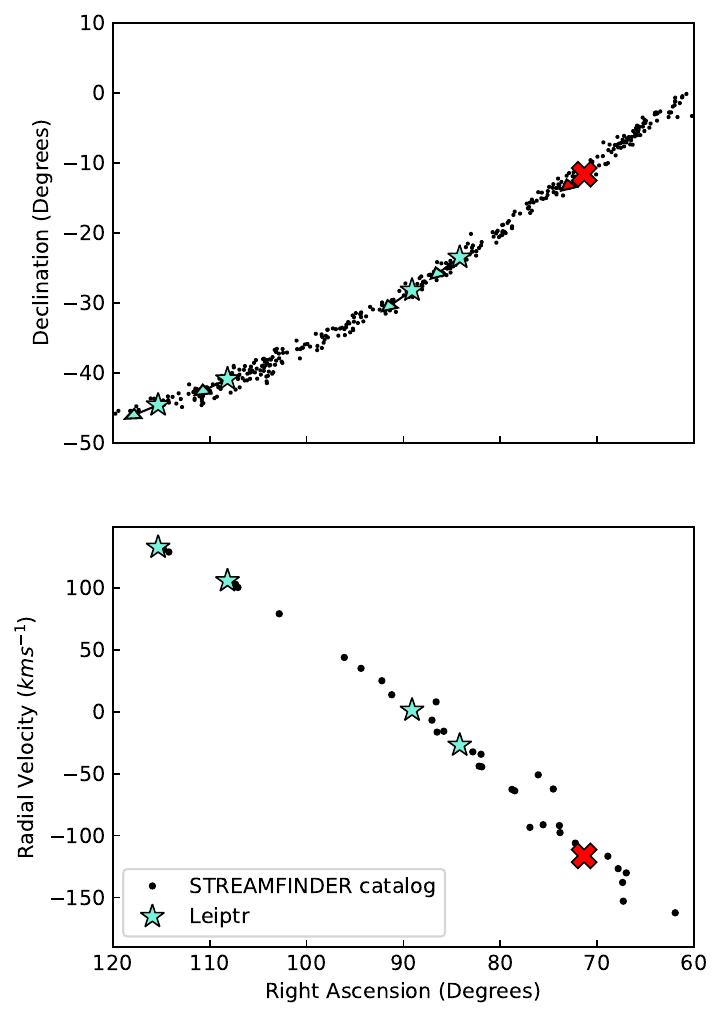}
    \caption{Positions and reflex-corrected radial velocities of Leiptr candidate stars.
    The members from the \texttt{STREAMFINDER} catalog are represented by black circles and four members analyzed in this work are represented by light-blue stars.
    The non-member, Leiptr-252, is emphasized as a red X.
    The reflex-corrected proper motions in degrees per Myr are indicated by arrows in the top panel.
    Any stars in the \texttt{STREAMFINDER} catalog without confirmed radial velocities were omitted.}  
    \label{fig:members}
\end{figure}

Of 67 original Leiptr candidate stars from the \Gaia DR2 \texttt{STREAMFINDER} catalog \citep{Ibata19}, we selected five of the brightest candidate stars to observe, giving them arbitrary ID numbers.
We obtained spectra of these stars on November 2, 2022 using the 6.5-meter Magellan Clay Telescope’s Magellan Inamori Kyocera Echelle (MIKE; \citealt{Bernstein03}) spectrograph at Las Campanas Observatory in Chile.
Figure \ref{fig:members} shows the declinations and radial velocities against right ascensions of these stars compared to the other Leiptr candidates from the \Gaia DR3 \texttt{STREAMFINDER} catalog in \citet{Ibata24}.
The arrows represent reflex-corrected proper motions.
One of these five stars, Leiptr-252, is no longer a kinematic Leiptr member star in the \Gaia DR3 \texttt{STREAMFINDER} catalog \citep{Ibata24}, which will also be apparent from its chemistry. This non-member star is shown as a red X in Figure~\ref{fig:members} and all subsequent figures.

\begin{figure}
    \centering
    \includegraphics[height=0.31\textheight]{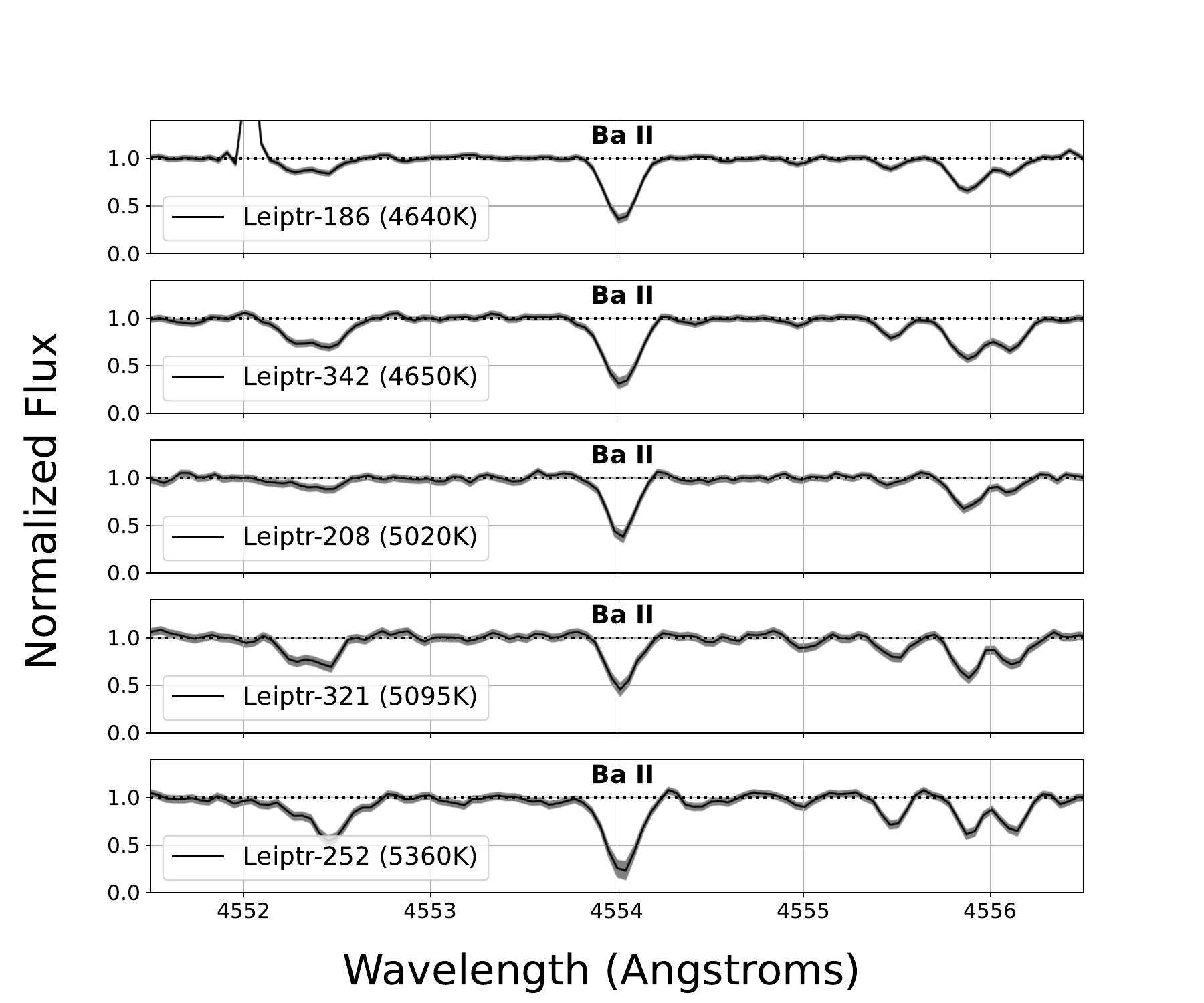}
    \includegraphics[height=0.31\textheight]{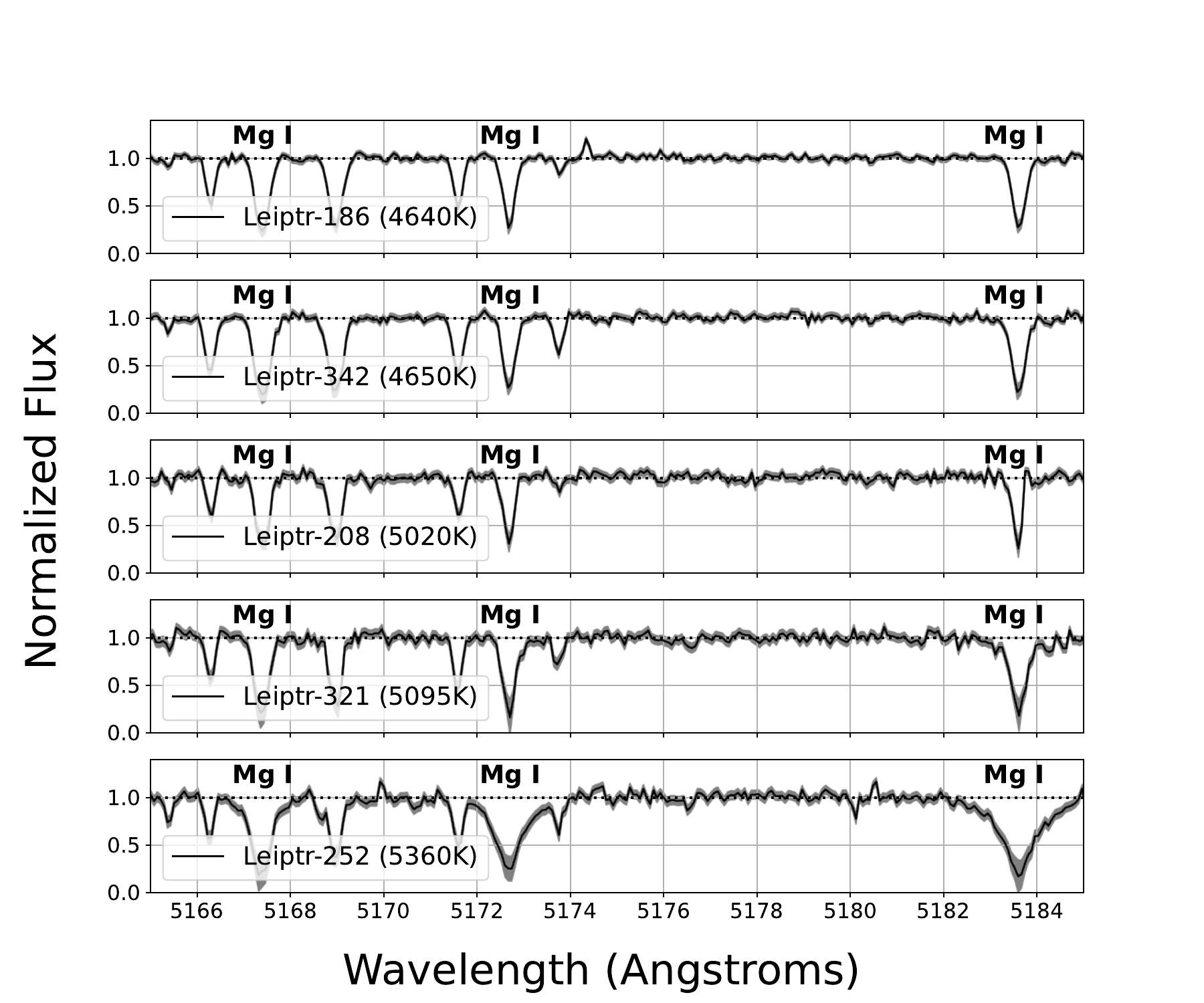}
    \includegraphics[height=0.31\textheight]{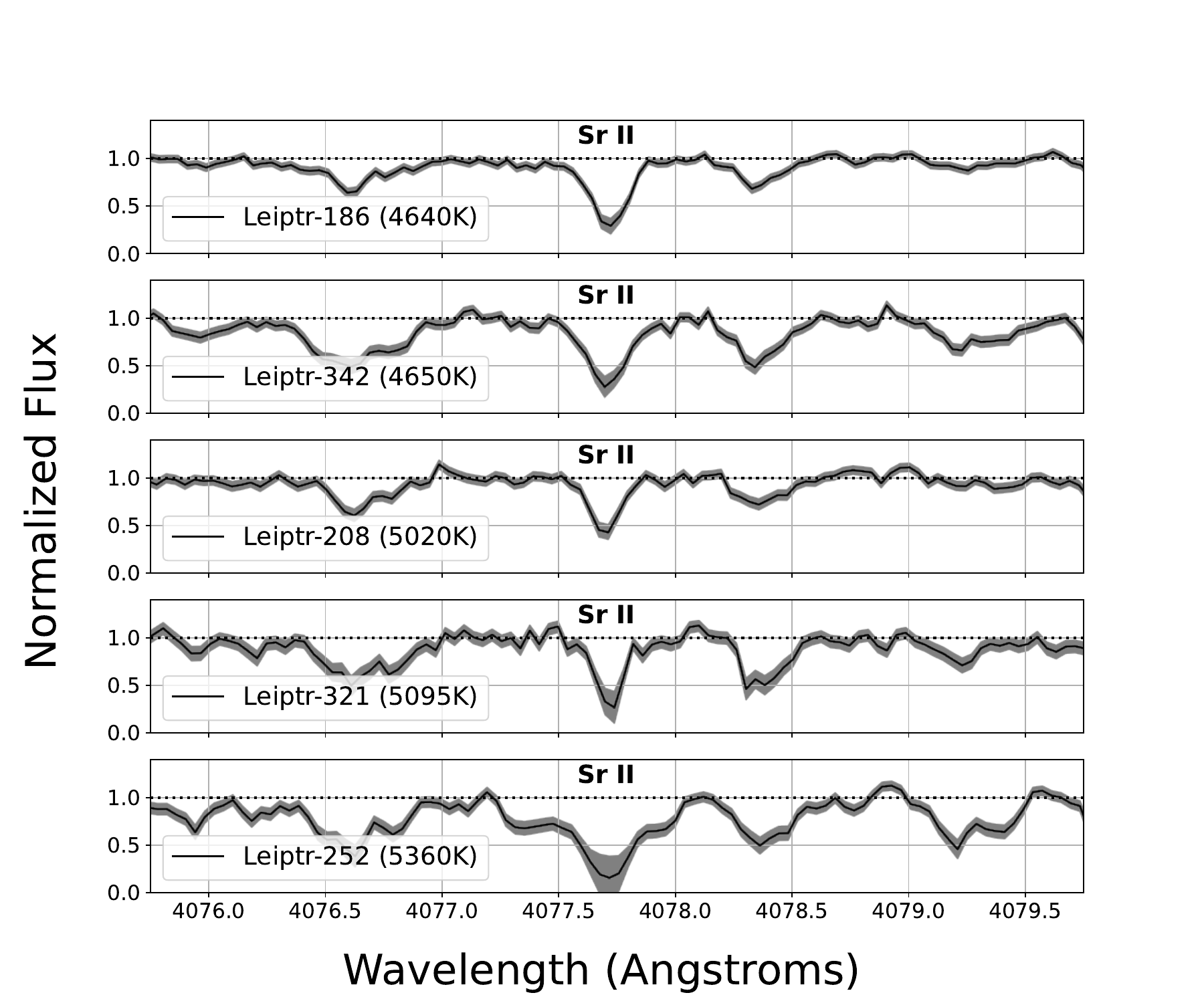}
    \caption{Spectra around the Ba II line at 4554.0~\AA; Mg I b triplet; and the Sr II line at 4077.7~\AA.
    The stars are ordered top to bottom from coolest to hottest.
    The shaded region around the spectra shows the uncertainty in normalized flux.
    Leiptr-252 is a clear outlier with stronger and broader lines.}
    \label{fig:spectra}
\end{figure}

We used the 0.7$^{\prime \prime}\times$5.0$^{\prime \prime}$ entrance slit resulting in $R \sim 35,000$ in the blue arm from $3300-5000$~\AA\ and $R \sim 28,000$ in the red arm from $5000-9000$~\AA.
The data were reduced using the CarPy software package’s MIKE reduction pipeline \citep{Kelson_2003}.
A metal-poor template spectrum of HD122563 was cross-correlated with the echelle order containing the Mg b triplet to determine radial velocities.
The uncertainty is dominated by wavelength calibration accuracy and is about 1 km s$^{-1}$ \citep{Ji20car}.
Table \ref{tab:leiptrstars} reports each star, exposure time, signal-to-noise per pixel at 4500~\AA\ and 6500~\AA, heliocentric radial velocity, right ascension, declination, and \Gaia magnitudes.
\Gaia DR3 RVS \citep{GAIA22} gives velocities of 190.43 $\pm$ 0.49 km s$^{-1}$ and 335.96 $\pm$ 0.43 km s$^{-1}$ for Leiptr-186 and Leiptr-342, respectively, which closely agree with our results.
Figure~\ref{fig:spectra} shows spectral regions around Ba II at 4554.0~\AA, the Mg b triplet, and Sr II at 4077.7~\AA.

\section{Abundance Analysis}\label{sec:Abundance Analysis}

The chemical abundance analysis was performed using a custom analysis code \code{LESSPayne} \citep{Ji25}, generally following \citet{Ji20}.
As described in \citet{Limberg2024}, \code{LESSPayne} is a combination of the PayneEchelle \citep{Ting19} and SMHR \citep{Casey14} programs.
First, we ran a full spectrum fit between $5000-6800$~\AA\ using a neural network emulator of high-resolution spectra.
We did not use the stellar parameters from this fit, but instead used the best-fit spectrum to locate and mask regions where absorption lines were expected to occur and initialize an SMHR analysis.
We manually inspected and corrected the normalization for every spectral order before stitching them together.
We then fit Gaussian profiles to determine equivalent widths using the line list from \citet{Ji20}, which is based on atomic data from \code{linemake} \citep{Placco2021}, mostly keeping to lines with 4000~\AA\ $< \lambda <$ 7000~\AA.
Some stronger lines required Voigt profile fits.
We also fit some lines using spectral synthesis with the same line lists as \citet{Ji20}.

We determined chemical abundances using the 1D local thermodynamic equilibrium (LTE) radiative transfer code \code{MOOG} \citep{Sneden73} and the ATLAS model atmospheres \citep{Castelli03}.
Stellar parameters were determined spectroscopically following \citet{Frebel13}, due to uncertain distances to the Leiptr stars.
Effective temperature was derived by balancing Fe I abundance against excitation potential, surface gravity by balancing Fe I and II abundances, and microturbulence by balancing Fe II abundance against reduced equivalent width.
Next, we applied the correction to a photometric temperature scale from \citet{Frebel13} and redetermined the other stellar parameters.
We estimate the microturbulence using Fe II lines because of the photometric correction: using Fe I lines computed in LTE with photometric temperatures results in artificially high microturbulence \citep{Ji20}.
This is possible as we used ${\sim}20$ Fe\,II lines per star.
The corrected effective temperature ($T_{\text{eff}}$), surface gravity (log g), microturbulence ($v_t$), metallicity ([Fe/H]), and their associated uncertainties ($e_{T}$, $e_{g}$, $e_{v}$, and $e_{Fe}$, respectively) are given in Table \ref{tab:stelparams}.
\teff systematic uncertainty was estimated at 150 K due to intrinsic scatter in the temperature correction from \citet{Frebel13}.
The corresponding systematic uncertainties for the remaining stellar parameters, \logg, $v_t$, and \feh, were set to 0.3 dex, 0.2 km s$^{-1}$, and 0.2 dex, respectively.
These are added in quadrature to the statistical uncertainties in the slopes and standard errors of the spectroscopic balance to determine the total stellar parameter uncertainties \citep[see][for details]{Ji20}.

\begin{deluxetable*}{cccccccccc}
\tablecolumns{9}
\tablecaption{\label{tab:stelparams}Leiptr Candidate Stellar Parameters and Uncertainties}
\tablehead{Name & $T_{\text{eff}}$ & $e_{T}$ & $T_{\text{eff, phot}}$ & \logg & $e_{g}$ & $v_t$ & $e_{v}$ & [Fe/H] & $e_{\text{Fe}}$}
\startdata
Leiptr-186 & 4640 & 152 & 4607 & 1.00 & 0.31 & 2.51 & 0.24 & $-2.54$ & 0.21 \\
Leiptr-208 & 5020 & 157 & 5126 & 1.90 & 0.31 & 1.85 & 0.24 & $-2.25$ & 0.21 \\
Leiptr-252 & 5360 & 163 & 5329 & 3.60 & 0.31 & 1.31 & 0.25 & $-1.20$ & 0.33 \\
Leiptr-321 & 5095 & 160 & 5059 & 2.40 & 0.31 & 1.95 & 0.27 & $-1.80$ & 0.22 \\
Leiptr-342 & 4650 & 153 & 4513 & 1.00 & 0.30 & 2.32 & 0.25 & $-1.93$ & 0.21 \\
\enddata
\tablecomments{Temperatures are given in K and microturbulence in km s$^{-1}$.
$T_{\text{eff}}$ is spectroscopic effective temperature with the correction from \citet{Frebel13}, log g is surface gravity, and $v_t$ is microturbulence. 
We also include the photometric temperature $T_{\text{eff, phot}}$, as calculated using \citet{Mucciarelli2017} color-temperature relations for \textit{Gaia} BP-RP colors.
The $e$ values represent the stellar parameters' respective uncertainties, which are a quadrature sum of the statistical uncertainty calculated from the iron lines and adopted systematic uncertainties for Teff, \logg, $v_t$, and \feh of 150 K, 0.3 dex, 0.2 km s$^{-1}$, and 0.2 dex, respectively. \vspace{-0.4cm}}
\end{deluxetable*}

Overall, we detected or put upper limits on 29 species of 24 elements, where CH, CN, Al I, Sc II, V I, V II, Mn I, Co I, Sr II, Y II, Ba II, La II, and Eu II were synthesized and O I, Na I, Mg I, K I, Ca I, Ti I, Ti II, Cr I, Cr II, Fe I, Fe II, Ni I, Zn I, Sr I, Zr II, and Dy II were determined with equivalent widths.
Mean abundances for each species are computed with a weighted average of individual absorption lines and compared to their respective solar abundances from \citet{Asplund2009}.

For a given absorption line $i$ for element $X$, we measure an abundance $\log \epsilon_i(X)$.
The measured abundance is uncertain, due to uncertainties in the stellar parameters and the measurement itself.
We therefore measure abundance uncertainties for each absorption line by independently propagating the stellar parameter uncertainties, denoted $e_{{T_{\text{eff}},i}}$, $e_{g,i}$, $e_{M,i}$, and $e_{v,i}$.
Additionally, uncertainty in the measured equivalent width propagates into an abundance uncertainty denoted $e_{stat,i}$.

To measure the overall abundance for the star, we use a weighted average of measured abundances from all absorption lines.
Weighting is introduced in order to de-prioritize measurements with high errors to most accurately represent the measured abundance of the star.
The weighted abundance $\log \epsilon_w(X)$ is calculated as:
\begin{equation}
    \log \epsilon_w(X) = \frac{\Sigma_i (\log \epsilon_i(X) \times w_i)}{\Sigma_i w_i}.
\end{equation}
The weight $w_i$ for a measured line $i$ reflects the total uncertainty of a measurement by combining all of the measured uncertainties described above.
The weight is calculated as the inverse of the quadrature sum of each contributing source of uncertainty:
\begin{equation}
    \frac{1}{w_i} = e_{{T_{\text{eff}}}, i}^2 + e_{{g,i}}^2 + e_{M, i}^2 + e_{{v}, i}^2 + e_{{stat}, i}^2.
\end{equation}

We also calculate the total standard error, $\sigma^2_{\text{stat}}$, which represents the contributed systematic and statistical uncertainties from each individual line measurement to the overall error of the element abundance, and is calculated as:
\begin{equation}\label{eq1}
\sigma^2_{\text{stat}} = \frac{1}{\Sigma_i \left(e^2_{\text{sys},i}+e^2_{\text{stat},i}\right)^{-1}}.
\end{equation}
We set the per-line systematic error floor $e_{{sys}, i} = 0.1$ to account for unforeseen systematics in our stellar models and inferred abundance measurements.

We calculate the variance of our measurements, and continue to weight the contribution of each measurement, to appropriately estimate the systematic uncertainty in line-to-line scatter.
The square of the weighted standard deviation $\sigma^2_w$ is calculated as:
\begin{equation}
    \sigma_w^2(X) = \frac{\Sigma_i \left( w_i \times (\log \epsilon_i(X) - \log \epsilon_w(X))^2\right)}{\Sigma_i w_i}.
\end{equation}

We also calculate the contribution of the uncertainty relative to each stellar parameter.
The propagated stellar parameter uncertainty, $\sigma^2_{\text{SP}}$, is:
\begin{equation} \label{eSP}
\begin{split}
\sigma^2_{\text{SP}} & =  \left(\frac{\Sigma_i (e_{\teff, i} \times w_i)}{\Sigma_i w_i}\right)^2 + \left(\frac{\Sigma_i (e_{g, i} \times w_i)}{\Sigma_i w_i}\right)^2 \\
 &~~~ + \left(\frac{\Sigma_i (e_{v, i} \times w_i)}{\Sigma_i w_i}\right)^2 + \left(\frac{\Sigma_i (e_{M, i} \times w_i)}{\Sigma_i w_i}\right)^{2}.\\
\end{split}
\end{equation}
Each of these terms is treated as the total uncertainty of the element $X$ due to the corresponding stellar parameter, e.g.:
\begin{equation}
    e_{\teff} = \frac{\Sigma_i (e_{\teff, i} \times w_i)}{\Sigma_i w_i}.
\end{equation}
Here, we neglect correlations between stellar parameters. 
\citet{Ji20} gave a formalism for how to propagate correlated stellar parameter uncertainties to abundances, but it required the assumption that abundances depend linearly on stellar parameters.
We have unfortunately found that this assumption often breaks down, which can result in systematic biases in the resulting abundances.
By ignoring the correlations, we find that we slightly overestimate the stellar parameter uncertainties, which we find preferable to obtaining biased abundances.

The total uncertainty for the abundance of element $X$ in the star, [\textit{X}/H], combines all previous uncertainties. 
$\sigma_{\text{[X/H]}}$ is the quadrature sum of the total standard error, the stellar parameter uncertainties, and the weighted standard deviation:
\begin{equation}
\sigma_{\text{[X/H]}} = \sqrt{\sigma^2_{\text{stat}} + \sigma^2_{\text{SP}} + \max\left(\sigma_w^2, 0.1^2 \right)}.
\end{equation}
We include a minimum floor of $0.1^2$ on the weighted variance of measurements to account for elements with too few lines to measure a standard deviation.
In our tables and figures, we also include the abundance uncertainty excluding this systematic uncertainty buffer and weighted variance:
\begin{equation}
\sigma_{\text{[X/H]}, \text{ no sys}} = \sqrt{\sigma^2_{\text{stat}} + \sigma^2_{\text{SP}}}.
\end{equation}
We present both sets of uncertainties, one that gives our best estimate of a total uncertainty, and the other to estimate abundance precision.

We use [Fe~I/H] to calculate [X/Fe].
The uncertainty of [X/Fe] needs to account for statistical uncertainties for both X and Fe, in addition to correlated abundance uncertainties from stellar parameters and the weighted variance with corresponding uncertainty floors:
\begin{equation}
\begin{split}
\sigma^2_{\text{[X/Fe]}} & = \sigma^2_{\text{stat}} + \sigma^2_{\text{stat,Fe}} \\
&~~~+(e_{\teff} - e_{\teff, \text{Fe}})^2+(e_{g} - e_{g, \text{Fe}})^2 \\
&~~~+(e_{v} - e_{v, \text{Fe}})^2+(e_M - e_{M, \text{Fe}})^{2} \\
&~~~+\max\left(\sigma_w^2, 0.1^2 \right) + \max\left(\sigma_{w,\text{Fe}}^2, 0.1^2 \right).
\end{split}
\end{equation}
Again, we exclude the uncertainty floor to give an approximate error without a conservative error buffer:
\begin{equation}
\begin{split}
\sigma^2_{\text{[X/Fe], no sys}} & = \sigma^2_{\text{stat}} + \sigma^2_{\text{stat,Fe}} \\
&~~~+(e_{\teff} - e_{\teff, \text{Fe}})^2+(e_{g} - e_{g, \text{Fe}})^2 \\
&~~~+(e_{v} - e_{v, \text{Fe}})^2+(e_M - e_{M, \text{Fe}})^{2}.
\end{split}
\end{equation}

\begin{figure}
    \centering
    \includegraphics[width=\linewidth]{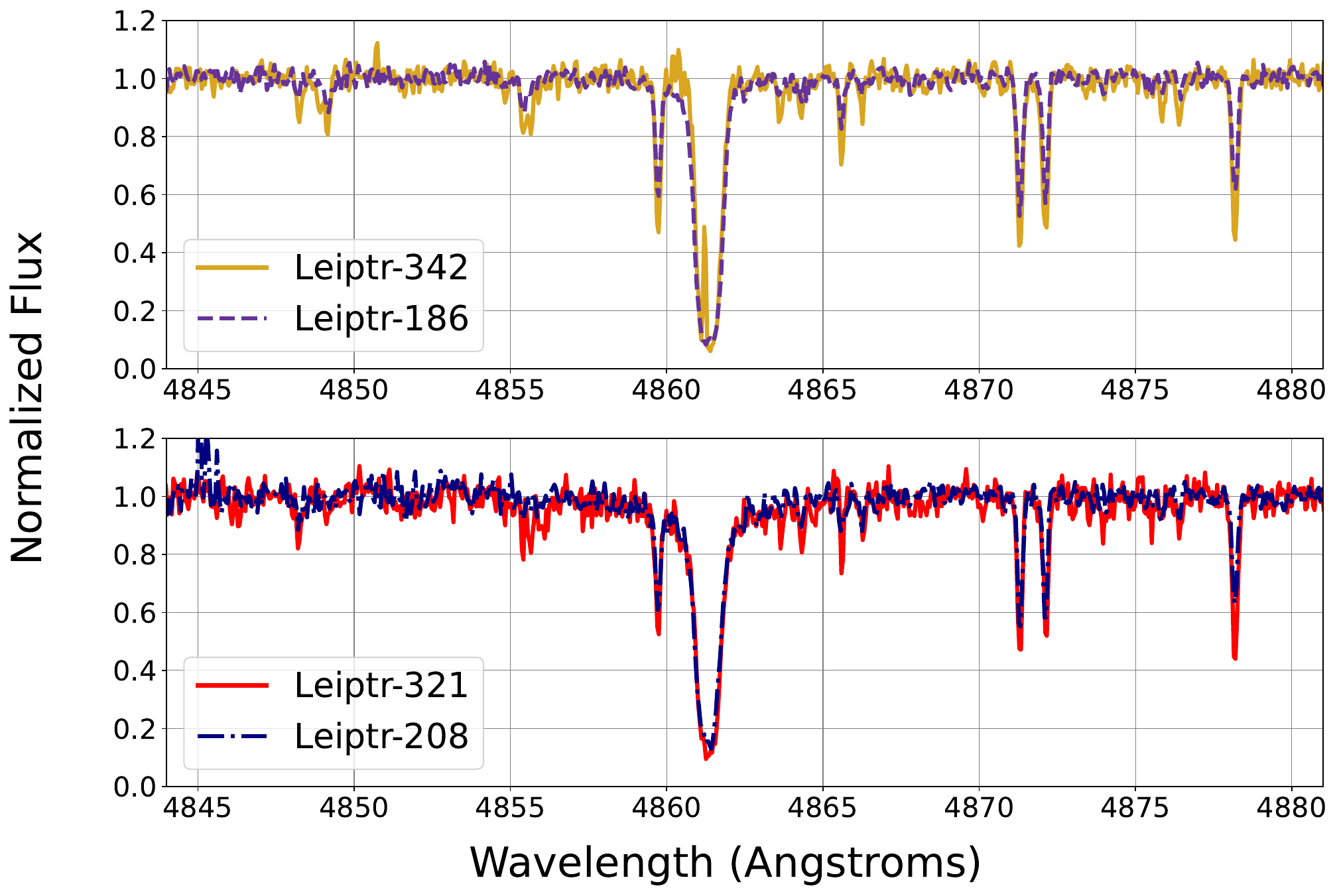}
    \caption{A comparison of normalized spectra of stars with similar stellar parameters around the H$\beta$ line at 4861 \AA.
    Additional absorption lines are shown adjacent to the central line.
    The top plot has the stars Leiptr-186 (purple dashed line) and Leiptr-342 (gold solid line), both with \teff around 4650 K.
    The bottom plot has stars Leiptr-208 (blue dot-dashed line) and Leiptr-321 (red solid line), both with \teff around 5050 K.
    The similarity of the hydrogen line profiles suggest each pair of stars has very similar temperatures; but Leiptr-321 and Leiptr-342 have stronger metal lines than Leiptr-208 and Leiptr-186, respectively, indicating their metallicities must differ.}
    \label{fig:hbeta_spec}
\end{figure}

\begin{figure}
    \centering
    \includegraphics[width=\linewidth]{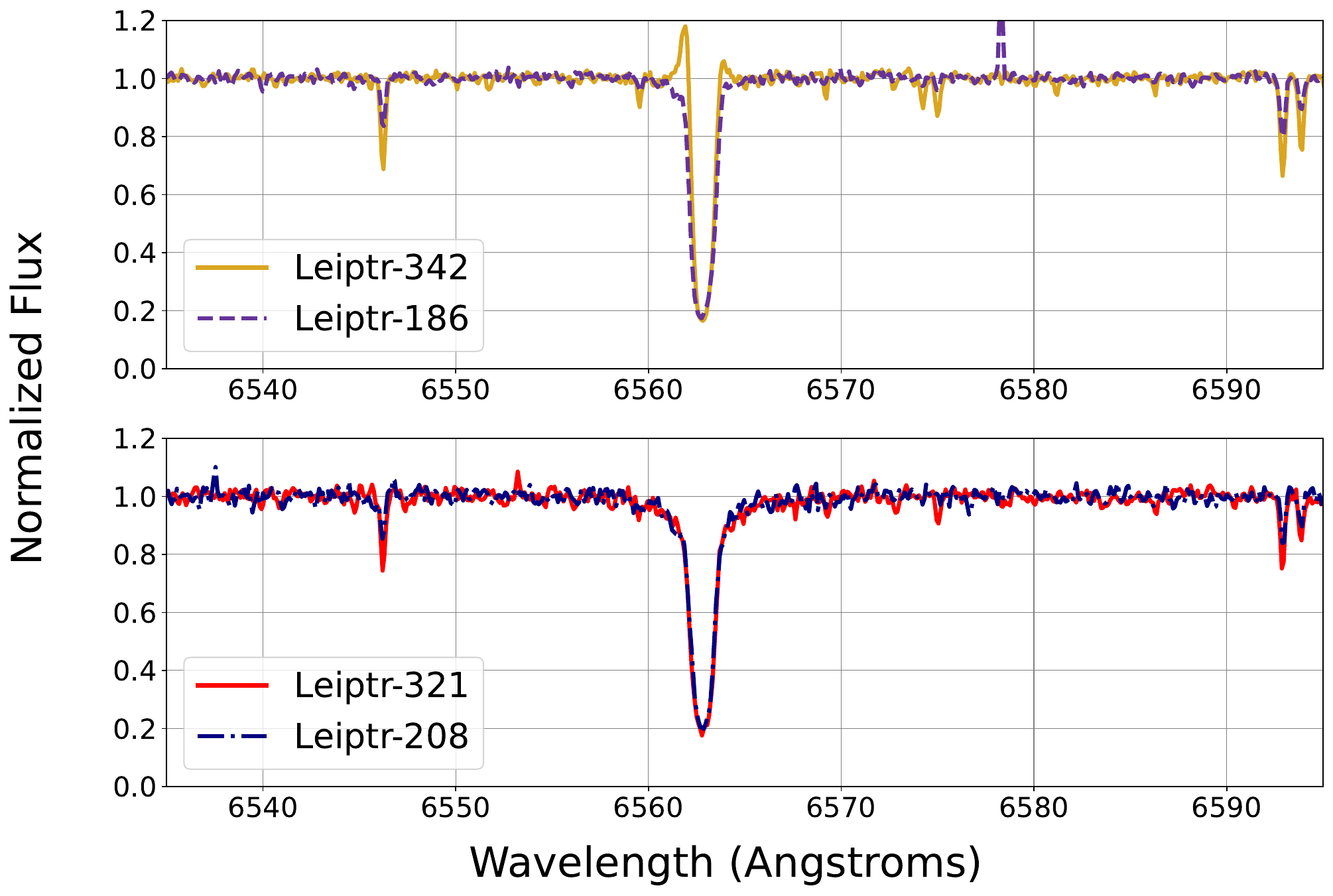}
    \caption{A comparison of normalized spectra of stars with similar stellar parameters around the H$\alpha$ line at 6563 \AA.
    Additional absorption lines are shown adjacent to the central line.
    The top plot has the stars Leiptr-186 (purple dashed line) and Leiptr-342 (gold solid line), both with \teff around 4650 K.
    The bottom plot has stars Leiptr-208 (blue dot-dashed line) and Leiptr-321 (red solid line), both with \teff around 5050 K.
    The similarity of the hydrogen line profiles suggest each pair of stars has very similar temperatures; but Leiptr-321 and Leiptr-342 have stronger metal lines than Leiptr-208 and Leiptr-186, respectively, indicating their metallicities must differ.}
    \label{fig:halpha_spec}
\end{figure}

\begin{figure*}
    \centering
    \includegraphics[width=\linewidth]{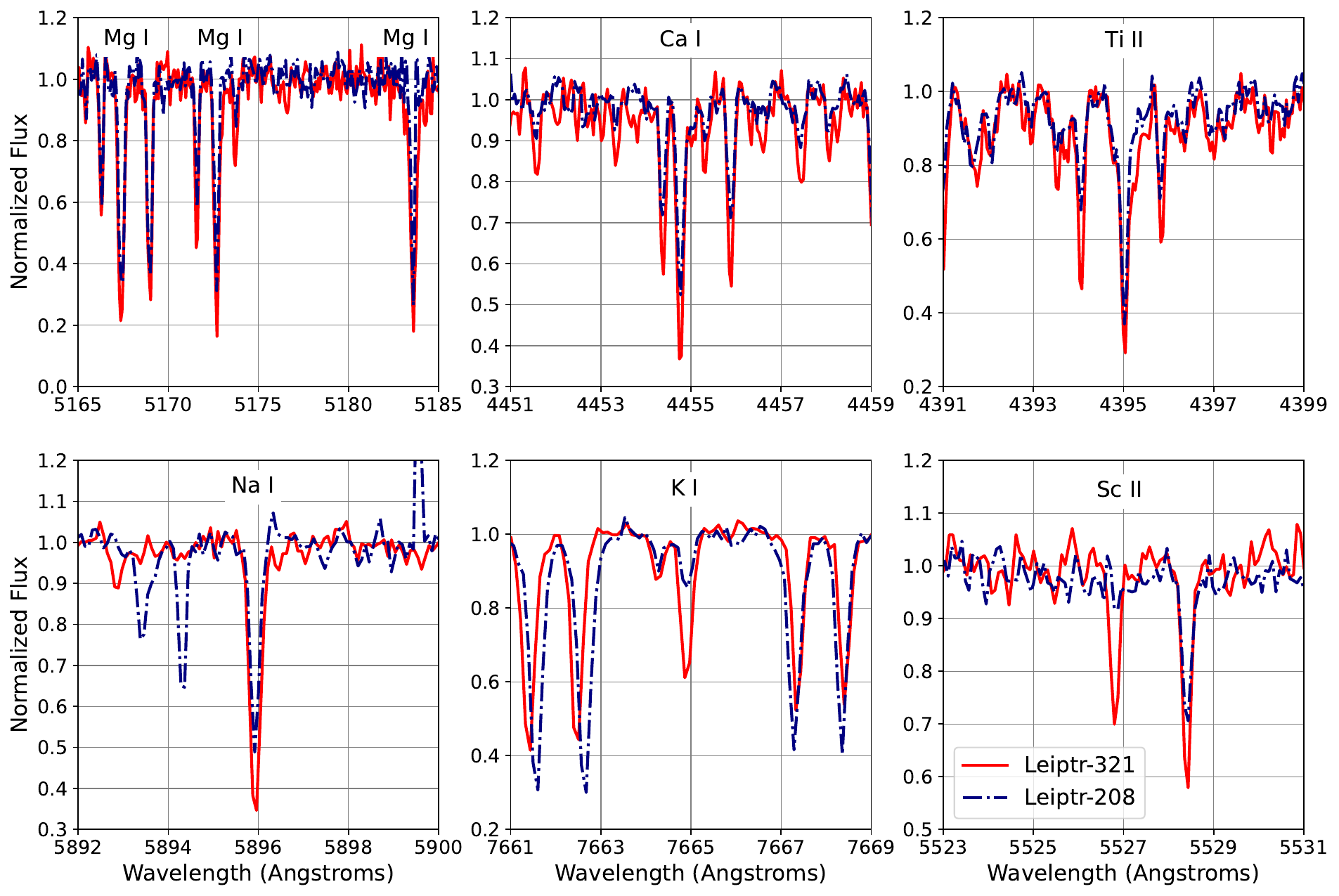}
    \caption{A comparison of normalized rest-frame spectra of Leiptr-321 and Leiptr-208 that have similar stellar parameters, centered around the Mg I b triplet; Ca I line at 4454.8 \AA; Ti II line at 4395.0 \AA; Na I line at 5895.9 \AA; K I line at 7664.9 \AA; and Sc II line at 5526.8 \AA.
    Additional absorption lines are shown adjacent to the central line, including interstellar medium and telluric features in the sodium and potassium panels, respectively.
    The telluric features around K I are offset because the stars are shifted to rest frame.
    The plots have stars Leiptr-208 (blue dot-dashed line) and Leiptr-321 (red solid line), both with \teff around 5050 K.
    It is clear that Leiptr-321 has overall higher metallicity, as well as much higher abundances of sodium, potassium, and scandium.}
    \label{fig:a+oddz_spec}
\end{figure*}

\begin{figure}
    \centering
    \includegraphics[width=\linewidth]{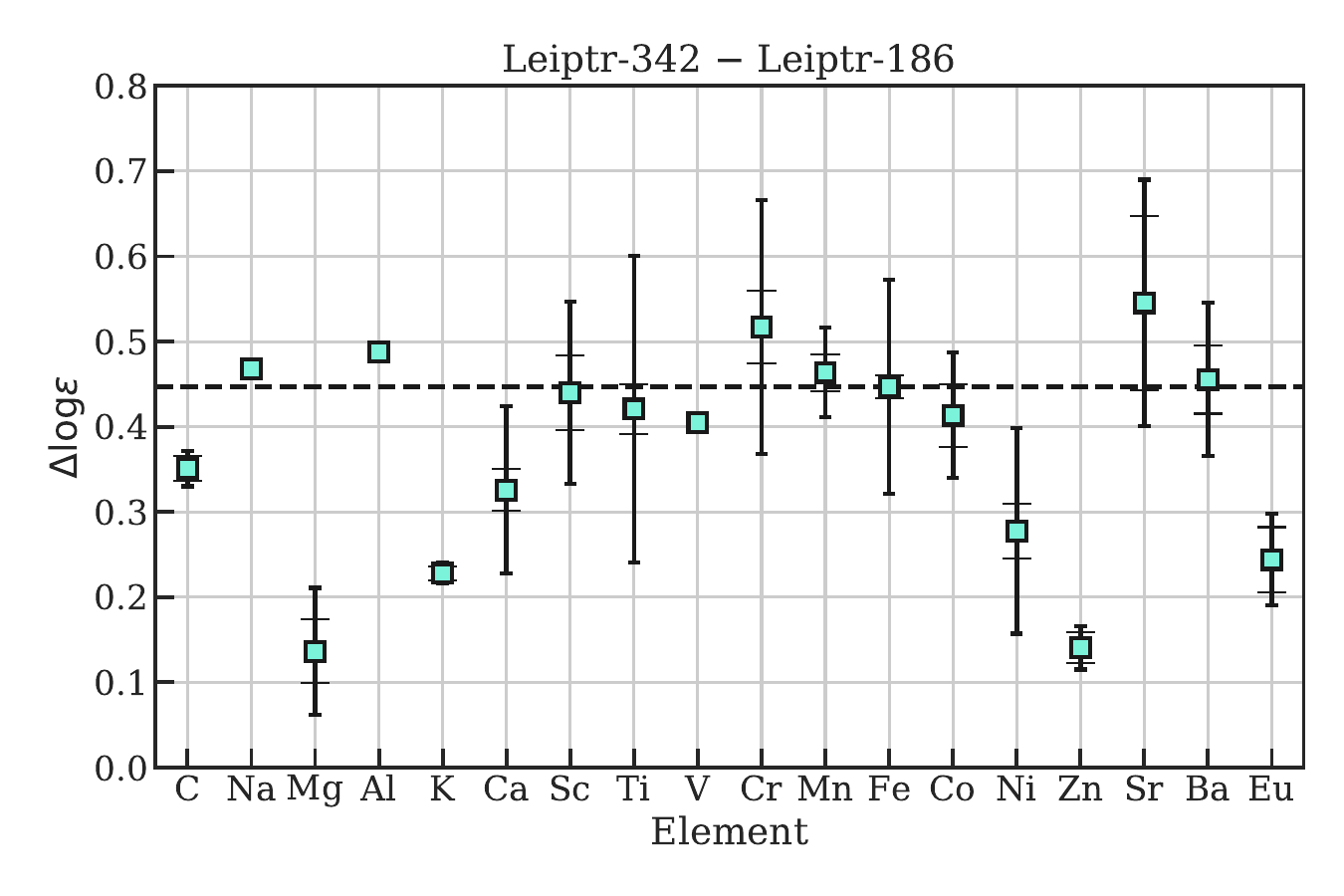}
    \includegraphics[width=\linewidth]{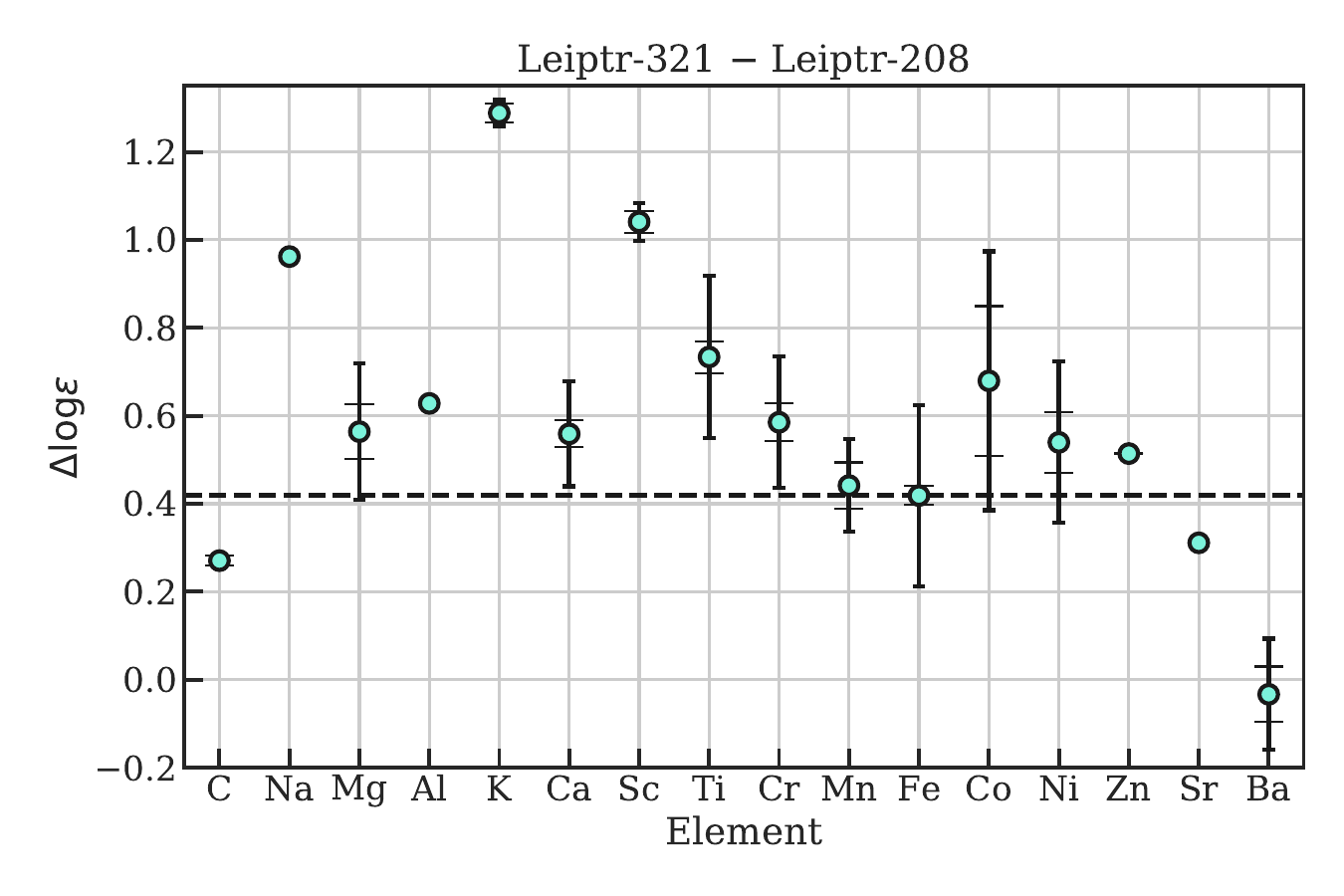}
    \caption{Abundance differences between matched pairs of stars with similar temperatures.
    The top panel shows the difference between the cooler pair, Leiptr-342 and Leiptr-186, indicated by light-blue squares in this and future figures.
    The bottom panel shows the difference between the warmer pair, Leiptr-321 and Leiptr-208, indicated by light-blue circles in this and future figures.
    The longer error bars with smaller cap sizes show the standard deviation. The shorter error bars with larger cap sizes show the standard error. No error bar is shown if just one line is used.
    A horizontal dashed line is shown corresponding to $\Delta \log \epsilon(\text{Fe})$; differences of other elements relative to this line indicate $\Delta\mbox{[X/Fe]}$.
    }
    \label{fig:diffabund}
\end{figure}

We present the start of a list of line measurements for the Leiptr candidate stars in Table \ref{tab:lines} which provides each line's wavelength, excitation potential, oscillator strength, equivalent width, and full width at half maximum. 
The full table is available in the arXiv source as a CSV file. 
The SMHR analysis files that include the spectra and line fits are available upon request to the first author.
The final abundances for our five stars are given in Tables \ref{tab:abunds186}-\ref{tab:abunds342}.
We record element, number of lines used, and log $\epsilon$, as well as [X/H] and [X/Fe] and their respective uncertainties, $\sigma_{\text{[X/H]}}$ and $\sigma_{\text{[X/Fe]}}$, for each star.
$\text{[X/Fe]}_{\text{N}}$ gives the non-LTE (NLTE) abundances for sodium, aluminum, potassium, and manganese (see Sections \ref{sec:oddz} and \ref{sec:fepeak}).
We comment on the NLTE effects for all significantly impacted elements, ignoring smaller contributions that are not informative to this analysis.

\begingroup
\setlength{\tabcolsep}{3pt} 

\begin{deluxetable*}{cccccccccccccccc}
\tablecolumns{16}
\tablecaption{\label{tab:lines}Leiptr Candidate Line List}
\tablehead{Star Name & Element & Species & Wavelength (\AA) & expot & loggf & logeps & $e_{stat}$ & $e_{sys}$ & eqw & $e_{eqw}$ & FWHM & $e_{T_{\text{eff}}}$ & $e_g$ & $e_v$ & $e_M$}
\startdata
Leiptr-186 & Mg I & 12.0 & 3986.75 & 4.35 & $-1.06$ & 5.27 & 0.06 & 0.10 & 36.5 & 3.7 & 0.14 & 0.09 & $-0.02$ & $-0.01$ & $-0.00$ \\
Leiptr-186 & Fe I & 26.0 & 4001.66 & 2.17 & $-1.90$ & 4.92 & 0.04 & 0.10 & 57.9 & 2.8 & 0.13 & 0.19 & $-0.03$ & $-0.03$ & $-0.02$ \\
Leiptr-186 & Ti I & 22.1 & 4012.38 & 0.57 & $-1.78$ & 2.58 & 0.08 & 0.10 & 118.2 & 3.8 & 0.20 & 0.07 & 0.08 & $-0.16$ & 0.03 \\
Leiptr-186 & Ti I & 22.1 & 4025.13 & 0.61 & $-2.11$ & 2.47 & 0.07 & 0.10 & 94.6 & 3.7 & 0.18 & 0.08 & 0.09 & $-0.09$ & 0.03 \\
... & ... & ... & ... & ... & ... & ... & ... & ... & ... & ... & ... & ... & ... & ... & ...\\
\enddata
\tablecomments{This table is available in its entirety in machine-readable form as part of the arXiv source (\code{leiptr\_lines.csv}).
expot is excitation potential; loggf is oscillator strength; logeps is the absolute stellar abundance to H; $e_{stat}$ is the statistical abundance uncertainty due to the equivalent width uncertainty; $e_{sys}$ is the systematic uncertainty, which we adopt to be 0.1 for all lines; eqw is equivalent width; $e_{eqw}$ is uncertainty in equivalent width; FWHM is the full width of the line at half maximum; $e_{T_{\text{eff}}}$ is the difference on the abundance due to $1\sigma$ uncertainties on effective temperature, and similarly for $e_g$ for the surface gravity, $e_{v}$ for the microturbulance, and $e_{M}$ for the metallicity.}
\end{deluxetable*}

\endgroup

\begingroup
\setlength{\tabcolsep}{3pt} 

\begin{deluxetable}{cccccccc}
\tablecolumns{8}
\tablecaption{\label{tab:abunds186}Leiptr-186 Abundances}
\tablehead{Element & N & log $\epsilon$ & [X/H] & $\sigma_{\text{[X/H]}}$ & [X/Fe] & $\text{[X/Fe]}_{\text{N}}$ & $\sigma_{\text{[X/Fe]}}$}
\startdata
CH & 2 & 5.38 & $-3.05$ & 0.44 & $-0.50$ & -- & 0.36 \\
CN & 1 & 6.12 &  $< -1.70$ & -- & $< 0.84$ & -- & -- \\
Na I & 2 & 3.52 & $-2.72$ & 0.33 & $-0.18$ & $-0.41$ & 0.27 \\
Mg I & 9 & 5.22 & $-2.38$ & 0.16 & 0.17 & -- & 0.18 \\
Al I & 1 & 3.06 & $-3.39$ & 0.40 & $-0.84$ & $-0.24$ & 0.41 \\
K I & 2 & 2.87 & $-2.16$ & 0.21 & 0.39 & $-0.11$ & 0.20 \\
Ca I & 23 & 3.92 & $-2.42$ & 0.16 & 0.13 & -- & 0.19 \\
Sc II & 8 & 0.35 & $-2.80$ & 0.17 & $-0.26$ & -- & 0.27 \\
Ti I & 29 & 2.38 & $-2.57$ & 0.25 & $-0.02$ & -- & 0.20 \\
Ti II & 49 & 2.54 & $-2.41$ & 0.17 & 0.13 & -- & 0.26 \\
V I & 2 & 1.13 & $-2.80$ & 0.28 & $-0.25$ & -- & 0.25 \\
V II & 2 & 1.28 & $-2.65$ & 0.19 & $-0.11$ & -- & 0.31 \\
Cr I & 14 & 2.85 & $-2.79$ & 0.23 & $-0.24$ & -- & 0.18 \\
Cr II & 4 & 2.96 & $-2.68$ & 0.14 & $-0.14$ & -- & 0.29 \\
Mn I & 6 & 2.51 & $-2.92$ & 0.18 & $-0.38$ & 0.05 & 0.18 \\
Fe I & 124 & 4.95 & $-2.55$ & 0.23 & 0.00 & -- & -- \\
Fe II & 21 & 4.96 & $-2.54$ & 0.14 & -- & -- & -- \\
Co I & 4 & 2.39 & $-2.60$ & 0.36 & $-0.05$ & -- & 0.31 \\
Ni I & 19 & 3.67 & $-2.55$ & 0.21 & $-0.00$ & -- & 0.20 \\
Zn I & 2 & 2.15 & $-2.41$ & 0.14 & 0.14 & -- & 0.22 \\
Sr II & 2 & $-0.73$ & $-3.60$ & 0.32 & $-1.05$ & -- & 0.34 \\
Ba II & 5 & $-1.26$ & $-3.44$ & 0.20 & $-0.90$ & -- & 0.26 \\
Eu II & 2 & $-2.30$ & $-2.82$ & 0.22 & $-0.28$ & -- & 0.30 \\
\enddata
\tablecomments{CN is an upper limit.
$\text{[X/Fe]}_{\text{N}}$ gives the NLTE abundances.}
\end{deluxetable}

\endgroup

\begingroup
\setlength{\tabcolsep}{3pt} 

\begin{deluxetable}{cccccccc}
\tablecolumns{8}
\tablecaption{\label{tab:abunds208}Leiptr-208 Abundances}
\tablehead{Element & N & log $\epsilon$ & [X/H] & $\sigma_{\text{[X/H]}}$ & [X/Fe] & $\text{[X/Fe]}_{\text{N}}$ & $\sigma_{\text{[X/Fe]}}$}
\startdata
CH & 2 & 6.4 & $-2.03$ & 0.26 & 0.22 & -- & 0.24 \\
CN & 1 & 6.08 & $< -1.75$ & -- & $< 0.51$ & -- & -- \\
Na I & 2 & 3.68 & $-2.56$ & 0.23 & $-0.30$ & $-0.70$ & 0.23 \\
Mg I & 6 & 5.29 & $-2.31$ & 0.15 & $-0.05$ & -- & 0.18 \\
Al I & 1 & 3.22 & $-3.23$ & 0.37 & $-0.97$ & $-0.37$ & 0.38 \\
K I & 2 & 2.86 & $-2.17$ & 0.13 & 0.09 & $-0.21$ & 0.17 \\
Ca I & 16 & 4.16 & $-2.18$ & 0.12 & 0.08 & -- & 0.16 \\
Sc II & 7 & 0.50 & $-2.65$ & 0.15 & $-0.39$ & -- & 0.21 \\
Ti I & 20 & 2.74 & $-2.21$ & 0.18 & 0.05 & -- & 0.20 \\
Ti II & 32 & 2.72 & $-2.23$ & 0.18 & 0.03 & -- & 0.23 \\
V II & 2 & 1.64 & $-2.29$ & 0.14 & $-0.03$ & -- & 0.23 \\
Cr I & 12 & 3.15 & $-2.49$ & 0.16 & $-0.23$ & -- & 0.18 \\
Cr II & 3 & 3.19 & $-2.45$ & 0.15 & $-0.19$ & -- & 0.23 \\
Mn I & 4 & 2.68 & $-2.75$ & 0.13 & $-0.49$ & $-0.12$ & 0.17 \\
Fe I & 111 & 5.24 & $-2.26$ & 0.16 & 0.00 & -- & -- \\
Fe II & 22 & 5.21 & $-2.29$ & 0.17 & -- & -- & -- \\
Co I & 3 & 2.68 & $-2.31$ & 0.21 & $-0.06$ & -- & 0.23 \\
Ni I & 11 & 3.85 & $-2.37$ & 0.14 & $-0.11$ & -- & 0.18 \\
Zn I & 2 & 2.23 & $-2.33$ & 0.11 & $-0.07$ & -- & 0.18 \\
Sr II & 2 & $-0.56$ & $-3.43$ & 0.45 & $-1.17$ & -- & 0.46 \\
Ba II & 5 & $-0.83$ & $-3.01$ & 0.18 & $-0.75$ & -- & 0.22 \\
Eu II & 1 & $-1.71$ & $< -2.23$ & -- & $< 0.03$ & -- & -- \\
\enddata
\tablecomments{CN and Eu II are upper limits.
$\text{[X/Fe]}_{\text{N}}$ gives the NLTE abundances.}
\end{deluxetable}

\endgroup

\subsection{Differential Abundance Analysis} \label{sec:diffabund}

Our overall abundance uncertainties are dominated by stellar parameter uncertainties, in particular the systematic 150 K temperature uncertainty set from the \citet{Frebel13} calibration.
However, abundance \textit{differences} between two stars of similar stellar parameters can be much more precise \citep[e.g.,][]{McWilliam2013,Nissen2018,Matsuno2022,McKenzie2022}. 
The four Leiptr member stars come in two pairs of stars with similar effective temperatures and surface gravities: Leiptr-186 and Leiptr-342 have \teff $\approx$ 4650 K and Leiptr-208 and Leiptr-321 have \teff $\approx$ 5050 K.
This similarity is clearly visible in the spectra: Figures~\ref{fig:hbeta_spec} and \ref{fig:halpha_spec} show the H$\beta$ and $\alpha$ lines.
The Balmer lines in Leiptr-186 are very similar to Leiptr-342 except that Leiptr-342 has some hydrogen emission; and Leiptr-321 and Leiptr-208 have identical Balmer line shapes.
It is also visually clear that the narrow metal absorption lines in Leiptr-321 and Leiptr-342 are substantially stronger than those in Leiptr-208 and Leiptr-186, respectively, suggesting each pair of stars in fact has different compositions.
Figure \ref{fig:a+oddz_spec} shows several significant differences between the spectra for Leiptr-321 and Leiptr-208 when we zoom in on metal absorption lines for Mg I, Ca I, Ti II, Na I, K I, and Sc II.

We thus conduct a differential abundance analysis between these two pairs of stars to examine whether a more precise analysis can distinguish between the star compositions.
Adopting the same stellar parameters as our main analysis, we calculate abundance differences for matched lines between the two pairs of stars using Table~\ref{tab:lines}\footnote{Note that differential abundance analyses in the literature usually use one reference star to compare to all other stars, but in this case we are simply looking at two pairs of stars with similar stellar parameters.}.
Unlike the initial abundance analysis, we do not apply any weighting for the differential analysis.
We estimate uncertainties by taking the standard deviation and standard error of the line abundance differences, and not propagating any stellar parameter uncertainties that we expect to largely cancel out in this differential analysis.

The results are shown in Figure~\ref{fig:diffabund}. Matching the visual impression from Figures~\ref{fig:hbeta_spec}, \ref{fig:halpha_spec}, and \ref{fig:a+oddz_spec}, each pair of stars differs by over 0.4 dex in iron abundance, clearly demonstrating that the Leiptr stars do not all have the same chemical composition.
There are also several other [X/Fe] differences (e.g., lower magnesium, europium, and zinc in Leiptr-342 vs Leiptr-186 and higher sodium, potassium, and scandium in Leiptr-321 vs Leiptr-208) that will be discussed in the next section.

\section{Abundance Results}\label{sec:Results}

We first discuss the overall metallicity dispersion in the Leiptr stream in Section \ref{sec:metallicity}.
Then, we discuss the abundance trends for halo, dwarf galaxy, and globular cluster stars in each element group (Sections \ref{sec:cn}-\ref{sec:nc}) and compare to our results for Leiptr.
Total uncertainty is shown in black for accuracy and abundance uncertainty without additional systematic errors is shown in the same color as the stars for statistical precision as described in Section \ref{sec:Abundance Analysis}.
Our literature sample is compiled based on references whose analysis process is very similar to ours (i.e., high-resolution optical spectroscopy of individual stars, mostly with spectroscopic stellar parameters).
Abundances for stars in the metal-poor Milky Way halo come from selected references in JINAbase (\citealt{Abohalima18}, from \citealt{Barklem2005,Cohen2013,Roederer14,Jacobson2015}) and abundances of GSE stars come from \citet{Carrillo22}.
We take stellar masses of dwarf galaxies from the compilation by \citet{Simon19}.
Massive dwarf galaxy ($M_\star \gtrsim 10^7 M_\odot$) abundances come from \citealt{Lemasle14} (Fornax), \citealt{Jonsson} (Sagittarius, APOGEE stars), and \citealt{Hill19} (Sculptor).
Intermediate-mass dwarf galaxy ($M_\star \gtrsim 10^6 M_\odot$) abundances come from \citealt{Norris17} (Carina), \citealt{Cohen09} (Draco), \citealt{Theler20} (Sextans), and \citealt{Cohen10} (Ursa Minor).
Ultra-faint dwarf galaxy ($M_\star \lesssim 10^5 M_\odot$) abundances come from \citet{Koch2008,Feltzing2009,Norris2010a,Norris2010b,Simon2010,Frebel2010,Frebel2014,Gilmore2013,Koch2013,Roederer2014segue,Ishigaki2014,Francois2016,Ji2016a,Ji2016c,Ji19,Ji20car,Venn2017,Kirby2017,Hansen2017,Hansen2020grus,Hansen2024,Nagasawa2018,Spite2018,Marshall2019,Chiti2018,Chiti2023,Waller2023}; and \citet{Webber2023}.
Chemical abundances for Milky Way globular clusters come from \citet{Carretta09}, \citet{Cohen12}, \citet{Kirby23}, \citet{Roederer11}, and \citet{Worley13}.
Stars from dwarf galaxy and globular cluster streams come from \citet{Ji20}.

Before we begin, it is clear that Leiptr-252 is a non-member star, and we have shown it as an X on all figures. Leiptr-252 was originally selected to be a member using \Gaia DR2 kinematics in \citet{Ibata19}, but our analysis suggests it is a chemical non-member, with a discrepantly high metallicity and abundance ratios all matching background Milky Way stars. After our chemical analysis was complete, \citet{Ibata24} released a new Leiptr kinematic selection using \Gaia DR3, and Leiptr-252 is no longer considered a member with the more precise astrometry. Thus Leiptr-252 is a clear chemodynamic non-member, and we do not discuss it further.

\begin{figure*}
    \centering
    \includegraphics[width=\linewidth]{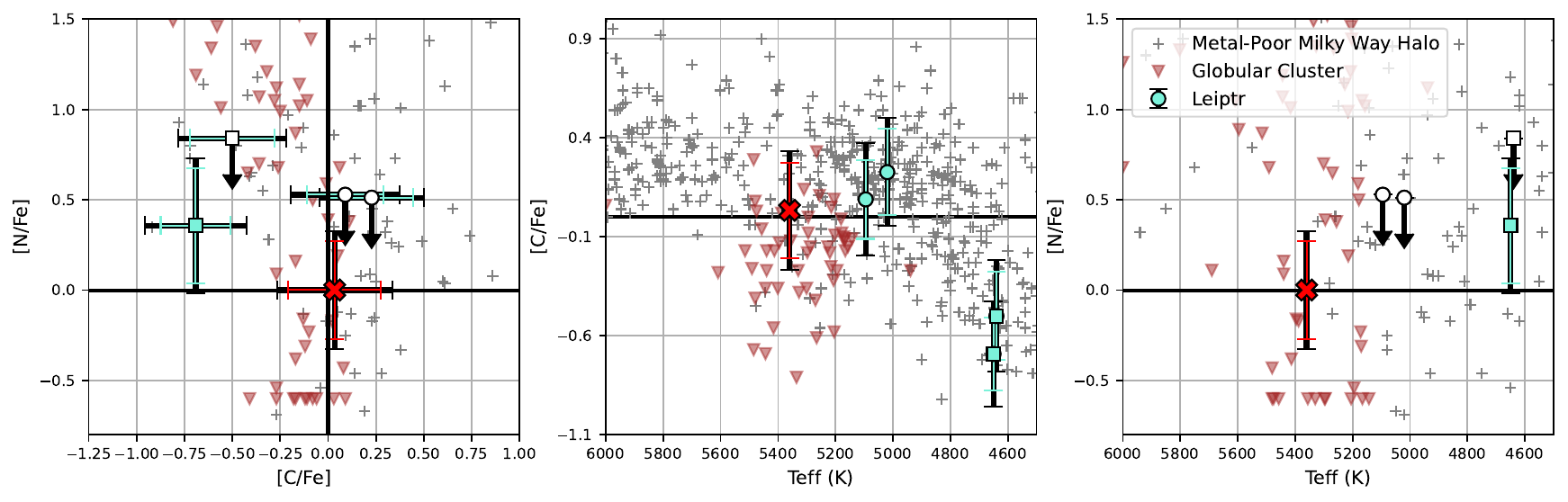}
    \caption{Comparison of Leiptr-208 and Leiptr-321 (light-blue circles) and Leiptr-186 and Leiptr-342 (light-blue squares) nitrogen abundances against carbon on the leftmost plot.
    The other two plots to the right illustrate carbon and nitrogen abundance in relation to effective temperature.
    The non-member Leiptr-252 is shown as a red X.
    Total uncertainty is shown in black for accuracy and abundance uncertainty without additional systematic errors is shown in the same color as the stars for statistical precision as described in Section \ref{sec:Abundance Analysis}.
    [X/Fe] = 0 is plotted as a bold black line.
    Abundances of metal-poor Milky Way halo stars appear as gray plus signs and abundances from the M15 globular cluster appear as dark-red upside-down triangles.}
    \label{fig:cno}
\end{figure*}

The four Leiptr member stars fall into two categories.
The three lowest-metallicity stars (Leiptr-186, Leiptr-208 and Leiptr-342) have similar element abundances: low $\alpha$, odd-z, and neutron-capture ratios.
The most metal-rich star (Leiptr-321) has similar abundances to the low-metallicity stars, but higher $\alpha$ and odd-z ratios.
We further investigate our pairs from Section \ref{sec:diffabund} by marking Leiptr-208 and Leiptr-321 as circles and Leiptr-186 and Leiptr-342 as squares in Figures~\ref{fig:cno}-\ref{fig:ncapture}.
Halo stars are marked as plus signs in gray if from the Milky Way or dark-blue if from GSE; dwarf galaxy stars are marked as triangles in yellow if massive (\mstar~\textgreater~$10^7$ \msun), pink if intermediate-mass ($10^5$ \msun~\textless~\mstar~\textless~$10^7$ \msun), or purple if ultra-faint (\mstar~\textless~$10^5$ \msun); globular clusters are marked as upside-down triangles in dark-red; and stellar streams are marked as stars in pink if from a dwarf galaxy or dark-red if from a globular cluster.

\subsection{Metallicity Mean and Dispersion}\label{sec:metallicity}

\begingroup
\setlength{\tabcolsep}{3pt} 

\begin{deluxetable}{cccccccc}
\tablecolumns{8}
\tablecaption{\label{tab:abunds252}Leiptr-252\tablenotemark{$*$} Abundances}
\tablehead{Element & N & log $\epsilon$ & [X/H] & $\sigma_{\text{[X/H]}}$ & [X/Fe] & $\text{[X/Fe]}_{\text{N}}$ & $\sigma_{\text{[X/Fe]}}$}
\startdata
CH & 2 & 7.26 & $-1.17$ & 0.33 & 0.03 & -- & 0.34 \\
CN & 1 & 6.63 & $-1.20$ & 0.40 & $-0.00$ & -- & 0.40 \\
Na I & 4 & 5.12 & $-1.12$ & 0.15 & 0.08 & $-0.09$ & 0.21 \\
Mg I & 4 & 6.83 & $-0.77$ & 0.18 & 0.43 & -- & 0.21 \\
Al I & 1 & 4.67 & $-1.78$ & 0.37 & $-0.58$ & 0.02 & 0.34 \\
K I & 2 & 4.73 & $-0.30$ & 0.21 & 0.89 & 0.30 & 0.22 \\
Ca I & 18 & 5.49 & $-0.85$ & 0.15 & 0.35 & -- & 0.18 \\
Sc II & 4 & 2.07 & $-1.08$ & 0.15 & 0.11 & -- & 0.23 \\
Ti I & 41 & 4.05 & $-0.90$ & 0.22 & 0.30 & -- & 0.25 \\
Ti II & 38 & 4.15 & $-0.80$ & 0.18 & 0.40 & -- & 0.25 \\
V I & 1 & 2.69 & $-1.24$ & 0.24 & $-0.04$ & -- & 0.29 \\
V II & 1 & 2.99 & $-0.94$ & 0.31 & 0.26 & -- & 0.39 \\
Cr I & 15 & 4.47 & $-1.17$ & 0.19 & 0.03 & -- & 0.20 \\
Cr II & 5 & 4.62 & $-1.02$ & 0.15 & 0.18 & -- & 0.25 \\
Mn I & 6 & 4.15 & $-1.28$ & 0.20 & $-0.08$ & 0.01 & 0.24 \\
Fe I & 105 & 6.3 & $-1.20$ & 0.20 & 0.00 & -- & -- \\
Fe II & 19 & 6.3 & $-1.20$ & 0.18 & -- & -- & -- \\
Co I & 4 & 3.86 & $-1.13$ & 0.27 & 0.07 & -- & 0.29 \\
Ni I & 19 & 5.04 & $-1.18$ & 0.17 & 0.02 & -- & 0.21 \\
Zn I & 2 & 3.51 & $-1.05$ & 0.15 & 0.15 & -- & 0.23 \\
Sr I & 1 & 1.83 & $-1.04$ & 0.21 & 0.16 & -- & 0.24 \\
Sr II & 2 & 1.94 & $-0.93$ & 0.21 & 0.27 & -- & 0.25 \\
Y II & 3 & 1.03 & $-1.18$ & 0.20 & 0.02 & -- & 0.27 \\
Zr II & 1 & 1.74 & $-0.84$ & 0.21 & 0.35 & -- & 0.31 \\
Ba II & 5 & 1.23 & $-0.95$ & 0.20 & 0.25 & -- & 0.23 \\
La II & 3 & 0.24 & $-0.86$ & 0.20 & 0.34 & -- & 0.29 \\
Eu II & 2 & -0.03 & $-0.55$ & 0.18 & 0.64 & -- & 0.27 \\
Dy II & 1 & 0.3 & $-0.80$ & 0.53 & 0.40 & -- & 0.54 \\
\enddata
\tablecomments{$\text{[X/Fe]}_{\text{N}}$ gives the NLTE abundances.}
\tablenotetext{$^*$}{Non-member}
\end{deluxetable}

\endgroup

The four Leiptr stars fall between [Fe/H] = $-2.54$ and $-1.80$.
We follow the method used in \citet{Usman24} to determine the mean metallicity and intrinsic dispersion, assuming a single Gaussian with heteroskedastic uncertainties,
sampled using the nested sampling algorithm \code{dynesty} \citep{dynesty, Higson2019, Koposov2022} to calculate the Bayesian posteriors of the means and dispersions.
The reported uncertainties are determined with 1$\sigma$ percentiles and 90\% upper limits.
We acknowledge that [Fe/H] may be underestimated by about 0.2 dex by ignoring NLTE effects on the measured iron lines \citep[e.g.,][]{Bergemann12a,Bergemann12c,Ezzeddine17}, but each star will shift a similar amount, so as a whole Leiptr will maintain the same metallicity dispersion.

Using our full abundance uncertainties, which are dominated by propagating both statistical and systematic uncertainties in stellar parameters, the mean [Fe\,I/H] metallicity of the four Leiptr stars is $-2.20 \pm 0.15$, with a dispersion 90\% upper limit less than 0.38.
If instead we use [Fe\,II/H], the per-star uncertainties are reduced to ${\sim}0.15$ dex for each star, as the effect of our surface gravity uncertainties on Fe\,II are much less than the effect of our effective temperature uncertainties on Fe\,I. We then resolve a [Fe\,II/H] spread to be 0.29$^{+0.23}_{-0.14}$, with ${>}$90\% confidence that the spread is larger than 0.1 dex.
We note the mean metallicity is 0.6 dex lower than the photometric estimate by \citet{Ibata19}. 

A resolved metallicity dispersion, indicating internal enrichment by supernovae, is crucial for classifying Leiptr as a dwarf galaxy \citep{Willman12}.
We thus performed a more sensitive test of internal metallicity variations by looking at our pairwise differential abundances (Section~\ref{sec:diffabund}, Figure~\ref{fig:diffabund}). 
These clearly show that our full uncertainties for [Fe\,I/H] are too conservative for intrinsic dispersion measurements. 
The four likely Leiptr member stars come in two pairs of stars with similar stellar parameters and clearly distinct chemical abundances.
Thus, the metallicity differences and [Fe\,II/H] metallicity dispersion strongly suggest that Leiptr was a dwarf galaxy, rather than a globular cluster.

\subsection{C and N}
\label{sec:cn}

\begingroup
\setlength{\tabcolsep}{3pt} 

\begin{deluxetable}{cccccccc}
\tablecolumns{8}
\tablecaption{\label{tab:abunds321}Leiptr-321 Abundances}
\tablehead{Element & N & log $\epsilon$ & [X/H] & $\sigma_{\text{[X/H]}}$ & [X/Fe] & $\text{[X/Fe]}_{\text{N}}$ & $\sigma_{\text{[X/Fe]}}$}
\startdata
CH & 2 & 6.67 & $-1.76$ & 0.26 & 0.09 & -- & 0.28 \\
CN & 1 & 6.51 & $< -1.32$ & -- & $< 0.53$ & -- & -- \\
Na I & 1 & 4.59 & $-1.65$ & 0.29 & 0.20 & $-0.29$ & 0.29 \\
Mg I & 8 & 5.87 & $-1.73$ & 0.17 & 0.12 & -- & 0.23 \\
Al I & 1 & 3.84 & $-2.61$ & 0.33 & -0.76 & $-0.15$ & 0.34 \\
K I & 2 & 4.13 & $-0.90$ & 0.22 & 0.95 & 0.25 & 0.24 \\
Ca I & 22 & 4.75 & $-1.59$ & 0.13 & 0.26 & -- & 0.20 \\
Sc II & 4 & 1.6 & $-1.55$ & 0.16 & 0.30 & -- & 0.25 \\
Ti I & 28 & 3.34 & $-1.61$ & 0.19 & 0.24 & -- & 0.23 \\
Ti II & 41 & 3.49 & $-1.46$ & 0.17 & 0.39 & -- & 0.25 \\
V I & 1 & 2.15 & $-1.78$ & 0.19 & 0.07 & -- & 0.25 \\
V II & 2 & 2.28 & $-1.65$ & 0.22 & 0.20 & -- & 0.31 \\
Cr I & 13 & 3.74 & $-1.90$ & 0.18 & $-0.05$ & -- & 0.22 \\
Cr II & 4 & 3.72 & $-1.92$ & 0.14 & $-0.07$ & -- & 0.26 \\
Mn I & 6 & 3.14 & $-2.29$ & 0.12 & $-0.44$ & $-0.20$ & 0.21 \\
Fe I & 109 & 5.65 & $-1.85$ & 0.22 & 0.00 & -- & -- \\
Fe II & 23 & 5.68 & $-1.82$ & 0.16 & -- & -- & -- \\
Co I & 3 & 3.31 & $-1.68$ & 0.27 & 0.17 & -- & 0.29 \\
Ni I & 16 & 4.39 & $-1.83$ & 0.15 & 0.02 & -- & 0.22 \\
Zn I & 2 & 2.75 & $-1.81$ & 0.12 & 0.04 & -- & 0.22 \\
Sr II & 1 & 0.07 & $-2.80$ & 0.33 & $-0.95$ & -- & 0.36 \\
Ba II & 4 & -0.89 & $-3.07$ & 0.16 & $-1.22$ & -- & 0.24 \\
Eu II & 1 & -1.32 & $< -1.84$ & -- & $< 0.01$ & -- & -- \\
\enddata
\tablecomments{CN and Eu II are upper limits.
$\text{[X/Fe]}_{\text{N}}$ gives the NLTE abundances.}
\end{deluxetable}

\endgroup

\begingroup
\setlength{\tabcolsep}{3pt} 

\begin{deluxetable}{cccccccc}
\tablecolumns{8}
\tablecaption{\label{tab:abunds342}Leiptr-342 Abundances}
\tablehead{Element & N & log $\epsilon$ & [X/H] & $\sigma_{\text{[X/H]}}$ & [X/Fe] & $\text{[X/Fe]}_{\text{N}}$ & $\sigma_{\text{[X/Fe]}}$}
\startdata
CH & 2 & 5.74 & $-2.69$ & 0.22 & $-0.70$ & -- & 0.27 \\
CN & 1 & 6.19 & $-1.64$ & 0.41 & 0.35 & -- & 0.42 \\
O I & 1 & 7.09 & $-1.60$ & 0.19 & 0.40 & -- & 0.28 \\
Na I & 1 & 4.16 & $-2.08$ & 0.33 & $-0.08$ & $-0.36$ & 0.32 \\
Mg I & 4 & 5.45 & $-2.15$ & 0.17 & $-0.15$ & -- & 0.21 \\
Al I & 1 & 3.55 & $-2.90$ & 0.35 & $-0.91$ & $-0.30$ & 0.38 \\
K I & 2 & 3.21 & $-1.82$ & 0.17 & 0.17 & $-0.42$ & 0.21 \\
Ca I & 17 & 4.32 & $-2.02$ & 0.13 & $-0.02$ & -- & 0.20 \\
Sc II & 6 & 0.8 & $-2.35$ & 0.14 & $-0.35$ & -- & 0.25 \\
Ti I & 28 & 2.8 & $-2.15$ & 0.18 & $-0.15$ & -- & 0.21 \\
Ti II & 41 & 2.94 & $-2.01$ & 0.17 & $-0.01$ & -- & 0.27 \\
V I & 1 & 1.46 & $-2.47$ & 0.21 & $-0.48$ & -- & 0.25 \\
V II & 1 & 1.94 & $-1.99$ & 0.19 & 0.00 & -- & 0.29 \\
Cr I & 14 & 3.53 & $-2.11$ & 0.19 & $-0.12$ & -- & 0.21 \\
Cr II & 6 & 3.55 & $-2.09$ & 0.14 & $-0.10$ & -- & 0.26 \\
Mn I & 6 & 2.97 & $-2.46$ & 0.13 & $-0.47$ & $-0.16$ & 0.19 \\
Fe I & 93 & 5.51 & $-1.99$ & 0.21 & 0.00 & -- & -- \\
Fe II & 20 & 5.56 & $-1.94$ & 0.15 & -- & -- & -- \\
Co I & 4 & 2.86 & $-2.13$ & 0.32 & $-0.14$ & -- & 0.34 \\
Ni I & 18 & 4.06 & $-2.16$ & 0.20 & $-0.17$ & -- & 0.24 \\
Zn I & 2 & 2.35 & $-2.21$ & 0.10 & $-0.22$ & -- & 0.22 \\
Sr II & 2 & -0.21 & $-3.08$ & 0.42 & $-1.08$ & -- & 0.43 \\
Ba II & 5 & -0.81 & $-2.99$ & 0.18 & $-0.99$ & -- & 0.26 \\
Eu II & 2 & -2.06 & $-2.58$ & 0.16 & $-0.58$ & -- & 0.28 \\
\enddata
\tablecomments{$\text{[X/Fe]}_{\text{N}}$ gives the NLTE abundances.}
\end{deluxetable}

\endgroup

\begin{figure*}
    \centering
    \includegraphics[width=\linewidth]{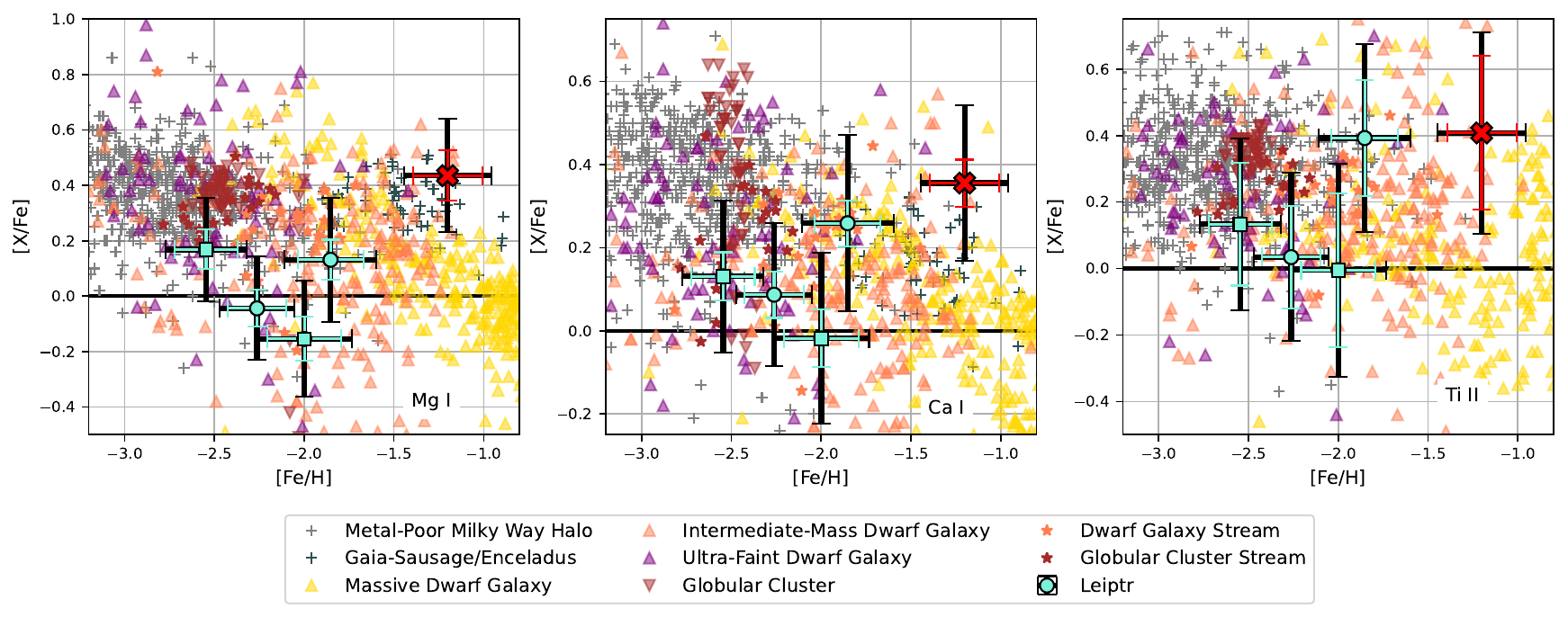}
    \caption{Comparison of Leiptr-208 and Leiptr-321 (light-blue circles) and Leiptr-186 and Leiptr-342 (light-blue squares) $\alpha$ abundances with other stellar populations.
    The non-member Leiptr-252 is shown as a red X.
    Total uncertainty is shown in black for accuracy and abundance uncertainty without additional systematic errors is shown in the same color as the stars for statistical precision as described in Section \ref{sec:Abundance Analysis}.
    [X/Fe] = 0 is plotted as a bold black line.
    Metal-poor Milky Way halo stars appear as gray plus signs and GSE stars appear as dark-blue plus signs.
    Abundances of stars from massive dwarf galaxies, intermediate-mass dwarf galaxies, and ultra-faint dwarf galaxies appear as yellow triangles, pink triangles, and purple triangles, respectively.
    Abundances of stars from the M92 and NGC 2419 Milky Way globular clusters appear as dark-red upside-down triangles. 
    Abundances from dwarf galaxy stream stars appear as pink stars and abundances from globular cluster stream stars appear as dark-red stars.}
    \label{fig:alpha}
\end{figure*}

In principle, carbon and nitrogen abundances can be used to distinguish between globular clusters and dwarf galaxies.
This is because the majority of stars in massive globular clusters display unusual carbon depletions and nitrogen enhancements, a phenomenon whose origin is not understood but is empirically well-documented \citep[e.g.,][]{Bastian18,Gratton19}.
We use the 4310 and 4323~\AA\ CH absorption bands to determine the carbon abundance and 3877~\AA\ CN band to find upper limits on nitrogen for Leiptr-186, Leiptr-208, and Leiptr-321.
The NLTE correction is negligible for carbon \citep{Amarsi19a,Amarsi19b} and nitrogen \citep{Amarsi20,Mashonkina24}.
We find that Leiptr-342 has high nitrogen and low carbon.
Thus, this star could potentially indicate that Leiptr is a disrupted globular cluster.

However, red giant branch stars also change composition as they ascend the red giant branch due to extra mixing that also depletes carbon and enhances nitrogen \citep[e.g.,][]{Shetrone19}, so we must also examine the relationship between chemical abundance and effective temperature.
We plot carbon and nitrogen abundance against \teff in the right two panels of Figure \ref{fig:cno}.
We expect to see higher nitrogen and lower carbon as \teff decreases due to extra mixing (e.g., Figure 6 of \citealt{Spite05}).
Enriched globular cluster stars would not follow this temperature dependence and instead show a distinct group of stars with high nitrogen and low carbon, irrespective of measured temperature.
Compared to \citet{Cohen05}, which specifically targeted stars in M15 that are on the lower red giant branch to avoid extra mixing, the Leiptr stars are cooler and within the regime of possible extra mixing effects.
Since we do observe that the coolest Leiptr stars have higher nitrogen and lower carbon, the trend with $T_{\text{eff}}$ suggests that extra mixing is most likely the reason for Leiptr's chemical abundance pattern.
Furthermore, the nitrogen increase is not as extreme as it is in globular clusters like M15 \citep{Cohen05}.
However, with only one nitrogen determination in the Leiptr member stars, more stars would be needed for this to be a strong constraint on whether Leiptr's progenitor is a globular cluster or dwarf galaxy.

\subsection{$\alpha$ elements (Mg, Ca, Ti)}
\label{sec:alpha}

\begin{figure*}
    \centering
    \includegraphics[width=\linewidth]{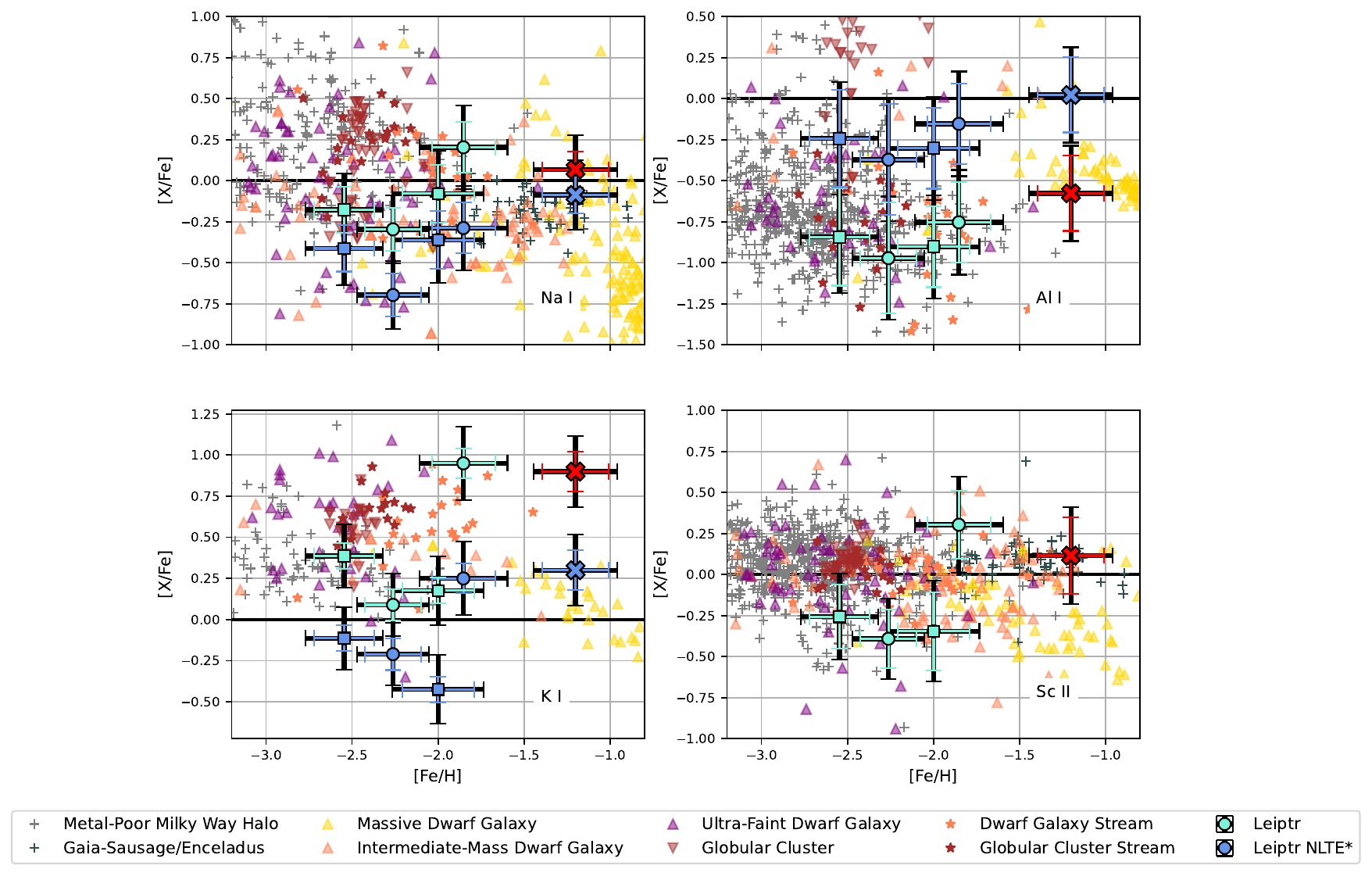}
    \caption{Comparison of Leiptr-208 and Leiptr-321 (light-blue circles) and Leiptr-186 and Leiptr-342 (light-blue squares) odd-z abundances with other stellar populations.
    The non-member Leiptr-252 is shown as a red X.
    Total uncertainty is shown in black for accuracy and abundance uncertainty without additional systematic errors is shown in the same color as the stars for statistical precision as described in Section \ref{sec:Abundance Analysis}.
    [X/Fe] = 0 is plotted as a bold black line.
    Metal-poor Milky Way halo stars appear as gray plus signs and GSE stars appear as dark-blue plus signs.
    Abundances of stars from massive dwarf galaxies, intermediate-mass dwarf galaxies, and ultra-faint dwarf galaxies appear as yellow triangles, pink triangles, and purple triangles, respectively.
    Abundances of stars from the M92 and NGC 2419 Milky Way globular clusters appear as dark-red upside-down triangles. 
    Abundances from dwarf galaxy stream stars appear as pink stars and abundances from globular cluster stream stars appear as dark-red stars.
    Leiptr-208 and Leiptr-321 (blue circles) and Leiptr-186 and Leiptr-342 (blue squares) abundances for sodium, aluminum, and potassium are additionally plotted to correct for especially significant NLTE effects.
    \\
    $^*$ approximate NLTE corrections are calculated as an average for aluminum and independently for sodium and potassium (see text for details)}
    \label{fig:oddz}
\end{figure*}

The [$\alpha$/Fe] ratio is an excellent indicator of a stream's progenitor due to galactic chemical evolution.
$\alpha$ elements are primarily produced by core-collapse supernovae, while iron is produced by both core-collapse and Type~Ia supernovae.
The [$\alpha$/Fe] ratio thus starts high early then declines to lower values later as a galaxy forms.
Efficiently star-forming galaxies with lower mass-loading factors like the Milky Way maintain a high [$\alpha$/Fe] until relatively high metallicities, while inefficiently star-forming dwarf galaxies with higher mass-loading factors have an $\alpha$ downturn at lower metallicities \citep[e.g.,][]{Tinsley1979,Matteucci1990,McWilliam97,Tolstoy2009,Sheffield12}.
This downturn, often called an $\alpha$ ``knee'', tends to occur at lower metallicities in lower-mass galaxies \citep[e.g.,][]{Tolstoy2009,Kirby2011,Kirby20,Frebel2012,Helmi18,Hill19,Theler20}.
Globular clusters can be understood in this context as inheriting the chemical patterns of their host galaxies at their birth metallicity, as they are the most massive star clusters to form in a galaxy \citep[e.g.,][]{Forbes2018,Choksi2018,ReinaCampos2019}.
Thus, globular clusters born in the Milky Way will tend to be high-$\alpha$, while those born in dwarf galaxies will be low-$\alpha$ \citep[e.g.,][]{Mucciarelli2017,Usman24}.

Figure~\ref{fig:alpha} shows that Leiptr's stars are overall low-$\alpha$.
The NLTE correction is negligible for magnesium \citep{Mashonkina13,Bergemann17}, calcium \citep{Mashonkina07,Spite12}, and titanium \citep{Bergemann11,Sitnova16}.
While Milky Way halo stars and stars in Milky Way globular clusters M92 and NGC 2419 have [$\alpha$/Fe] $\approx 0.3-0.4$ (NGC 2419 has some low-magnesium stars, but this is from magnesium depletion due to multiple populations),
the three lowest-metallicity Leiptr stars' magnesium, calcium, and titanium abundances are within $1 \sigma$ of [$\alpha$/Fe] = 0.
Thus, the Milky Way halo and its globular clusters' high-$\alpha$ abundances are overall inconsistent with Leiptr.
However, several dwarf galaxies have low [$\alpha$/Fe] ratios similar to Leiptr.
At $\mbox{[Fe/H]} \approx -2$, lower-mass classical dwarf galaxies ($M_\star \lesssim 10^6 M_\odot$) like Draco, Ursa Minor, and Sextans, as well as the dwarf galaxy stream Elqui, have low [$\alpha$/Fe] abundances similar to Leiptr.
The lowest-mass ultra-faint dwarf galaxies like Carina~II, Reticulum II, and Horologium~I ($M_\star \lesssim 10^5 M_\odot$) also mostly have low [$\alpha$/Fe] abundances at $\mbox{[Fe/H]} \approx -2$.
In contrast, higher-mass dwarf galaxies like Sculptor, Fornax, and GSE ($M_\star \gtrsim 10^7 M_\odot$) still have relatively high [$\alpha$/Fe] ratios at $\mbox{[Fe/H]}=-2$, not seeing a significant $\alpha$ decline until higher metallicity, $\mbox{[Fe/H]} \gtrsim -1.5$.
Thus, Leiptr's $\alpha$ element abundances are most similar to lower-mass dwarf galaxies.

Examining the differential abundances allows us to interpret possible trends with metallicity in the $\alpha$ elements.
Comparing Leiptr-186 and Leiptr-342 (light-blue squares in Figure~\ref{fig:alpha}, top panel of Figure~\ref{fig:diffabund}), there is a significant decline in [Mg/Fe] and potentially small decline in [Ca/Fe] as the metallicity increases by 0.45 dex between these two stars. This decline in [$\alpha$/Fe] with metallicity would be expected in a dwarf galaxy due to increased Type~Ia supernova enrichment.
The most metal-rich star, Leiptr-321, is potentially somewhat different than its matched star Leiptr-208 (light-blue circles in Figure~\ref{fig:alpha}). It has a similar low [Mg/Fe] abundance as the other three Leiptr stars, but it also has slightly higher [Ca/Fe] and [Ti/Fe] abundances. The bottom panel of Figure~\ref{fig:diffabund} shows the differential abundances give only a 0.1-0.2 dex increase in [Mg,Ca,Ti/Fe] relative to Leiptr-208, which is not very significant compared to the line-to-line scatter, but it will be interesting to interpret this in the context of the odd-Z elements discussed next.

\subsection{Odd-Z elements (Na, Al, K, Sc)}
\label{sec:oddz}

\begin{figure*}
    \centering
    \includegraphics[width=\linewidth]{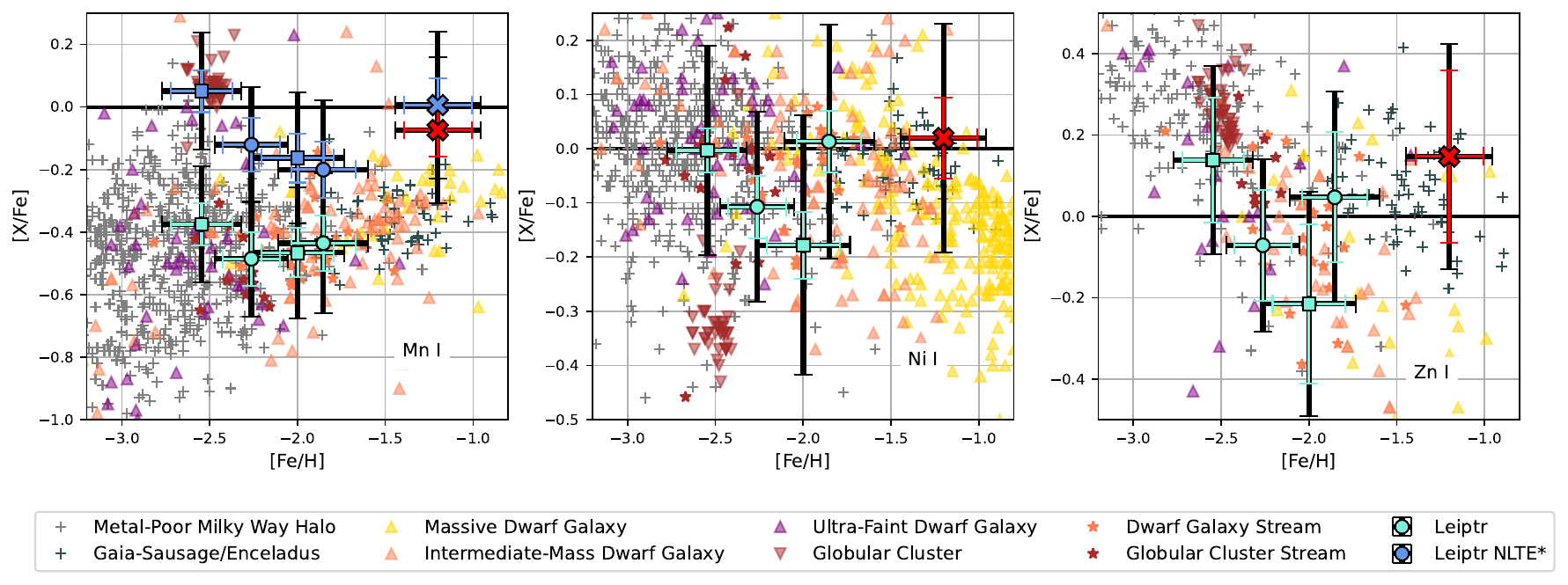}
    \caption{Comparison of Leiptr-208 and Leiptr-321 (light-blue circles) and Leiptr-186 and Leiptr-342 (light-blue squares) iron-peak abundances with other stellar populations.
    The non-member Leiptr-252 is shown as a red X.
    Total uncertainty is shown in black for accuracy and abundance uncertainty without additional systematic errors is shown in the same color as the stars for statistical precision as described in Section \ref{sec:Abundance Analysis}.
    [X/Fe] = 0 is plotted as a bold black line.
    Metal-poor Milky Way halo stars appear as gray plus signs and GSE stars appear as dark-blue plus signs.
    Abundances of stars from massive dwarf galaxies, intermediate-mass dwarf galaxies, and ultra-faint dwarf galaxies appear as yellow triangles, pink triangles, and purple triangles, respectively.
    Abundances of stars from the M92 Milky Way globular cluster appear as dark-red upside-down triangles. 
    Abundances from dwarf galaxy stream stars appear as pink stars and abundances from globular cluster stream stars appear as dark-red stars.
    Leiptr-208 and Leiptr-321 (blue circles) and Leiptr-186 and Leiptr-342 (blue squares) abundances for manganese are additionally plotted to correct for especially significant NLTE effects.
    \\
    $^*$ approximate NLTE corrections are calculated independently for manganese (see text for details)}
    \label{fig:fepeak}
\end{figure*}

Figure~\ref{fig:oddz} shows the abundances for the odd-Z elements.
In blue points, we show the effect of applying corrections to sodium, aluminum, and potassium for NLTE effects on the lines where the radiation field is partially decoupled from local conditions.
Using the calculations for neutral sodium from \citet{Lind11}, we compute the average of the corrections from each of the used lines for each star. 
All five stars include the 5895~\AA\ sodium line and some of the five also use the 5682, 5688, and 5889~\AA\ lines. 
The typical size of the correction is $-0.35$ dex.
Aluminum is determined using the 3961~\AA~resonance line, and we adopt a constant correction for all stars of adding 0.6 to [Al/Fe], following \citet{Roederer19}.
For potassium, we use the two resonance lines at 7664~\AA\ and 7669~\AA. 
The two lines always agree excellently in abundance, and we note the 7664~\AA~line is not affected by tellurics given Leiptr’s range of radial velocities (Figure~\ref{fig:a+oddz_spec}).
We identify stars in Table A.1 of \citet{Reggiani19} that have the most comparable stellar parameters as the Leiptr stars and use the corresponding potassium NLTE corrections. 
We use the corrections from stars HD 186478, CS 29513-014, BD +19 1185A, HD 45282, and HD 126238 with a range from $-0.7$ to $-0.3$ dex.
The NLTE correction is negligible for scandium \citep{Zhang14}.
The LTE and NLTE abundances are both reported in Tables~\ref{tab:abunds186}-\ref{tab:abunds342}.

In Figure~\ref{fig:oddz}, compared to halo stars (whose abundances are mostly computed in LTE and should be compared to the light-blue points), Leiptr lies on the lower end of the abundance scatter for sodium, potassium, and scandium.
Abundances of aluminum in the Milky Way globular clusters, M92 and NGC 2419, are also substantially higher than in all four Leiptr stars. 
The globular clusters should be compared to the NLTE aluminum abundances (blue points): the M92 stars use the same aluminum 3961~\AA\ line and were corrected using \citet{Nordlander2017} while the NGC 2419 stars use the subordinate doublet at $6696-6698$~\AA\ that are minimally affected by NLTE. 
These aluminum abundances further disprove that Leiptr may be an accreted globular cluster with multiple populations, which should result in aluminum enhancements as shown by the globular clusters in Figure~\ref{fig:oddz} \citep[see][for details]{Bastian18,Gratton19}.

However, Leiptr is consistent with some odd-z element abundances from ultra-faint and intermediate-mass dwarf galaxies.
Overlap between the halo scatter, globular clusters, and dwarf galaxy stars make it difficult to distinguish how Leiptr compares to these systems in most of the elements, but the potassium abundances from stars in ultra-faint dwarf galaxies are the only others as low as Leiptr's three low-metallicity stars.
The most metal-rich star, Leiptr-321, appears to be an outlier in the odd-Z elements compared to the other stars, with clearly higher potassium and scandium, and possibly higher sodium, than the three lower-metallicity stars. 
This is corroborated (in LTE) by comparing to Leiptr-208 in the bottom panel of Figure~\ref{fig:diffabund} and visually apparent in Figure~\ref{fig:a+oddz_spec}. 
As noted previously, this star also has a mild enhancement in the $\alpha$ elements.

\subsection{Iron-peak elements (Cr, Mn, Co, Ni, Zn)}
\label{sec:fepeak}

\begin{figure*}
    \centering
    \includegraphics[width=\linewidth]{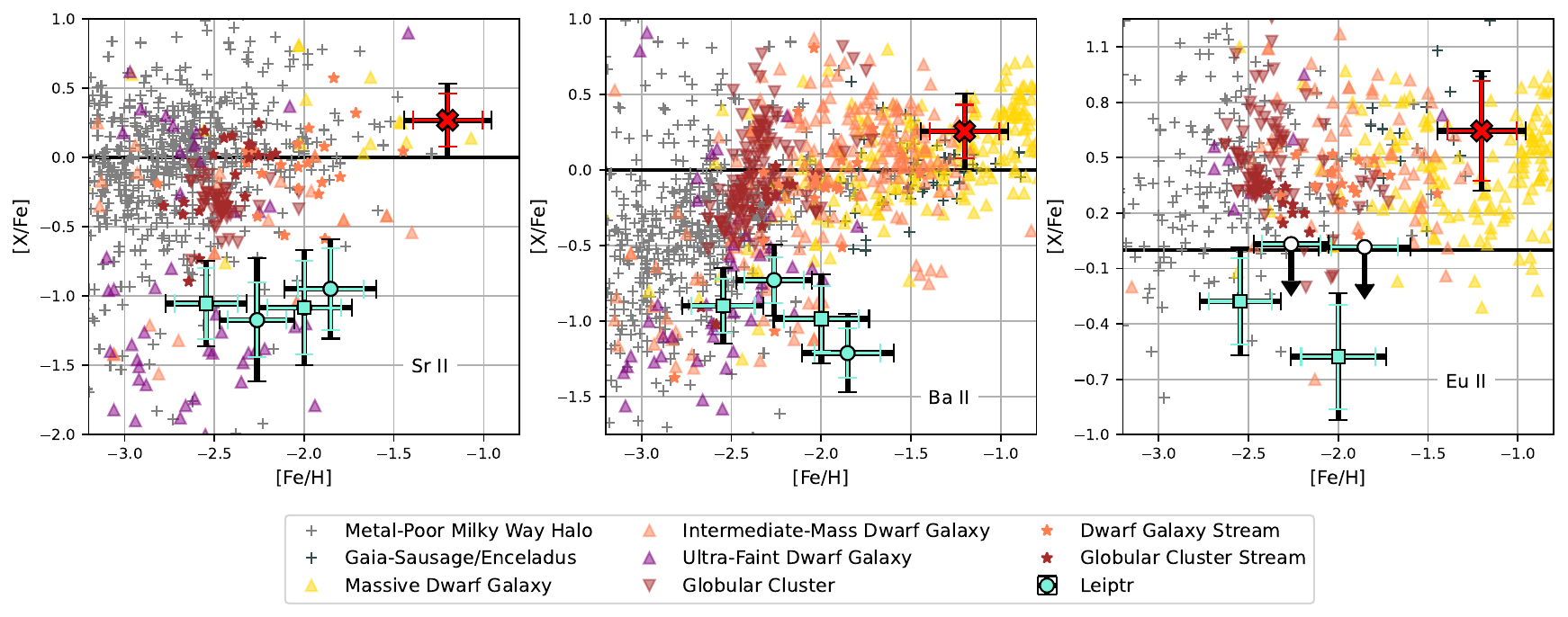}
    \caption{Comparison of Leiptr-208 and Leiptr-321 (light-blue circles) and Leiptr-186 and Leiptr-342 (light-blue squares) neutron-capture abundances with other stellar populations.
    The non-member Leiptr-252 is shown as a red X.
    Total uncertainty is shown in black for accuracy and abundance uncertainty without additional systematic errors is shown in the same color as the stars for statistical precision as described in Section \ref{sec:Abundance Analysis}.
    [X/Fe] = 0 is plotted as a bold black line.
    Metal-poor Milky Way halo stars appear as gray plus signs and GSE stars appear as dark-blue plus signs.
    Abundances of stars from massive dwarf galaxies, intermediate-mass dwarf galaxies, and ultra-faint dwarf galaxies appear as yellow triangles, pink triangles, and purple triangles, respectively.
    Abundances of stars from the M15, M22, M92, and NGC 2419 Milky Way globular clusters appear as dark-red upside-down triangles. 
    Abundances from dwarf galaxy stream stars appear as pink stars and abundances from globular cluster stream stars appear as dark-red stars.}
    \label{fig:ncapture}
\end{figure*}

Figure~\ref{fig:fepeak} shows some of the iron-peak abundances in Leiptr. 
Manganese deserves special discussion due to NLTE effects. 
We reject the resonance lines at 4030~\AA\ and only use weaker lines around 4800~\AA. This was motivated due to expecting relatively small NLTE corrections \citep{Bergemann08}, but more recent calculations show that the corrections for the weaker lines can also be $0.2-0.4$ dex \citep{Bergemann19}, so we apply an individual NLTE correction from \citet{Bergemann19} to each star from the average of the corrections for each line, which appear as blue points in Figure \ref{fig:fepeak}. 
We note that the comparison stars in Figure~\ref{fig:fepeak} do not have NLTE corrections applied, so these should be compared to the LTE abundances. 
The NLTE correction is negligible for nickel \citep{Eitner23} and zinc \citep{Takeda05,Ezzeddine19}.

Overall, the iron-peak elements are not that informative for understanding Leiptr's origin, because the [X/Fe] ratios do not vary significantly as a function of environment in the metallicity range of Leiptr stars.
In addition to manganese, nickel, and zinc, we record chromium and cobalt abundances for each star in Tables \ref{tab:abunds186}-\ref{tab:abunds342}.
All of Leiptr's stars have approximately equal manganese, cobalt, and chromium abundances, similar to other stars in the Milky Way halo, globular clusters, and dwarf galaxies at $\mbox{[Fe/H]} \approx -2$. The nickel and zinc abundances also match the background halo stars, but there are some differences comparing the most metal-poor star, Leiptr-186, to its similar-temperature counterpart, Leiptr-342, which has substantially lower [Ni/Fe] and [Zn/Fe] (light-blue squares in Figure~\ref{fig:fepeak} and top panel of  Figure~\ref{fig:diffabund}).

\subsection{Neutron-capture elements (Sr, Ba, Eu)}
\label{sec:nc}

Figure~\ref{fig:ncapture} shows that Leiptr has very low neutron-capture abundances, [Sr/Fe] ${\approx}$ [Ba/Fe] $\approx -1$.
The NLTE correction is negligible for strontium \citep{Mashonkina01,Bergemann12b}, barium \citep{Mashonkina99,Mashonkina01}, and europium \citep{Mashonkina01,Guo25}.
The four Leiptr member stars have significantly lower [Sr/Fe] and [Ba/Fe] than halo stars, most dwarf galaxy stars, or any of the Milky Way globular clusters.
For example, globular clusters M92 and NGC 2419 have higher [Sr/Fe] between $-0.5$ and 0, and Leiptr's [Sr/Fe] only falls within the extreme tail end of the halo scatter in Figure \ref{fig:ncapture}.
The systems with the most similar neutron-capture element abundances, especially for strontium, are the ultra-faint dwarf galaxies with $\mstar < 10^5 \msun$, which typically have low [Sr/Fe] and [Ba/Fe] between $-2$ and $-1$ \citep{Frebel15,Kirby2017,Ji19}.

We detect [Eu/Fe] in the two cooler Leiptr members at quite low values, $\mbox{[Eu/Fe]} \sim -0.4$.
These two stars have [Ba/Eu] ${\sim}-0.5$ compared to a pure heavy $r$-process ratio of $-0.8$ \citep[e.g.,][]{Simmerer2004}, suggesting the neutron-capture element abundances in Leiptr are dominated by production via the $r$-process but potentially with some contribution from the $s$-process (slow neutron-capture process) in asymptotic giant branch stars.

The most chemically similar dwarf galaxy stream is Elqui \citep{Ji20}, which also has low [$\alpha$/Fe] at $\mbox{[Fe/H]} \approx -2$.
Based on its low mean metallicity similar to Leiptr, \citet{Li22} argued Elqui was an ultra-faint dwarf galaxy using the mass-metallicity relation (as well as the Turranburra stream).
But Elqui has sharply rising [Ba/Fe] with increasing metallicity and solar [Sr/Fe], which is more similar to the faintest classical dwarf spheroidal galaxies like Draco, while Leiptr has constant or declining [Sr/Fe] and [Ba/Fe] with metallicity, which is most similar to the ultra-faint dwarf galaxies. 
This suggests Leiptr is the stream of an even lower-mass dwarf galaxy than Elqui.
\newline
\newline

\section{Discussion}
\label{sec:Discussion}

\begin{figure*}[t]
    \centering
    \includegraphics[width=\linewidth]{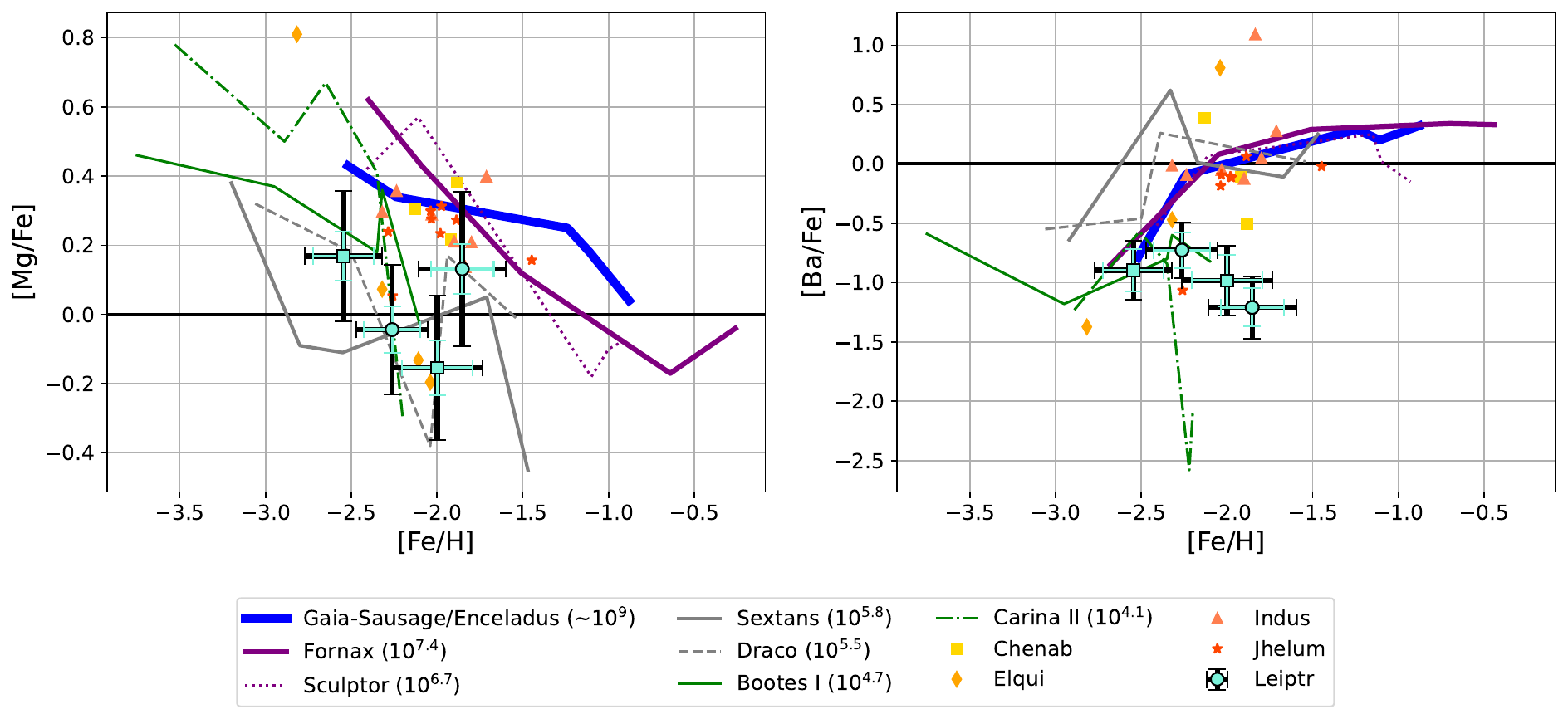}
    \caption{[Mg/Fe] and [Ba/Fe] for known dwarf galaxies compared to Leiptr-208 and Leiptr-321 (light-blue circles) and Leiptr-186 and Leiptr-342 (light-blue squares).
    Total uncertainty is shown in black for accuracy and abundance uncertainty without additional systematic errors is shown in the same color as the stars for statistical precision as described in Section \ref{sec:Abundance Analysis}.
    [X/Fe] = 0 is plotted as a bold black line.
    In order from highest to lowest mass using LOESS smoothing, abundances are plotted for GSE (thick blue line), Fornax (purple line), Sculptor (dotted purple line), Sextans (thin gray line), Draco (dashed gray line), Bo\"{o}tes I (thin green line), and Carina II (dot-dashed green line).
    Solar masses are given in the legend.
    Abundances are also plotted for the dwarf galaxy streams Chenab (yellow square), Elqui (orange diamond), Indus (pink triangle), and Jhelum (dark-orange star).
    Leiptr's low-$\alpha$ abundances are most similar to dwarf galaxies of mass $M_\star \lesssim 10^6 M_\odot$.
    It's neutron-capture abundances are even lower, only consistent with dwarf galaxies of mass $M_\star \lesssim 10^5 M_\odot$.
    Unlike Elqui, the previous lowest-mass dwarf galaxy stream, Leiptr's neutron-capture abundances are constant or declining with increasing metallicity which suggests that it is even lower-mass.}
    \label{fig:dgsummary}
\end{figure*}

We have observed and analyzed four member stars from the Leiptr stellar stream with the goal of classifying its progenitor, as well as identifying one non-member star Leiptr-252.
We now discuss the four main conclusions we draw from these abundances, starting from the most certain and ending with the least certain.
Overall, the chemistry of these four stars suggests that Leiptr is likely the first known ultra-faint dwarf galaxy stream without an intact progenitor.

In Figure \ref{fig:dgsummary}, we compare [Mg/Fe] and [Ba/Fe] for known dwarf galaxies to Leiptr-208 and Leiptr-321 marked as light-blue circles and Leiptr-186 and Leiptr-342 marked as light-blue squares. 
From highest to lowest mass using LOESS smoothing we include GSE as a thick blue line \citep{Ou24}, Fornax as a purple line \citep{Lemasle14}, Sculptor as a dotted purple line \citep{Hill19}, Sextans as a thin gray line \citep{Theler20}, Draco as a dashed gray line \citep{Cohen09}, Bo\"{o}tes I as a thin green line \citep{Mashonkina17}, and Carina
II as a dot-dashed green line \citep{Ji20car}. 
Additionally, we mark Chenab as a yellow square,
Elqui as an orange diamond, Indus as a pink triangle, and Jhelum as a dark-orange star \citep{Ji20}.

\subsection{Extragalactic Origin}\label{sec:exsitu}

Leiptr's eccentric and retrograde orbit is indicative of the progenitor's \textit{ex-situ} formation \citep{Ibata19,Ibata24}.
The orbit has most recently been determined by \citet{Malhan2022} to have a pericenter of 12.3 $\pm$ 0.1 kpc and the apocenter to be 45.1 $\pm$ 0.2 kpc, with a resulting eccentricity of 0.57.
The stream is also the second-most retrograde system in that sample.
Accreted systems can sometimes be linked together by looking at their energy and angular momentum, indicating that the systems have fallen into the Milky Way together, perhaps having formed with each other.
\citet{Bonaca21} suggested Leiptr could be associated with other retrograde streams, including GD-1, Phlegethon, Ylgr, and especially Gj\"oll and Wambelong. 
They suggest this group of streams originated from the retrograde Sequoia/I'itoi group using the average isochrone [Fe/H] = $-1.6$ from \citet{Ibata19}.
While they did not explicitly address Leiptr, \citet{Malhan2022} also associated these retrograde streams with the Sequoia/I'itoi group\footnote{The more metal-rich Arjuna system is also kinematically associated with this group \citep{Naidu20}, but it is likely debris from an early orbit of GSE \citep{Naidu2021}.}.
Given the updated metallicity from this work, we further note that Leiptr has clearly distinct chemical abundances from Sequoia \citep{Matsuno2022}, with similarly low [$\alpha$/Fe] $\sim 0$ but at lower-metallicity, $\feh = -2.2$, compared to Sequoia at $\feh = -1.5$ and with lower strontium and barium abundances.
Therefore, it is unlikely that Leiptr originated from a globular cluster in this dwarf galaxy.
Instead it is expected to be the last stream of a fully disrupted object.

An extragalactic formation would also be consistent with the $\alpha$ elements derived in Leiptr.
At a given metallicity, dwarf galaxies have lower values of [$\alpha$/Fe] than Milky Way \textit{in-situ} stars and globular clusters.
Thus, globular clusters that formed \textit{ex-situ} in dwarf galaxies and their remnants show similarly low levels of $\alpha$ elements.
For example, the globular cluster stellar stream 300S was originally mistaken for a dwarf galaxy stream due to its low $\alpha$ element abundances \citep{Usman24}.
With only this information, Leiptr's progenitor could either be a dwarf galaxy, or a globular cluster that formed in an $\alpha$ poor dwarf galaxy and later accreted onto the Milky Way.

\subsection{A Dwarf Galaxy, Not a Globular Cluster}\label{sec:DG}

Stellar metallicity dispersion is the primary chemical distinguisher between disrupted globular clusters and dwarf galaxies.
Dwarf galaxy stars have a large spread in their metallicities due to extended chemical evolution, while stars in globular clusters usually have the same metallicity \citep{Willman12}.
Our stars’ metallicities span a wide range, $\feh = -2.54$ to $-1.80$. 
Our conservative metallicity uncertainties for [Fe\,I/H] only allow a 90\% confidence upper limit of 0.38 dex on the metallicity dispersion, but our more precise [Fe\,II/H] abundances show a resolved metallicity dispersion of 0.29$^{+0.23}_{-0.14}$ dex.
Additionally, we can clearly see a metallicity difference within Leiptr using differential abundances, as described in Section~\ref{sec:metallicity}.
The spectra shown in Figures~\ref{fig:hbeta_spec} and~\ref{fig:halpha_spec} demonstrate that the pairs of stars have similar temperatures (based on their matching hydrogen lines), while also showing qualitatively different strengths of the nearby metal lines, indicating that the chemical compositions between the matching stars differ significantly. 
In Figure~\ref{fig:diffabund}, the changes in element abundance between these pairs of stars are quantified.
The iron abundance clearly shows a large discrepancy between both pairs of stars, with a change of $\Delta \feh = 0.42 \pm 0.21$ between Leiptr-321 and Leiptr-208 and $\Delta \feh = 0.45 \pm 0.13$ between Leiptr-342 and Leiptr-186.
These disparities are inconsistent with the uniform metallicities found in globular cluster stars, indicating Leiptr’s progenitor was a dwarf galaxy.

For completeness, we note that detailed chemical abundances are less useful at classifying a stream as a dwarf galaxy or globular cluster. 
This is because globular clusters track the overall composition of their birth galaxies \citep[e.g.,][]{Mucciarelli2017,Usman24}, the [X/Fe] ratios just help confirm whether a stream is accreted or not (as discussed in Section~\ref{sec:exsitu}).
However, the detection of multiple stellar populations, e.g. high nitrogen, sodium, and aluminum abundances, would be strong evidence that a stellar stream is a globular cluster (\citealt{Bastian18,Gratton19}; or at least contained a disrupted globular cluster, \citealt{Hansen2021,Limberg2024}).
This is more difficult in low-mass stellar streams, because the fraction of 2P enriched stars with high nitrogen, sodium, and aluminum decreases at lower globular cluster masses \citep{Milone2017,Milone2020,Gratton19,Usman24}.
Thus the lack of stars with enhanced nitrogen, sodium, or aluminum in Leiptr prefers a dwarf galaxy, but it is not a strong constraint.

\subsection{Low-Mass, Possibly Ultra-Faint}\label{sec:UFDG}

Based on the chemistry of our stars, we conclude Leiptr's progenitor was probably a low-mass dwarf galaxy with $M_\star \lesssim 10^5 M_\odot$, possibly small enough to be considered an ultra-faint dwarf galaxy.
First, according to the empirical galaxy mass-metallicity relation \citep{Kirby13}, Leiptr would have a stellar mass of 10$^{4.4 \pm 0.5}$ \msun given its mean $\feh = -2.2 \pm 0.15$.
Second, Leiptr’s $\alpha$ abundances are also consistent with low-mass galaxies. This is shown in Figure~\ref{fig:alpha}, where the low magnesium, calcium, and titanium in Leiptr are consistent with ultra-faint or intermediate-mass galaxies with $M_\star \lesssim 10^6 M_\odot$, but not with higher-mass galaxies like Sculptor, Sagittarius, or Fornax with $M_\star \gtrsim 10^7 M_\odot$.
This is also shown with [Mg/Fe] in comparison to intact dwarf galaxies and dwarf galaxy streams in Figure~\ref{fig:dgsummary}, showing Leiptr's [Mg/Fe] is most comparable to Sextans, Draco, Bo\"{o}tes I, Carina II, and Elqui.
These dwarf galaxies have masses between $10^{4.1}$ and $10^{6} \msun$.
Leiptr’s $\alpha$ chemical evolution therefore most closely resembles a low-mass galaxy with mass $\lesssim 10^6$ \msun.

The strongest progenitor mass constraint is from the neutron-capture abundances derived in the Leiptr stars.
Ultra-faint dwarf galaxies have $M_\star < 10^5 M_\odot$, meaning that a typical ultra-faint dwarf galaxy is enriched by fewer than one rare and prolific $r$-process event such as a binary neutron star merger, which typically occurs once per $10^5 M_\odot$ of stars formed \citep{Ji2016b}. 
This explains the empirical result that the vast majority (${\sim}90\%$) of ultra-faint dwarf galaxies have such low strontium, barium, and europium abundances ($\mbox{[Sr,Ba/Fe]} \lesssim -1$; \citealt{Ji19,Frebel23}) that can be used as a chemical tag to identify stars from these low-mass galaxies \citep[e.g.,][]{Casey2017,Roederer2017}. 
Figures~\ref{fig:ncapture} and \ref{fig:dgsummary} highlight that the differences are largest at $\mbox{[Fe/H]} = -2$, where all dwarf galaxies with $M_\star \gtrsim 10^6 M_\odot$ have high abundances of strontium and barium, while those with $M_\star \lesssim 10^5 M_\odot$, like Bo\"otes I and Carina~II, keep low strontium and barium abundances below $-1$.
Leiptr’s neutron-capture abundances are also lower than the previous lowest-mass dwarf galaxy stream documented, Elqui, whose chemical evolution is similar to the dwarf galaxy Draco \citep{Ji20}, suggesting Elqui had a more massive progenitor than Leiptr despite their similar mean metallicity.
This suggests that Leiptr was sufficiently low-mass to still be in the stochastic neutron-capture element regime, or stellar mass $\lesssim 10^5 \msun$.
We note that the origin of the low but non-zero neutron-capture element abundances in most ultra-faint dwarf galaxies remains unknown, though it could be attributed to neutrino-driven winds, magnetorotationally driven jets, spinstars, or other unknown low-yield $r$-process sources \citep{Ji19}.

Leiptr's low mean metallicity of $\feh = -2.2$, low [$\alpha$/Fe]$\sim0$, and low [Sr, Ba/Fe] $\sim -1$ all point toward the progenitor of Leiptr being a very low-mass dwarf galaxy. Given that the intact galaxies with the most similar chemical compositions are ultra-faint dwarf galaxies with $\mstar \lesssim 10^5 \msun$, we conclude that Leiptr's progenitor was likely an ultra-faint dwarf galaxy.
Leiptr is thus the first confirmed spatially and kinematically coherent stellar stream without a known intact progenitor from the myriad ultra-faint dwarf galaxies that should have contributed to the stellar halo \citep{Brauer2019,Brauer2022}.

\subsection{Inhomogeneous Core-Collapse Supernova Enrichment?}\label{sec:SN}

The four Leiptr stars are very close in composition, but they display some interesting chemical differences.
Focusing on the $\alpha$ and odd-Z elements, there is a somewhat decreasing [$\alpha$/Fe] ratio with increasing \feh for the three most metal-poor stars, followed by a sharp increase of both the $\alpha$ and odd-Z elements in the most metal-rich star, Leiptr-321.
These differences are visible in Figures~\ref{fig:alpha} and \ref{fig:oddz}.
The differential analysis in Figure~\ref{fig:diffabund} emphasizes the significant decrease in magnesium, potassium, and calcium [X/Fe] ratios from the metal-poor Leiptr-342 to the more metal-rich Leiptr-186; and a small increase in magnesium, calcium, and titanium [X/Fe] ratios along with a large increase in sodium, aluminum, potassium, and scandium ratios going from the metal-poor Leiptr-208 to the metal-rich Leiptr-321.
This is unusual because [$\alpha$/Fe] is usually expected to monotonically decrease with increasing \feh, and if the odd-Z elements are also produced hydrostatically they should follow this broad trend \citep[e.g.,][]{McWilliam2013}.

We suggest Leiptr-321 may show a signature of inhomogeneous enrichment from a core-collapse supernova.
Inhomogeneous Type~Ia supernova enrichment has previously been suggested as a way to explain unusual magnesium and calcium scatter in Carina \citep{Venn12} and an ``iron-rich'' star in Ursa Minor \citep{McWilliam18}.
In contrast, in Leiptr we see an enhancement of both $\alpha$ and odd-Z elements, which are synthesized by core-collapse supernovae.
If this is inhomogeneous core-collapse supernova enrichment, it would allow probing the nucleosynthesis output of material dominated by a single core-collapse supernova in the same manner that \citet{McWilliam18} did for a Type~Ia supernova.
However, with just four member stars, we believe more observations are needed in Leiptr before such an exploration is warranted.

An alternative explanation is that the increase in $\alpha$ and odd-Z elements is due to a late-time starburst.
Spikes in star formation can cause rising [X/Fe] \citep{Colavitti09} and such a late-time starburst has been suggested to explain a slight [$\alpha$/Fe] increase in the highest metallicity stars in the Magellanic Clouds \citep{Nidever20APOGEE,Hasselquist2021}.
While certainly a valid explanation, we consider this a less likely scenario than inhomogeneous enrichment, as it seems like most stars should then be in this higher-$\alpha$ starburst component.
However, more stars in Leiptr are needed to make any clear conclusions.

\section{Conclusion}\label{sec:Conclusions}
Prior to this work, it was not known whether Leiptr's progenitor was a dwarf galaxy or globular cluster, though its thin morphology and orbit suggested it might be an \textit{ex situ} globular cluster.
We have collected new high-resolution spectroscopic follow-up of four member stars, as well as one non-member star.
The chemical abundances of this stream suggest Leiptr's progenitor was likely a very low-mass galaxy, possibly an ultra-faint dwarf galaxy.
We have four main conclusions:
\begin{itemize}
	\item \emph{Extragalactic Origin:} Leiptr’s extremely retrograde and eccentric orbit makes it clear the progenitor formed \textit{ex-situ} and was later accreted onto the Milky Way. The stars’ low [Mg/Fe] reinforces this, as they are significantly lower than the Milky Way halo stars and globular clusters (see Figure~\ref{fig:alpha}).

	\item \emph{Dwarf Galaxy Progenitor:} The metallicity range between the four probable member stars spans \feh~=~--2.54 to --1.80. We identify a clear metallicity difference between stars with similar stellar parameters (Figure~\ref{fig:diffabund}) and measure a $\sim$0.29 metallicity dispersion using [Fe II/H]. This large spread in metallicity shows that Leiptr is more likely a dwarf galaxy than a globular cluster. 

    \item \emph{Ultra-Faint Dwarf Galaxy Progenitor:} Leiptr's chemical pattern is most similar to ultra-faint dwarf galaxies with $\mstar \lesssim 10^5 \msun$ (Figure~\ref{fig:dgsummary}). The mean metallicity of $\feh = -2.2$, low [Mg/Fe] and other $\alpha$ abundances (Figure~\ref{fig:alpha}), and low strontium and barium abundances (Figure~\ref{fig:ncapture}) all suggest a low-mass galaxy.
    
	\item \emph{Inhomogeneous Mixing?:} We find an intriguing and unusual enrichment in the highest-metallicity star, Leiptr-321, which shows a significant increase in $\alpha$ and odd-Z elements (Figures~\ref{fig:a+oddz_spec}, \ref{fig:alpha}, and \ref{fig:oddz}). This abundance difference could occur due to inhomogeneously mixed material from an individual core-collapse supernova. However, another possible explanation is that Leiptr experienced a late-time starburst.
\end{itemize}

Leiptr is thus the lowest-mass dwarf galaxy stream known, and likely the first identified ultra-faint dwarf galaxy stream without a known intact progenitor.
However, due to our small sample of member stars, further high-resolution spectroscopic follow-up is needed for a more confident mass estimate, as well as being able to verify evidence for inhomogeneous supernova enrichment.

\section*{Acknowledgments}

This paper includes data gathered with the 6.5 m Magellan Telescopes, located at Las Campanas Observatory, Chile.
We thank the referees for their comments, which have helped improve this paper, and Joss Bland-Hawthorn and other members of the S5 collaboration for helpful discussions.
KRA also thanks Donald Terndrup for thoughtful feedback on an earlier version of this paper submitted to the Ohio State University Libraries' Knowledge Bank Repository.
KRA acknowledges this material is based upon work supported by the National Science Foundation Graduate Research Fellowship under Grant No. 2234693.
SAU acknowledges support from the American Association of University Women through the American Dissertation Fellowship.
SAU and APJ acknowledge support from the National Science Foundation under grants AST-2206264 and AST-2307599.
TTH acknowledges support from the Swedish Research Council (VR 2021-05556). 
GFL, SLM, and DBZ acknowledge support from the Australian Research Council through Discovery Project grant DP220102254.

This work has made use of data from the European Space Agency (ESA) mission {\it Gaia} (\url{https://www.cosmos.esa.int/gaia}), processed by the {\it Gaia} Data Processing and Analysis Consortium (DPAC, \url{https://www.cosmos.esa.int/web/gaia/dpac/consortium}).
Funding for the DPAC has been provided by national institutions, in particular the institutions participating in the {\it Gaia} Multilateral Agreement.

This work was supported by the Australian Research Council Centre of Excellence for All Sky Astrophysics in 3 Dimensions (ASTRO 3D), through project number CE170100013.

{\it Software:} 
{\code{numpy} \citep{numpy}, 
\code{scipy} \citep{scipy}, 
\code{matplotlib} \citep{matplotlib}, 
\code{seaborn} \citep{seaborn},
\code{astropy} \citep{astropy,astropy:2018},
\code{pandas} \citep{pandas},
\code{dynesty} \citep{Higson2019,dynesty,Koposov2022},
\code{MOOG} \citep{Sneden73,Sobeck11}
}

\bibliography{main}{}

\begin{thebibliography}{}
\makeatletter
\relax
\def\mn@urlcharsother{\let\do\@makeother \do\$\do\&\do\#\do\^\do\_\do\%\do\~}
\def\mn@doi{\begingroup\mn@urlcharsother \@ifnextchar [ {\mn@doi@} {\mn@doi@[]}}
\def\mn@doi@[#1]#2{\def\@tempa{#1}\ifx\@tempa\@empty \href {http://dx.doi.org/#2} {doi:#2}\else \href {http://dx.doi.org/#2} {#1}\fi \endgroup}
\def\mn@eprint#1#2{\mn@eprint@#1:#2::\@nil}
\def\mn@eprint@arXiv#1{\href {http://arxiv.org/abs/#1} {{\tt arXiv:#1}}}
\def\mn@eprint@dblp#1{\href {http://dblp.uni-trier.de/rec/bibtex/#1.xml} {dblp:#1}}
\def\mn@eprint@#1:#2:#3:#4\@nil{\def\@tempa {#1}\def\@tempb {#2}\def\@tempc {#3}\ifx \@tempc \@empty \let \@tempc \@tempb \let \@tempb \@tempa \fi \ifx \@tempb \@empty \def\@tempb {arXiv}\fi \@ifundefined {mn@eprint@\@tempb}{\@tempb:\@tempc}{\expandafter \expandafter \csname mn@eprint@\@tempb\endcsname \expandafter{\@tempc}}}

\bibitem[\protect\citeauthoryear{{Abohalima} \& {Frebel}}{{Abohalima} \& {Frebel}}{2018}]{Abohalima18}
{Abohalima} A.,  {Frebel} A.,  2018, \mn@doi [\apjs] {10.3847/1538-4365/aadfe9}, \href {https://ui.adsabs.harvard.edu/abs/2018ApJS..238...36A} {238, 36}

\bibitem[\protect\citeauthoryear{{Amarsi}, {Nissen}, {Asplund}, {Lind}  \& {Barklem}}{{Amarsi} et~al.}{2019a}]{Amarsi19a}
{Amarsi} A.~M.,  {Nissen} P.~E.,  {Asplund} M.,  {Lind} K.,   {Barklem} P.~S.,  2019a, \mn@doi [\aap] {10.1051/0004-6361/201834480}, \href {https://ui.adsabs.harvard.edu/abs/2019A&A...622L...4A} {622, L4}

\bibitem[\protect\citeauthoryear{{Amarsi}, {Nissen}  \& {Sk{\'u}lad{\'o}ttir}}{{Amarsi} et~al.}{2019b}]{Amarsi19b}
{Amarsi} A.~M.,  {Nissen} P.~E.,   {Sk{\'u}lad{\'o}ttir} {\'A}.,  2019b, \mn@doi [\aap] {10.1051/0004-6361/201936265}, \href {https://ui.adsabs.harvard.edu/abs/2019A&A...630A.104A} {630, A104}

\bibitem[\protect\citeauthoryear{{Amarsi}, {Grevesse}, {Grumer}, {Asplund}, {Barklem}  \& {Collet}}{{Amarsi} et~al.}{2020}]{Amarsi20}
{Amarsi} A.~M.,  {Grevesse} N.,  {Grumer} J.,  {Asplund} M.,  {Barklem} P.~S.,   {Collet} R.,  2020, \mn@doi [\aap] {10.1051/0004-6361/202037890}, \href {https://ui.adsabs.harvard.edu/abs/2020A&A...636A.120A} {636, A120}

\bibitem[\protect\citeauthoryear{{Asplund}, {Grevesse}, {Sauval}  \& {Scott}}{{Asplund} et~al.}{2009}]{Asplund2009}
{Asplund} M.,  {Grevesse} N.,  {Sauval} A.~J.,   {Scott} P.,  2009, \mn@doi [\araa] {10.1146/annurev.astro.46.060407.145222}, \href {https://ui.adsabs.harvard.edu/abs/2009ARA&A..47..481A} {47, 481}

\bibitem[\protect\citeauthoryear{{Astropy Collaboration} et~al.,}{{Astropy Collaboration} et~al.}{2013}]{astropy}
{Astropy Collaboration} et~al., 2013, \mn@doi [\aap] {10.1051/0004-6361/201322068}, \href {http://adsabs.harvard.edu/abs/2013A%26A...558A..33A} {558, A33}

\bibitem[\protect\citeauthoryear{{Awad} et~al.,}{{Awad} et~al.}{2024}]{Awad2024}
{Awad} P.,  et~al., 2024, \mn@doi [\aap] {10.1051/0004-6361/202347848}, \href {https://ui.adsabs.harvard.edu/abs/2024A&A...683A..14A} {683, A14}

\bibitem[\protect\citeauthoryear{{Barklem} et~al.,}{{Barklem} et~al.}{2005}]{Barklem2005}
{Barklem} P.~S.,  et~al., 2005, \mn@doi [\aap] {10.1051/0004-6361:20052967}, \href {https://ui.adsabs.harvard.edu/abs/2005A&A...439..129B} {439, 129}

\bibitem[\protect\citeauthoryear{{Bastian} \& {Lardo}}{{Bastian} \& {Lardo}}{2018}]{Bastian18}
{Bastian} N.,  {Lardo} C.,  2018, \mn@doi [\araa] {10.1146/annurev-astro-081817-051839}, \href {https://ui.adsabs.harvard.edu/abs/2018ARA&A..56...83B} {56, 83}

\bibitem[\protect\citeauthoryear{{Belokurov} et~al.,}{{Belokurov} et~al.}{2007}]{Belokurov2007}
{Belokurov} V.,  et~al., 2007, \mn@doi [\apj] {10.1086/511302}, \href {https://ui.adsabs.harvard.edu/abs/2007ApJ...658..337B} {658, 337}

\bibitem[\protect\citeauthoryear{{Belokurov}, {Erkal}, {Evans}, {Koposov}  \& {Deason}}{{Belokurov} et~al.}{2018}]{Belokurov18}
{Belokurov} V.,  {Erkal} D.,  {Evans} N.~W.,  {Koposov} S.~E.,   {Deason} A.~J.,  2018, \mn@doi [\mnras] {10.1093/mnras/sty982}, \href {https://ui.adsabs.harvard.edu/abs/2018MNRAS.478..611B} {478, 611}

\bibitem[\protect\citeauthoryear{{Bergemann}}{{Bergemann}}{2011}]{Bergemann11}
{Bergemann} M.,  2011, \mn@doi [\mnras] {10.1111/j.1365-2966.2011.18295.x}, \href {https://ui.adsabs.harvard.edu/abs/2011MNRAS.413.2184B} {413, 2184}

\bibitem[\protect\citeauthoryear{{Bergemann} \& {Gehren}}{{Bergemann} \& {Gehren}}{2008}]{Bergemann08}
{Bergemann} M.,  {Gehren} T.,  2008, \mn@doi [\aap] {10.1051/0004-6361:200810098}, \href {https://ui.adsabs.harvard.edu/abs/2008A&A...492..823B} {492, 823}

\bibitem[\protect\citeauthoryear{{Bergemann}, {Lind}, {Collet}, {Magic}  \& {Asplund}}{{Bergemann} et~al.}{2012a}]{Bergemann12c}
{Bergemann} M.,  {Lind} K.,  {Collet} R.,  {Magic} Z.,   {Asplund} M.,  2012a, \mn@doi [\mnras] {10.1111/j.1365-2966.2012.21687.x}, \href {https://ui.adsabs.harvard.edu/abs/2012MNRAS.427...27B} {427, 27}

\bibitem[\protect\citeauthoryear{{Bergemann}, {Hansen}, {Bautista}  \& {Ruchti}}{{Bergemann} et~al.}{2012b}]{Bergemann12b}
{Bergemann} M.,  {Hansen} C.~J.,  {Bautista} M.,   {Ruchti} G.,  2012b, \mn@doi [\aap] {10.1051/0004-6361/201219406}, \href {https://ui.adsabs.harvard.edu/abs/2012A&A...546A..90B} {546, A90}

\bibitem[\protect\citeauthoryear{{Bergemann}, {Kudritzki}, {Plez}, {Davies}, {Lind}  \& {Gazak}}{{Bergemann} et~al.}{2012c}]{Bergemann12a}
{Bergemann} M.,  {Kudritzki} R.-P.,  {Plez} B.,  {Davies} B.,  {Lind} K.,   {Gazak} Z.,  2012c, \mn@doi [\apj] {10.1088/0004-637X/751/2/156}, \href {https://ui.adsabs.harvard.edu/abs/2012ApJ...751..156B} {751, 156}

\bibitem[\protect\citeauthoryear{{Bergemann}, {Collet}, {Amarsi}, {Kovalev}, {Ruchti}  \& {Magic}}{{Bergemann} et~al.}{2017}]{Bergemann17}
{Bergemann} M.,  {Collet} R.,  {Amarsi} A.~M.,  {Kovalev} M.,  {Ruchti} G.,   {Magic} Z.,  2017, \mn@doi [\apj] {10.3847/1538-4357/aa88cb}, \href {https://ui.adsabs.harvard.edu/abs/2017ApJ...847...15B} {847, 15}

\bibitem[\protect\citeauthoryear{{Bergemann} et~al.,}{{Bergemann} et~al.}{2019}]{Bergemann19}
{Bergemann} M.,  et~al., 2019, \mn@doi [\aap] {10.1051/0004-6361/201935811}, \href {https://ui.adsabs.harvard.edu/abs/2019A&A...631A..80B} {631, A80}

\bibitem[\protect\citeauthoryear{{Bernstein}, {Shectman}, {Gunnels}, {Mochnacki}  \& {Athey}}{{Bernstein} et~al.}{2003}]{Bernstein03}
{Bernstein} R.,  {Shectman} S.~A.,  {Gunnels} S.~M.,  {Mochnacki} S.,   {Athey} A.~E.,  2003, in {Iye} M.,  {Moorwood} A. F.~M.,  eds,  Society of Photo-Optical Instrumentation Engineers (SPIE) Conference Series Vol. 4841, Instrument Design and Performance for Optical/Infrared Ground-based Telescopes. pp 1694--1704, \mn@doi{10.1117/12.461502}

\bibitem[\protect\citeauthoryear{{Bonaca} \& {Price-Whelan}}{{Bonaca} \& {Price-Whelan}}{2025}]{Bonaca24}
{Bonaca} A.,  {Price-Whelan} A.~M.,  2025, \mn@doi [\nar] {10.1016/j.newar.2024.101713}, \href {https://ui.adsabs.harvard.edu/abs/2025NewAR.10001713B} {100, 101713}

\bibitem[\protect\citeauthoryear{{Bonaca} et~al.,}{{Bonaca} et~al.}{2021}]{Bonaca21}
{Bonaca} A.,  et~al., 2021, \mn@doi [\apjl] {10.3847/2041-8213/abeaa9}, \href {https://ui.adsabs.harvard.edu/abs/2021ApJ...909L..26B} {909, L26}

\bibitem[\protect\citeauthoryear{{Brauer}, {Ji}, {Frebel}, {Dooley}, {G{\'o}mez}  \& {O'Shea}}{{Brauer} et~al.}{2019}]{Brauer2019}
{Brauer} K.,  {Ji} A.~P.,  {Frebel} A.,  {Dooley} G.~A.,  {G{\'o}mez} F.~A.,   {O'Shea} B.~W.,  2019, \mn@doi [\apj] {10.3847/1538-4357/aafafb}, \href {https://ui.adsabs.harvard.edu/abs/2019ApJ...871..247B} {871, 247}

\bibitem[\protect\citeauthoryear{{Brauer}, {Andales}, {Ji}, {Frebel}, {Mardini}, {G{\'o}mez}  \& {O'Shea}}{{Brauer} et~al.}{2022}]{Brauer2022}
{Brauer} K.,  {Andales} H.~D.,  {Ji} A.~P.,  {Frebel} A.,  {Mardini} M.~K.,  {G{\'o}mez} F.~A.,   {O'Shea} B.~W.,  2022, \mn@doi [\apj] {10.3847/1538-4357/ac85b9}, \href {https://ui.adsabs.harvard.edu/abs/2022ApJ...937...14B} {937, 14}

\bibitem[\protect\citeauthoryear{{Bullock} \& {Boylan-Kolchin}}{{Bullock} \& {Boylan-Kolchin}}{2017}]{Bullock2017}
{Bullock} J.~S.,  {Boylan-Kolchin} M.,  2017, \mn@doi [\araa] {10.1146/annurev-astro-091916-055313}, \href {https://ui.adsabs.harvard.edu/abs/2017ARA&A..55..343B} {55, 343}

\bibitem[\protect\citeauthoryear{{Bullock} \& {Johnston}}{{Bullock} \& {Johnston}}{2005}]{Bullock05}
{Bullock} J.~S.,  {Johnston} K.~V.,  2005, \mn@doi [\apj] {10.1086/497422}, \href {https://ui.adsabs.harvard.edu/abs/2005ApJ...635..931B} {635, 931}

\bibitem[\protect\citeauthoryear{{Carlberg}}{{Carlberg}}{2009}]{Carlberg09}
{Carlberg} R.~G.,  2009, \mn@doi [\apjl] {10.1088/0004-637X/705/2/L223}, \href {https://ui.adsabs.harvard.edu/abs/2009ApJ...705L.223C} {705, L223}

\bibitem[\protect\citeauthoryear{{Carlberg}}{{Carlberg}}{2018}]{Carlberg2018}
{Carlberg} R.~G.,  2018, \mn@doi [\apj] {10.3847/1538-4357/aac88a}, \href {https://ui.adsabs.harvard.edu/abs/2018ApJ...861...69C} {861, 69}

\bibitem[\protect\citeauthoryear{{Carretta}, {Bragaglia}, {Gratton}, {D'Orazi}  \& {Lucatello}}{{Carretta} et~al.}{2009}]{Carretta09}
{Carretta} E.,  {Bragaglia} A.,  {Gratton} R.,  {D'Orazi} V.,   {Lucatello} S.,  2009, \mn@doi [\aap] {10.1051/0004-6361/200913003}, \href {https://ui.adsabs.harvard.edu/abs/2009A&A...508..695C} {508, 695}

\bibitem[\protect\citeauthoryear{{Carrillo}, {Hawkins}, {Jofr{\'e}}, {de Brito Silva}, {Das}  \& {Lucey}}{{Carrillo} et~al.}{2022}]{Carrillo22}
{Carrillo} A.,  {Hawkins} K.,  {Jofr{\'e}} P.,  {de Brito Silva} D.,  {Das} P.,   {Lucey} M.,  2022, \mn@doi [\mnras] {10.1093/mnras/stac518}, \href {https://ui.adsabs.harvard.edu/abs/2022MNRAS.513.1557C} {513, 1557}

\bibitem[\protect\citeauthoryear{{Casey}}{{Casey}}{2014}]{Casey14}
{Casey} A.~R.,  2014, PhD thesis, Australian National University, Canberra

\bibitem[\protect\citeauthoryear{{Casey} \& {Schlaufman}}{{Casey} \& {Schlaufman}}{2017}]{Casey2017}
{Casey} A.~R.,  {Schlaufman} K.~C.,  2017, \mn@doi [\apj] {10.3847/1538-4357/aa9079}, \href {https://ui.adsabs.harvard.edu/abs/2017ApJ...850..179C} {850, 179}

\bibitem[\protect\citeauthoryear{{Casey}, {Da Costa}, {Keller}  \& {Maunder}}{{Casey} et~al.}{2013}]{Casey2013}
{Casey} A.~R.,  {Da Costa} G.,  {Keller} S.~C.,   {Maunder} E.,  2013, \mn@doi [\apj] {10.1088/0004-637X/764/1/39}, \href {https://ui.adsabs.harvard.edu/abs/2013ApJ...764...39C} {764, 39}

\bibitem[\protect\citeauthoryear{{Casey} et~al.,}{{Casey} et~al.}{2021}]{Casey2021}
{Casey} A.~R.,  et~al., 2021, \mn@doi [\apj] {10.3847/1538-4357/ac1346}, \href {https://ui.adsabs.harvard.edu/abs/2021ApJ...921...67C} {921, 67}

\bibitem[\protect\citeauthoryear{{Castelli} \& {Kurucz}}{{Castelli} \& {Kurucz}}{2003}]{Castelli03}
{Castelli} F.,  {Kurucz} R.~L.,  2003, in {Piskunov} N.,  {Weiss} W.~W.,   {Gray} D.~F.,  eds,  IAU Symposium Vol. 210, Modelling of Stellar Atmospheres. p.~A20 (\mn@eprint {arXiv} {astro-ph/0405087}), \mn@doi{10.48550/arXiv.astro-ph/0405087}

\bibitem[\protect\citeauthoryear{{Chandra} et~al.,}{{Chandra} et~al.}{2022}]{Chandra2022}
{Chandra} V.,  et~al., 2022, \mn@doi [\apj] {10.3847/1538-4357/ac9b4b}, \href {https://ui.adsabs.harvard.edu/abs/2022ApJ...940..127C} {940, 127}

\bibitem[\protect\citeauthoryear{{Chiti}, {Frebel}, {Ji}, {Jerjen}, {Kim}  \& {Norris}}{{Chiti} et~al.}{2018}]{Chiti2018}
{Chiti} A.,  {Frebel} A.,  {Ji} A.~P.,  {Jerjen} H.,  {Kim} D.,   {Norris} J.~E.,  2018, \mn@doi [\apj] {10.3847/1538-4357/aab4fc}, \href {https://ui.adsabs.harvard.edu/abs/2018ApJ...857...74C} {857, 74}

\bibitem[\protect\citeauthoryear{{Chiti} et~al.,}{{Chiti} et~al.}{2023}]{Chiti2023}
{Chiti} A.,  et~al., 2023, \mn@doi [\aj] {10.3847/1538-3881/aca416}, \href {https://ui.adsabs.harvard.edu/abs/2023AJ....165...55C} {165, 55}

\bibitem[\protect\citeauthoryear{{Choksi}, {Gnedin}  \& {Li}}{{Choksi} et~al.}{2018}]{Choksi2018}
{Choksi} N.,  {Gnedin} O.~Y.,   {Li} H.,  2018, \mn@doi [\mnras] {10.1093/mnras/sty1952}, \href {https://ui.adsabs.harvard.edu/abs/2018MNRAS.480.2343C} {480, 2343}

\bibitem[\protect\citeauthoryear{{Cohen}}{{Cohen}}{2009}]{Cohen09}
{Cohen} J.,  2009, {Detailed Abundances for the Draco Carbon Stars and for Low Metallicity Sculptor Stars}, Keck Observatory Archive HIRES, id.C218Hr

\bibitem[\protect\citeauthoryear{{Cohen} \& {Huang}}{{Cohen} \& {Huang}}{2010}]{Cohen10}
{Cohen} J.~G.,  {Huang} W.,  2010, \mn@doi [\apj] {10.1088/0004-637X/719/1/931}, \href {https://ui.adsabs.harvard.edu/abs/2010ApJ...719..931C} {719, 931}

\bibitem[\protect\citeauthoryear{{Cohen} \& {Kirby}}{{Cohen} \& {Kirby}}{2012}]{Cohen12}
{Cohen} J.~G.,  {Kirby} E.~N.,  2012, \mn@doi [\apj] {10.1088/0004-637X/760/1/86}, \href {https://ui.adsabs.harvard.edu/abs/2012ApJ...760...86C} {760, 86}

\bibitem[\protect\citeauthoryear{{Cohen}, {Briley}  \& {Stetson}}{{Cohen} et~al.}{2005}]{Cohen05}
{Cohen} J.~G.,  {Briley} M.~M.,   {Stetson} P.~B.,  2005, \mn@doi [\aj] {10.1086/431974}, \href {https://ui.adsabs.harvard.edu/abs/2005AJ....130.1177C} {130, 1177}

\bibitem[\protect\citeauthoryear{{Cohen}, {Christlieb}, {Thompson}, {McWilliam}, {Shectman}, {Reimers}, {Wisotzki}  \& {Kirby}}{{Cohen} et~al.}{2013}]{Cohen2013}
{Cohen} J.~G.,  {Christlieb} N.,  {Thompson} I.,  {McWilliam} A.,  {Shectman} S.,  {Reimers} D.,  {Wisotzki} L.,   {Kirby} E.,  2013, \mn@doi [\apj] {10.1088/0004-637X/778/1/56}, \href {https://ui.adsabs.harvard.edu/abs/2013ApJ...778...56C} {778, 56}

\bibitem[\protect\citeauthoryear{{Colavitti}, {Pipino}  \& {Matteucci}}{{Colavitti} et~al.}{2009}]{Colavitti09}
{Colavitti} E.,  {Pipino} A.,   {Matteucci} F.,  2009, \mn@doi [\aap] {10.1051/0004-6361/200811379}, \href {https://ui.adsabs.harvard.edu/abs/2009A&A...499..409C} {499, 409}

\bibitem[\protect\citeauthoryear{{Coppi}, {Zinn}, {Baltay}, {Rabinowitz}, {Girard}, {Howard}, {Ment}  \& {Rahman}}{{Coppi} et~al.}{2024}]{Coppi2024}
{Coppi} P.~S.,  {Zinn} R.,  {Baltay} C.,  {Rabinowitz} D.,  {Girard} T.,  {Howard} R.,  {Ment} K.,   {Rahman} R.,  2024, \mn@doi [\mnras] {10.1093/mnras/stae488}, \href {https://ui.adsabs.harvard.edu/abs/2024MNRAS.tmp..533C} {}

\bibitem[\protect\citeauthoryear{{Dark Energy Survey Collaboration} et~al.,}{{Dark Energy Survey Collaboration} et~al.}{2016}]{DES16}
{Dark Energy Survey Collaboration} et~al., 2016, \mn@doi [\mnras] {10.1093/mnras/stw641}, \href {https://ui.adsabs.harvard.edu/abs/2016MNRAS.460.1270D} {460, 1270}

\bibitem[\protect\citeauthoryear{{Deason} \& {Belokurov}}{{Deason} \& {Belokurov}}{2024}]{Deason2024}
{Deason} A.~J.,  {Belokurov} V.,  2024, \mn@doi [\nar] {10.1016/j.newar.2024.101706}, \href {https://ui.adsabs.harvard.edu/abs/2024NewAR..9901706D} {99, 101706}

\bibitem[\protect\citeauthoryear{{Dodd}, {Callingham}, {Helmi}, {Matsuno}, {Ruiz-Lara}, {Balbinot}  \& {L{\"o}vdal}}{{Dodd} et~al.}{2023}]{Dodd2023}
{Dodd} E.,  {Callingham} T.~M.,  {Helmi} A.,  {Matsuno} T.,  {Ruiz-Lara} T.,  {Balbinot} E.,   {L{\"o}vdal} S.,  2023, \mn@doi [\aap] {10.1051/0004-6361/202244546}, \href {https://ui.adsabs.harvard.edu/abs/2023A&A...670L...2D} {670, L2}

\bibitem[\protect\citeauthoryear{{Drlica-Wagner} et~al.,}{{Drlica-Wagner} et~al.}{2015}]{DrlicaWagner15}
{Drlica-Wagner} A.,  et~al., 2015, \mn@doi [\apj] {10.1088/0004-637X/813/2/109}, \href {https://ui.adsabs.harvard.edu/abs/2015ApJ...813..109D} {813, 109}

\bibitem[\protect\citeauthoryear{{Eitner}, {Bergemann}, {Ruiter}, {Avril}, {Seitenzahl}, {Gent}  \& {C{\^o}t{\'e}}}{{Eitner} et~al.}{2023}]{Eitner23}
{Eitner} P.,  {Bergemann} M.,  {Ruiter} A.~J.,  {Avril} O.,  {Seitenzahl} I.~R.,  {Gent} M.~R.,   {C{\^o}t{\'e}} B.,  2023, \mn@doi [\aap] {10.1051/0004-6361/202244286}, \href {https://ui.adsabs.harvard.edu/abs/2023A&A...677A.151E} {677, A151}

\bibitem[\protect\citeauthoryear{{Erkal}, {Sanders}  \& {Belokurov}}{{Erkal} et~al.}{2016a}]{Erkal2016}
{Erkal} D.,  {Sanders} J.~L.,   {Belokurov} V.,  2016a, \mn@doi [\mnras] {10.1093/mnras/stw1400}, \href {https://ui.adsabs.harvard.edu/abs/2016MNRAS.461.1590E} {461, 1590}

\bibitem[\protect\citeauthoryear{{Erkal}, {Belokurov}, {Bovy}  \& {Sanders}}{{Erkal} et~al.}{2016b}]{Erkal16}
{Erkal} D.,  {Belokurov} V.,  {Bovy} J.,   {Sanders} J.~L.,  2016b, \mn@doi [\mnras] {10.1093/mnras/stw1957}, \href {https://ui.adsabs.harvard.edu/abs/2016MNRAS.463..102E} {463, 102}

\bibitem[\protect\citeauthoryear{{Erkal} et~al.,}{{Erkal} et~al.}{2019}]{Erkal19}
{Erkal} D.,  et~al., 2019, \mn@doi [\mnras] {10.1093/mnras/stz1371}, \href {https://ui.adsabs.harvard.edu/abs/2019MNRAS.487.2685E} {487, 2685}

\bibitem[\protect\citeauthoryear{{Ezzeddine}, {Frebel}  \& {Plez}}{{Ezzeddine} et~al.}{2017}]{Ezzeddine17}
{Ezzeddine} R.,  {Frebel} A.,   {Plez} B.,  2017, \mn@doi [\apj] {10.3847/1538-4357/aa8875}, \href {https://ui.adsabs.harvard.edu/abs/2017ApJ...847..142E} {847, 142}

\bibitem[\protect\citeauthoryear{{Ezzeddine} et~al.,}{{Ezzeddine} et~al.}{2019}]{Ezzeddine19}
{Ezzeddine} R.,  et~al., 2019, \mn@doi [\apj] {10.3847/1538-4357/ab14e7}, \href {https://ui.adsabs.harvard.edu/abs/2019ApJ...876...97E} {876, 97}

\bibitem[\protect\citeauthoryear{{Feltzing}, {Eriksson}, {Kleyna}  \& {Wilkinson}}{{Feltzing} et~al.}{2009}]{Feltzing2009}
{Feltzing} S.,  {Eriksson} K.,  {Kleyna} J.,   {Wilkinson} M.~I.,  2009, \mn@doi [\aap] {10.1051/0004-6361/200912833}, \href {https://ui.adsabs.harvard.edu/abs/2009A&A...508L...1F} {508, L1}

\bibitem[\protect\citeauthoryear{{Forbes} et~al.,}{{Forbes} et~al.}{2018}]{Forbes2018}
{Forbes} D.~A.,  et~al., 2018, \mn@doi [Proceedings of the Royal Society of London Series A] {10.1098/rspa.2017.0616}, \href {https://ui.adsabs.harvard.edu/abs/2018RSPSA.47470616F} {474, 20170616}

\bibitem[\protect\citeauthoryear{{Fran{\c{c}}ois}, {Monaco}, {Bonifacio}, {Moni Bidin}, {Geisler}  \& {Sbordone}}{{Fran{\c{c}}ois} et~al.}{2016}]{Francois2016}
{Fran{\c{c}}ois} P.,  {Monaco} L.,  {Bonifacio} P.,  {Moni Bidin} C.,  {Geisler} D.,   {Sbordone} L.,  2016, \mn@doi [\aap] {10.1051/0004-6361/201527181}, \href {https://ui.adsabs.harvard.edu/abs/2016A&A...588A...7F} {588, A7}

\bibitem[\protect\citeauthoryear{{Frebel} \& {Bromm}}{{Frebel} \& {Bromm}}{2012}]{Frebel2012}
{Frebel} A.,  {Bromm} V.,  2012, \mn@doi [\apj] {10.1088/0004-637X/759/2/115}, \href {https://ui.adsabs.harvard.edu/abs/2012ApJ...759..115F} {759, 115}

\bibitem[\protect\citeauthoryear{{Frebel} \& {Ji}}{{Frebel} \& {Ji}}{2023}]{Frebel23}
{Frebel} A.,  {Ji} A.~P.,  2023, \mn@doi [arXiv e-prints] {10.48550/arXiv.2302.09188}, \href {https://ui.adsabs.harvard.edu/abs/2023arXiv230209188F} {p. arXiv:2302.09188}

\bibitem[\protect\citeauthoryear{{Frebel} \& {Norris}}{{Frebel} \& {Norris}}{2015}]{Frebel15}
{Frebel} A.,  {Norris} J.~E.,  2015, \mn@doi [\araa] {10.1146/annurev-astro-082214-122423}, \href {https://ui.adsabs.harvard.edu/abs/2015ARA&A..53..631F} {53, 631}

\bibitem[\protect\citeauthoryear{{Frebel}, {Simon}, {Geha}  \& {Willman}}{{Frebel} et~al.}{2010}]{Frebel2010}
{Frebel} A.,  {Simon} J.~D.,  {Geha} M.,   {Willman} B.,  2010, \mn@doi [\apj] {10.1088/0004-637X/708/1/560}, \href {https://ui.adsabs.harvard.edu/abs/2010ApJ...708..560F} {708, 560}

\bibitem[\protect\citeauthoryear{{Frebel}, {Casey}, {Jacobson}  \& {Yu}}{{Frebel} et~al.}{2013a}]{Frebel13}
{Frebel} A.,  {Casey} A.~R.,  {Jacobson} H.~R.,   {Yu} Q.,  2013a, \mn@doi [\apj] {10.1088/0004-637X/769/1/57}, \href {https://ui.adsabs.harvard.edu/abs/2013ApJ...769...57F} {769, 57}

\bibitem[\protect\citeauthoryear{{Frebel}, {Lunnan}, {Casey}, {Norris}, {Wyse}  \& {Gilmore}}{{Frebel} et~al.}{2013b}]{Frebel2013}
{Frebel} A.,  {Lunnan} R.,  {Casey} A.~R.,  {Norris} J.~E.,  {Wyse} R. F.~G.,   {Gilmore} G.,  2013b, \mn@doi [\apj] {10.1088/0004-637X/771/1/39}, \href {https://ui.adsabs.harvard.edu/abs/2013ApJ...771...39F} {771, 39}

\bibitem[\protect\citeauthoryear{{Frebel}, {Simon}  \& {Kirby}}{{Frebel} et~al.}{2014}]{Frebel2014}
{Frebel} A.,  {Simon} J.~D.,   {Kirby} E.~N.,  2014, \mn@doi [\apj] {10.1088/0004-637X/786/1/74}, \href {https://ui.adsabs.harvard.edu/abs/2014ApJ...786...74F} {786, 74}

\bibitem[\protect\citeauthoryear{{Freeman} \& {Bland-Hawthorn}}{{Freeman} \& {Bland-Hawthorn}}{2002}]{Freeman02}
{Freeman} K.,  {Bland-Hawthorn} J.,  2002, \mn@doi [\araa] {10.1146/annurev.astro.40.060401.093840}, \href {https://ui.adsabs.harvard.edu/abs/2002ARA&A..40..487F} {40, 487}

\bibitem[\protect\citeauthoryear{{Fu} et~al.,}{{Fu} et~al.}{2018}]{Fu2018}
{Fu} S.~W.,  et~al., 2018, \mn@doi [\apj] {10.3847/1538-4357/aad9f9}, \href {https://ui.adsabs.harvard.edu/abs/2018ApJ...866...42F} {866, 42}

\bibitem[\protect\citeauthoryear{{Gaia Collaboration} et~al.,}{{Gaia Collaboration} et~al.}{2016}]{GAIA16}
{Gaia Collaboration} et~al., 2016, \mn@doi [\aap] {10.1051/0004-6361/201629272}, \href {https://ui.adsabs.harvard.edu/abs/2016A&A...595A...1G} {595, A1}

\bibitem[\protect\citeauthoryear{{Gaia Collaboration} et~al.,}{{Gaia Collaboration} et~al.}{2023}]{GAIA22}
{Gaia Collaboration} et~al., 2023, \mn@doi [\aap] {10.1051/0004-6361/202243940}, \href {https://ui.adsabs.harvard.edu/abs/2023A&A...674A...1G} {674, A1}

\bibitem[\protect\citeauthoryear{{Gandhi} et~al.,}{{Gandhi} et~al.}{2024}]{Gandhi2024}
{Gandhi} P.~J.,  et~al., 2024, \mn@doi [\mnras] {10.1093/mnras/stae1584}, \href {https://ui.adsabs.harvard.edu/abs/2024MNRAS.533.1059G} {533, 1059}

\bibitem[\protect\citeauthoryear{{Gilmore}, {Norris}, {Monaco}, {Yong}, {Wyse}  \& {Geisler}}{{Gilmore} et~al.}{2013}]{Gilmore2013}
{Gilmore} G.,  {Norris} J.~E.,  {Monaco} L.,  {Yong} D.,  {Wyse} R. F.~G.,   {Geisler} D.,  2013, \mn@doi [\apj] {10.1088/0004-637X/763/1/61}, \href {https://ui.adsabs.harvard.edu/abs/2013ApJ...763...61G} {763, 61}

\bibitem[\protect\citeauthoryear{{G{\'o}mez}, {Helmi}, {Cooper}, {Frenk}, {Navarro}  \& {White}}{{G{\'o}mez} et~al.}{2013}]{Gomez2013}
{G{\'o}mez} F.~A.,  {Helmi} A.,  {Cooper} A.~P.,  {Frenk} C.~S.,  {Navarro} J.~F.,   {White} S. D.~M.,  2013, \mn@doi [\mnras] {10.1093/mnras/stt1838}, \href {https://ui.adsabs.harvard.edu/abs/2013MNRAS.436.3602G} {436, 3602}

\bibitem[\protect\citeauthoryear{{Gratton}, {Bragaglia}, {Carretta}, {D'Orazi}, {Lucatello}  \& {Sollima}}{{Gratton} et~al.}{2019}]{Gratton19}
{Gratton} R.,  {Bragaglia} A.,  {Carretta} E.,  {D'Orazi} V.,  {Lucatello} S.,   {Sollima} A.,  2019, \mn@doi [\aapr] {10.1007/s00159-019-0119-3}, \href {https://ui.adsabs.harvard.edu/abs/2019A&ARv..27....8G} {27, 8}

\bibitem[\protect\citeauthoryear{{Grillmair} \& {Carlin}}{{Grillmair} \& {Carlin}}{2016}]{Grillmair16}
{Grillmair} C.~J.,  {Carlin} J.~L.,  2016, in {Newberg} H.~J.,  {Carlin} J.~L.,  eds,  Astrophysics and Space Science Library Vol. 420, Tidal Streams in the Local Group and Beyond. p.~87 (\mn@eprint {arXiv} {1603.08936}), \mn@doi{10.1007/978-3-319-19336-6_4}

\bibitem[\protect\citeauthoryear{{Grillmair} \& {Dionatos}}{{Grillmair} \& {Dionatos}}{2006}]{Grillmair2006}
{Grillmair} C.~J.,  {Dionatos} O.,  2006, \mn@doi [\apjl] {10.1086/505111}, \href {https://ui.adsabs.harvard.edu/abs/2006ApJ...643L..17G} {643, L17}

\bibitem[\protect\citeauthoryear{{Gudin} et~al.,}{{Gudin} et~al.}{2021}]{Gudin2021}
{Gudin} D.,  et~al., 2021, \mn@doi [\apj] {10.3847/1538-4357/abd7ed}, \href {https://ui.adsabs.harvard.edu/abs/2021ApJ...908...79G} {908, 79}

\bibitem[\protect\citeauthoryear{{Gull}, {Frebel}, {Hinojosa}, {Roederer}, {Ji}  \& {Brauer}}{{Gull} et~al.}{2021}]{Gull2021}
{Gull} M.,  {Frebel} A.,  {Hinojosa} K.,  {Roederer} I.~U.,  {Ji} A.~P.,   {Brauer} K.,  2021, \mn@doi [\apj] {10.3847/1538-4357/abea1a}, \href {https://ui.adsabs.harvard.edu/abs/2021ApJ...912...52G} {912, 52}

\bibitem[\protect\citeauthoryear{{Guo} et~al.,}{{Guo} et~al.}{2025}]{Guo25}
{Guo} Y.,  et~al., 2025, \mn@doi [\aap] {10.1051/0004-6361/202451536}, \href {https://ui.adsabs.harvard.edu/abs/2025A&A...693A.211G} {693, A211}

\bibitem[\protect\citeauthoryear{{Hansen} et~al.,}{{Hansen} et~al.}{2017}]{Hansen2017}
{Hansen} T.~T.,  et~al., 2017, \mn@doi [\apj] {10.3847/1538-4357/aa634a}, \href {https://ui.adsabs.harvard.edu/abs/2017ApJ...838...44H} {838, 44}

\bibitem[\protect\citeauthoryear{{Hansen} et~al.,}{{Hansen} et~al.}{2020a}]{Hansen2020grus}
{Hansen} T.~T.,  et~al., 2020a, \mn@doi [\apj] {10.3847/1538-4357/ab9643}, \href {https://ui.adsabs.harvard.edu/abs/2020ApJ...897..183H} {897, 183}

\bibitem[\protect\citeauthoryear{{Hansen}, {Riley}, {Strigari}, {Marshall}, {Ferguson}, {Zepeda}  \& {Sneden}}{{Hansen} et~al.}{2020b}]{Hansen2020}
{Hansen} T.~T.,  {Riley} A.~H.,  {Strigari} L.~E.,  {Marshall} J.~L.,  {Ferguson} P.~S.,  {Zepeda} J.,   {Sneden} C.,  2020b, \mn@doi [\apj] {10.3847/1538-4357/ababa5}, \href {https://ui.adsabs.harvard.edu/abs/2020ApJ...901...23H} {901, 23}

\bibitem[\protect\citeauthoryear{{Hansen} et~al.,}{{Hansen} et~al.}{2021}]{Hansen2021}
{Hansen} T.~T.,  et~al., 2021, \mn@doi [\apj] {10.3847/1538-4357/abfc54}, \href {https://ui.adsabs.harvard.edu/abs/2021ApJ...915..103H} {915, 103}

\bibitem[\protect\citeauthoryear{{Hansen}, {Simon}, {Li}, {Sharkey}, {Ji}, {Thompson}, {Reggiani}  \& {Galarza}}{{Hansen} et~al.}{2024}]{Hansen2024}
{Hansen} T.~T.,  {Simon} J.~D.,  {Li} T.~S.,  {Sharkey} D.,  {Ji} A.~P.,  {Thompson} I.~B.,  {Reggiani} H.~M.,   {Galarza} J.~Y.,  2024, \mn@doi [\apj] {10.3847/1538-4357/ad3a52}, \href {https://ui.adsabs.harvard.edu/abs/2024ApJ...968...21H} {968, 21}

\bibitem[\protect\citeauthoryear{{Hasselquist} et~al.,}{{Hasselquist} et~al.}{2021}]{Hasselquist2021}
{Hasselquist} S.,  et~al., 2021, \mn@doi [\apj] {10.3847/1538-4357/ac25f9}, \href {https://ui.adsabs.harvard.edu/abs/2021ApJ...923..172H} {923, 172}

\bibitem[\protect\citeauthoryear{{Hawkins} et~al.,}{{Hawkins} et~al.}{2023}]{Hawkins23}
{Hawkins} K.,  et~al., 2023, \mn@doi [\apj] {10.3847/1538-4357/acb698}, \href {https://ui.adsabs.harvard.edu/abs/2023ApJ...948..123H} {948, 123}

\bibitem[\protect\citeauthoryear{{Haywood}, {Di Matteo}, {Lehnert}, {Snaith}, {Khoperskov}  \& {G{\'o}mez}}{{Haywood} et~al.}{2018}]{Haywood18}
{Haywood} M.,  {Di Matteo} P.,  {Lehnert} M.~D.,  {Snaith} O.,  {Khoperskov} S.,   {G{\'o}mez} A.,  2018, \mn@doi [\apj] {10.3847/1538-4357/aad235}, \href {https://ui.adsabs.harvard.edu/abs/2018ApJ...863..113H} {863, 113}

\bibitem[\protect\citeauthoryear{{Helmi}}{{Helmi}}{2020}]{Helmi20}
{Helmi} A.,  2020, \mn@doi [\araa] {10.1146/annurev-astro-032620-021917}, \href {https://ui.adsabs.harvard.edu/abs/2020ARA&A..58..205H} {58, 205}

\bibitem[\protect\citeauthoryear{{Helmi} \& {White}}{{Helmi} \& {White}}{1999}]{Helmi99}
{Helmi} A.,  {White} S. D.~M.,  1999, \mn@doi [\mnras] {10.1046/j.1365-8711.1999.02616.x}, \href {https://ui.adsabs.harvard.edu/abs/1999MNRAS.307..495H} {307, 495}

\bibitem[\protect\citeauthoryear{{Helmi}, {White}, {de Zeeuw}  \& {Zhao}}{{Helmi} et~al.}{1999}]{Helmi1999}
{Helmi} A.,  {White} S. D.~M.,  {de Zeeuw} P.~T.,   {Zhao} H.,  1999, \mn@doi [\nat] {10.1038/46980}, \href {https://ui.adsabs.harvard.edu/abs/1999Natur.402...53H} {402, 53}

\bibitem[\protect\citeauthoryear{{Helmi}, {Babusiaux}, {Koppelman}, {Massari}, {Veljanoski}  \& {Brown}}{{Helmi} et~al.}{2018}]{Helmi18}
{Helmi} A.,  {Babusiaux} C.,  {Koppelman} H.~H.,  {Massari} D.,  {Veljanoski} J.,   {Brown} A. G.~A.,  2018, \mn@doi [\nat] {10.1038/s41586-018-0625-x}, \href {https://ui.adsabs.harvard.edu/abs/2018Natur.563...85H} {563, 85}

\bibitem[\protect\citeauthoryear{{Higson}, {Handley}, {Hobson}  \& {Lasenby}}{{Higson} et~al.}{2019}]{Higson2019}
{Higson} E.,  {Handley} W.,  {Hobson} M.,   {Lasenby} A.,  2019, \mn@doi [Statistics and Computing] {10.1007/s11222-018-9844-0}, \href {https://ui.adsabs.harvard.edu/abs/2019S&C....29..891H} {29, 891}

\bibitem[\protect\citeauthoryear{{Hill} et~al.,}{{Hill} et~al.}{2019}]{Hill19}
{Hill} V.,  et~al., 2019, \mn@doi [\aap] {10.1051/0004-6361/201833950}, \href {https://ui.adsabs.harvard.edu/abs/2019A&A...626A..15H} {626, A15}

\bibitem[\protect\citeauthoryear{Hunter}{Hunter}{2007}]{matplotlib}
Hunter J.~D.,  2007, \mn@doi [Computing in Science \& Engineering] {http://dx.doi.org/10.1109/MCSE.2007.55}, 9, 90

\bibitem[\protect\citeauthoryear{{Ibata}, {Irwin}, {Lewis}  \& {Stolte}}{{Ibata} et~al.}{2001}]{Ibata2001}
{Ibata} R.,  {Irwin} M.,  {Lewis} G.~F.,   {Stolte} A.,  2001, \mn@doi [\apjl] {10.1086/318894}, \href {https://ui.adsabs.harvard.edu/abs/2001ApJ...547L.133I} {547, L133}

\bibitem[\protect\citeauthoryear{{Ibata}, {Lewis}, {Irwin}  \& {Quinn}}{{Ibata} et~al.}{2002}]{Ibata02}
{Ibata} R.~A.,  {Lewis} G.~F.,  {Irwin} M.~J.,   {Quinn} T.,  2002, \mn@doi [\mnras] {10.1046/j.1365-8711.2002.05358.x}, \href {https://ui.adsabs.harvard.edu/abs/2002MNRAS.332..915I} {332, 915}

\bibitem[\protect\citeauthoryear{{Ibata}, {Malhan}  \& {Martin}}{{Ibata} et~al.}{2019}]{Ibata19}
{Ibata} R.~A.,  {Malhan} K.,   {Martin} N.~F.,  2019, \mn@doi [\apj] {10.3847/1538-4357/ab0080}, \href {https://ui.adsabs.harvard.edu/abs/2019ApJ...872..152I} {872, 152}

\bibitem[\protect\citeauthoryear{{Ibata} et~al.,}{{Ibata} et~al.}{2024}]{Ibata24}
{Ibata} R.,  et~al., 2024, \mn@doi [\apj] {10.3847/1538-4357/ad382d}, \href {https://ui.adsabs.harvard.edu/abs/2024ApJ...967...89I} {967, 89}

\bibitem[\protect\citeauthoryear{{Ishigaki}, {Aoki}, {Arimoto}  \& {Okamoto}}{{Ishigaki} et~al.}{2014}]{Ishigaki2014}
{Ishigaki} M.~N.,  {Aoki} W.,  {Arimoto} N.,   {Okamoto} S.,  2014, \mn@doi [\aap] {10.1051/0004-6361/201322796}, \href {https://ui.adsabs.harvard.edu/abs/2014A&A...562A.146I} {562, A146}

\bibitem[\protect\citeauthoryear{{Jacobson} et~al.,}{{Jacobson} et~al.}{2015}]{Jacobson2015}
{Jacobson} H.~R.,  et~al., 2015, \mn@doi [\apj] {10.1088/0004-637X/807/2/171}, \href {https://ui.adsabs.harvard.edu/abs/2015ApJ...807..171J} {807, 171}

\bibitem[\protect\citeauthoryear{{Jethwa}, {Erkal}  \& {Belokurov}}{{Jethwa} et~al.}{2018}]{Jethwa2018}
{Jethwa} P.,  {Erkal} D.,   {Belokurov} V.,  2018, \mn@doi [\mnras] {10.1093/mnras/stx2330}, \href {https://ui.adsabs.harvard.edu/abs/2018MNRAS.473.2060J} {473, 2060}

\bibitem[\protect\citeauthoryear{{Ji}, {Frebel}, {Chiti}  \& {Simon}}{{Ji} et~al.}{2016a}]{Ji2016b}
{Ji} A.~P.,  {Frebel} A.,  {Chiti} A.,   {Simon} J.~D.,  2016a, \mn@doi [\nat] {10.1038/nature17425}, \href {https://ui.adsabs.harvard.edu/abs/2016Natur.531..610J} {531, 610}

\bibitem[\protect\citeauthoryear{{Ji}, {Frebel}, {Simon}  \& {Geha}}{{Ji} et~al.}{2016b}]{Ji2016a}
{Ji} A.~P.,  {Frebel} A.,  {Simon} J.~D.,   {Geha} M.,  2016b, \mn@doi [\apj] {10.3847/0004-637X/817/1/41}, \href {https://ui.adsabs.harvard.edu/abs/2016ApJ...817...41J} {817, 41}

\bibitem[\protect\citeauthoryear{{Ji}, {Frebel}, {Simon}  \& {Chiti}}{{Ji} et~al.}{2016c}]{Ji2016c}
{Ji} A.~P.,  {Frebel} A.,  {Simon} J.~D.,   {Chiti} A.,  2016c, \mn@doi [\apj] {10.3847/0004-637X/830/2/93}, \href {https://ui.adsabs.harvard.edu/abs/2016ApJ...830...93J} {830, 93}

\bibitem[\protect\citeauthoryear{{Ji}, {Simon}, {Frebel}, {Venn}  \& {Hansen}}{{Ji} et~al.}{2019}]{Ji19}
{Ji} A.~P.,  {Simon} J.~D.,  {Frebel} A.,  {Venn} K.~A.,   {Hansen} T.~T.,  2019, \mn@doi [\apj] {10.3847/1538-4357/aaf3bb}, \href {https://ui.adsabs.harvard.edu/abs/2019ApJ...870...83J} {870, 83}

\bibitem[\protect\citeauthoryear{{Ji} et~al.,}{{Ji} et~al.}{2020a}]{Ji20}
{Ji} A.~P.,  et~al., 2020a, \mn@doi [\aj] {10.3847/1538-3881/abacb6}, \href {https://ui.adsabs.harvard.edu/abs/2020AJ....160..181J} {160, 181}

\bibitem[\protect\citeauthoryear{{Ji} et~al.,}{{Ji} et~al.}{2020b}]{Ji20car}
{Ji} A.~P.,  et~al., 2020b, \mn@doi [\apj] {10.3847/1538-4357/ab6213}, \href {https://ui.adsabs.harvard.edu/abs/2020ApJ...889...27J} {889, 27}

\bibitem[\protect\citeauthoryear{{Ji} et~al.,}{{Ji} et~al.}{2021}]{Ji2021}
{Ji} A.~P.,  et~al., 2021, \mn@doi [\apj] {10.3847/1538-4357/ac1869}, \href {https://ui.adsabs.harvard.edu/abs/2021ApJ...921...32J} {921, 32}

\bibitem[\protect\citeauthoryear{{Ji} et~al.,}{{Ji} et~al.}{2025}]{Ji25}
{Ji} A.~P.,  et~al., 2025, {LESSPayne: Labeling Echelle Spectra with SMHR and Payne}, Astrophysics Source Code Library, record ascl:2503.025

\bibitem[\protect\citeauthoryear{{Johnston}}{{Johnston}}{1998}]{Johnston98}
{Johnston} K.~V.,  1998, \mn@doi [\apj] {10.1086/305273}, \href {https://ui.adsabs.harvard.edu/abs/1998ApJ...495..297J} {495, 297}

\bibitem[\protect\citeauthoryear{{Johnston}, {Spergel}  \& {Haydn}}{{Johnston} et~al.}{2002}]{Johnston02}
{Johnston} K.~V.,  {Spergel} D.~N.,   {Haydn} C.,  2002, \mn@doi [\apj] {10.1086/339791}, \href {https://ui.adsabs.harvard.edu/abs/2002ApJ...570..656J} {570, 656}

\bibitem[\protect\citeauthoryear{{Johnston}, {Bullock}, {Sharma}, {Font}, {Robertson}  \& {Leitner}}{{Johnston} et~al.}{2008}]{Johnston2008}
{Johnston} K.~V.,  {Bullock} J.~S.,  {Sharma} S.,  {Font} A.,  {Robertson} B.~E.,   {Leitner} S.~N.,  2008, \mn@doi [\apj] {10.1086/592228}, \href {https://ui.adsabs.harvard.edu/abs/2008ApJ...689..936J} {689, 936}

\bibitem[\protect\citeauthoryear{Jones, Oliphant, Peterson  et~al.}{Jones et~al.}{2001}]{scipy}
Jones E.,  Oliphant T.,  Peterson P.,   et~al., 2001, {SciPy}: Open source scientific tools for {Python}, \url {http://www.scipy.org/}

\bibitem[\protect\citeauthoryear{{J{\"o}nsson} et~al.,}{{J{\"o}nsson} et~al.}{2020}]{Jonsson}
{J{\"o}nsson} H.,  et~al., 2020, \mn@doi [\aj] {10.3847/1538-3881/aba592}, \href {https://ui.adsabs.harvard.edu/abs/2020AJ....160..120J} {160, 120}

\bibitem[\protect\citeauthoryear{Kelson}{Kelson}{2003}]{Kelson_2003}
Kelson D.~D.,  2003, \mn@doi [Publications of the Astronomical Society of the Pacific] {10.1086/375502}, 115, 688

\bibitem[\protect\citeauthoryear{{Kirby}, {Cohen}, {Smith}, {Majewski}, {Sohn}  \& {Guhathakurta}}{{Kirby} et~al.}{2011}]{Kirby2011}
{Kirby} E.~N.,  {Cohen} J.~G.,  {Smith} G.~H.,  {Majewski} S.~R.,  {Sohn} S.~T.,   {Guhathakurta} P.,  2011, \mn@doi [\apj] {10.1088/0004-637X/727/2/79}, \href {https://ui.adsabs.harvard.edu/abs/2011ApJ...727...79K} {727, 79}

\bibitem[\protect\citeauthoryear{{Kirby}, {Cohen}, {Guhathakurta}, {Cheng}, {Bullock}  \& {Gallazzi}}{{Kirby} et~al.}{2013}]{Kirby13}
{Kirby} E.~N.,  {Cohen} J.~G.,  {Guhathakurta} P.,  {Cheng} L.,  {Bullock} J.~S.,   {Gallazzi} A.,  2013, \mn@doi [\apj] {10.1088/0004-637X/779/2/102}, \href {https://ui.adsabs.harvard.edu/abs/2013ApJ...779..102K} {779, 102}

\bibitem[\protect\citeauthoryear{{Kirby}, {Cohen}, {Simon}, {Guhathakurta}, {Thygesen}  \& {Duggan}}{{Kirby} et~al.}{2017}]{Kirby2017}
{Kirby} E.~N.,  {Cohen} J.~G.,  {Simon} J.~D.,  {Guhathakurta} P.,  {Thygesen} A.~O.,   {Duggan} G.~E.,  2017, \mn@doi [\apj] {10.3847/1538-4357/aa6570}, \href {https://ui.adsabs.harvard.edu/abs/2017ApJ...838...83K} {838, 83}

\bibitem[\protect\citeauthoryear{{Kirby}, {Gilbert}, {Escala}, {Wojno}, {Guhathakurta}, {Majewski}  \& {Beaton}}{{Kirby} et~al.}{2020}]{Kirby20}
{Kirby} E.~N.,  {Gilbert} K.~M.,  {Escala} I.,  {Wojno} J.,  {Guhathakurta} P.,  {Majewski} S.~R.,   {Beaton} R.~L.,  2020, \mn@doi [\aj] {10.3847/1538-3881/ab5f0f}, \href {https://ui.adsabs.harvard.edu/abs/2020AJ....159...46K} {159, 46}

\bibitem[\protect\citeauthoryear{{Kirby}, {Ji}  \& {Kovalev}}{{Kirby} et~al.}{2023}]{Kirby23}
{Kirby} E.~N.,  {Ji} A.~P.,   {Kovalev} M.,  2023, \mn@doi [\apj] {10.3847/1538-4357/acf309}, \href {https://ui.adsabs.harvard.edu/abs/2023ApJ...958...45K} {958, 45}

\bibitem[\protect\citeauthoryear{{Koch}, {McWilliam}, {Grebel}, {Zucker}  \& {Belokurov}}{{Koch} et~al.}{2008}]{Koch2008}
{Koch} A.,  {McWilliam} A.,  {Grebel} E.~K.,  {Zucker} D.~B.,   {Belokurov} V.,  2008, \mn@doi [\apjl] {10.1086/595001}, \href {https://ui.adsabs.harvard.edu/abs/2008ApJ...688L..13K} {688, L13}

\bibitem[\protect\citeauthoryear{{Koch}, {Feltzing}, {Ad{\'e}n}  \& {Matteucci}}{{Koch} et~al.}{2013}]{Koch2013}
{Koch} A.,  {Feltzing} S.,  {Ad{\'e}n} D.,   {Matteucci} F.,  2013, \mn@doi [\aap] {10.1051/0004-6361/201220742}, \href {https://ui.adsabs.harvard.edu/abs/2013A&A...554A...5K} {554, A5}

\bibitem[\protect\citeauthoryear{{Koposov} et~al.,}{{Koposov} et~al.}{2019}]{Koposov2019}
{Koposov} S.~E.,  et~al., 2019, \mn@doi [\mnras] {10.1093/mnras/stz457}, \href {https://ui.adsabs.harvard.edu/abs/2019MNRAS.485.4726K} {485, 4726}

\bibitem[\protect\citeauthoryear{{Koposov} et~al.,}{{Koposov} et~al.}{2022}]{Koposov2022}
{Koposov} S.,  et~al., 2022, {joshspeagle/dynesty: v2.0.3}, Zenodo, \mn@doi{10.5281/zenodo.7388523}

\bibitem[\protect\citeauthoryear{{Koppelman}, {Helmi}, {Massari}, {Price-Whelan}  \& {Starkenburg}}{{Koppelman} et~al.}{2019}]{Koppelman2019}
{Koppelman} H.~H.,  {Helmi} A.,  {Massari} D.,  {Price-Whelan} A.~M.,   {Starkenburg} T.~K.,  2019, \mn@doi [\aap] {10.1051/0004-6361/201936738}, \href {https://ui.adsabs.harvard.edu/abs/2019A&A...631L...9K} {631, L9}

\bibitem[\protect\citeauthoryear{{Lee}, {Beers}  \& {Kim}}{{Lee} et~al.}{2019}]{Lee19}
{Lee} Y.~S.,  {Beers} T.~C.,   {Kim} Y.~K.,  2019, \mn@doi [\apj] {10.3847/1538-4357/ab4791}, \href {https://ui.adsabs.harvard.edu/abs/2019ApJ...885..102L} {885, 102}

\bibitem[\protect\citeauthoryear{{Lemasle} et~al.,}{{Lemasle} et~al.}{2014}]{Lemasle14}
{Lemasle} B.,  et~al., 2014, \mn@doi [\aap] {10.1051/0004-6361/201423919}, \href {https://ui.adsabs.harvard.edu/abs/2014A&A...572A..88L} {572, A88}

\bibitem[\protect\citeauthoryear{{Li} et~al.,}{{Li} et~al.}{2018}]{Li2018}
{Li} T.~S.,  et~al., 2018, \mn@doi [\apj] {10.3847/1538-4357/aadf91}, \href {https://ui.adsabs.harvard.edu/abs/2018ApJ...866...22L} {866, 22}

\bibitem[\protect\citeauthoryear{{Li} et~al.,}{{Li} et~al.}{2019}]{Li19}
{Li} T.~S.,  et~al., 2019, \mn@doi [\mnras] {10.1093/mnras/stz2731}, \href {https://ui.adsabs.harvard.edu/abs/2019MNRAS.490.3508L} {490, 3508}

\bibitem[\protect\citeauthoryear{{Li} et~al.,}{{Li} et~al.}{2021}]{Li2021}
{Li} T.~S.,  et~al., 2021, \mn@doi [\apj] {10.3847/1538-4357/abeb18}, \href {https://ui.adsabs.harvard.edu/abs/2021ApJ...911..149L} {911, 149}

\bibitem[\protect\citeauthoryear{{Li} et~al.,}{{Li} et~al.}{2022}]{Li22}
{Li} T.~S.,  et~al., 2022, \mn@doi [\apj] {10.3847/1538-4357/ac46d3}, \href {https://ui.adsabs.harvard.edu/abs/2022ApJ...928...30L} {928, 30}

\bibitem[\protect\citeauthoryear{{Limberg} et~al.,}{{Limberg} et~al.}{2021a}]{Limberg2021UFD}
{Limberg} G.,  et~al., 2021a, \mn@doi [\apj] {10.3847/1538-4357/abcb87}, \href {https://ui.adsabs.harvard.edu/abs/2021ApJ...907...10L} {907, 10}

\bibitem[\protect\citeauthoryear{{Limberg} et~al.,}{{Limberg} et~al.}{2021b}]{Limberg2021}
{Limberg} G.,  et~al., 2021b, \mn@doi [\apjl] {10.3847/2041-8213/ac0056}, \href {https://ui.adsabs.harvard.edu/abs/2021ApJ...913L..28L} {913, L28}

\bibitem[\protect\citeauthoryear{{Limberg} et~al.,}{{Limberg} et~al.}{2024}]{Limberg2024}
{Limberg} G.,  et~al., 2024, \mn@doi [\mnras] {10.1093/mnras/stae969}, \href {https://ui.adsabs.harvard.edu/abs/2024MNRAS.530.2512L} {530, 2512}

\bibitem[\protect\citeauthoryear{{Lind}, {Asplund}, {Barklem}  \& {Belyaev}}{{Lind} et~al.}{2011}]{Lind11}
{Lind} K.,  {Asplund} M.,  {Barklem} P.~S.,   {Belyaev} A.~K.,  2011, \mn@doi [\aap] {10.1051/0004-6361/201016095}, \href {https://ui.adsabs.harvard.edu/abs/2011A&A...528A.103L} {528, A103}

\bibitem[\protect\citeauthoryear{{Lindegren} et~al.,}{{Lindegren} et~al.}{2021}]{Lindegren21}
{Lindegren} L.,  et~al., 2021, \mn@doi [\aap] {10.1051/0004-6361/202039709}, \href {https://ui.adsabs.harvard.edu/abs/2021A&A...649A...2L} {649, A2}

\bibitem[\protect\citeauthoryear{{Majewski}, {Skrutskie}, {Weinberg}  \& {Ostheimer}}{{Majewski} et~al.}{2003}]{Majewski2003}
{Majewski} S.~R.,  {Skrutskie} M.~F.,  {Weinberg} M.~D.,   {Ostheimer} J.~C.,  2003, \mn@doi [\apj] {10.1086/379504}, \href {https://ui.adsabs.harvard.edu/abs/2003ApJ...599.1082M} {599, 1082}

\bibitem[\protect\citeauthoryear{{Malhan}, {Valluri}  \& {Freese}}{{Malhan} et~al.}{2021}]{Malhan2021}
{Malhan} K.,  {Valluri} M.,   {Freese} K.,  2021, \mn@doi [\mnras] {10.1093/mnras/staa3597}, \href {https://ui.adsabs.harvard.edu/abs/2021MNRAS.501..179M} {501, 179}

\bibitem[\protect\citeauthoryear{{Malhan} et~al.,}{{Malhan} et~al.}{2022}]{Malhan2022}
{Malhan} K.,  et~al., 2022, \mn@doi [\apj] {10.3847/1538-4357/ac4d2a}, \href {https://ui.adsabs.harvard.edu/abs/2022ApJ...926..107M} {926, 107}

\bibitem[\protect\citeauthoryear{{Marshall} et~al.,}{{Marshall} et~al.}{2019}]{Marshall2019}
{Marshall} J.~L.,  et~al., 2019, \mn@doi [\apj] {10.3847/1538-4357/ab3653}, \href {https://ui.adsabs.harvard.edu/abs/2019ApJ...882..177M} {882, 177}

\bibitem[\protect\citeauthoryear{{Martin} et~al.,}{{Martin} et~al.}{2022}]{Martin2022}
{Martin} N.~F.,  et~al., 2022, \mn@doi [\nat] {10.1038/s41586-021-04162-2}, \href {https://ui.adsabs.harvard.edu/abs/2022Natur.601...45M} {601, 45}

\bibitem[\protect\citeauthoryear{{Mashonkina}}{{Mashonkina}}{2013}]{Mashonkina13}
{Mashonkina} L.,  2013, \mn@doi [\aap] {10.1051/0004-6361/201220761}, \href {https://ui.adsabs.harvard.edu/abs/2013A&A...550A..28M} {550, A28}

\bibitem[\protect\citeauthoryear{{Mashonkina} \& {Gehren}}{{Mashonkina} \& {Gehren}}{2001}]{Mashonkina01}
{Mashonkina} L.,  {Gehren} T.,  2001, \mn@doi [\aap] {10.1051/0004-6361:20010965}, \href {https://ui.adsabs.harvard.edu/abs/2001A&A...376..232M} {376, 232}

\bibitem[\protect\citeauthoryear{{Mashonkina} \& {Ryabchikova}}{{Mashonkina} \& {Ryabchikova}}{2024}]{Mashonkina24}
{Mashonkina} L.~I.,  {Ryabchikova} T.~A.,  2024, \mn@doi [arXiv e-prints] {10.48550/arXiv.2406.11367}, \href {https://ui.adsabs.harvard.edu/abs/2024arXiv240611367M} {p. arXiv:2406.11367}

\bibitem[\protect\citeauthoryear{{Mashonkina}, {Gehren}  \& {Bikmaev}}{{Mashonkina} et~al.}{1999}]{Mashonkina99}
{Mashonkina} L.,  {Gehren} T.,   {Bikmaev} I.,  1999, \aap, \href {https://ui.adsabs.harvard.edu/abs/1999A&A...343..519M} {343, 519}

\bibitem[\protect\citeauthoryear{{Mashonkina}, {Korn}  \& {Przybilla}}{{Mashonkina} et~al.}{2007}]{Mashonkina07}
{Mashonkina} L.,  {Korn} A.~J.,   {Przybilla} N.,  2007, \mn@doi [\aap] {10.1051/0004-6361:20065999}, \href {https://ui.adsabs.harvard.edu/abs/2007A&A...461..261M} {461, 261}

\bibitem[\protect\citeauthoryear{{Mashonkina}, {Jablonka}, {Sitnova}, {Pakhomov}  \& {North}}{{Mashonkina} et~al.}{2017}]{Mashonkina17}
{Mashonkina} L.,  {Jablonka} P.,  {Sitnova} T.,  {Pakhomov} Y.,   {North} P.,  2017, \mn@doi [\aap] {10.1051/0004-6361/201731582}, \href {https://ui.adsabs.harvard.edu/abs/2017A&A...608A..89M} {608, A89}

\bibitem[\protect\citeauthoryear{{Mateu}}{{Mateu}}{2023}]{Mateu2023}
{Mateu} C.,  2023, \mn@doi [\mnras] {10.1093/mnras/stad321}, \href {https://ui.adsabs.harvard.edu/abs/2023MNRAS.520.5225M} {520, 5225}

\bibitem[\protect\citeauthoryear{{Mateu}, {Read}  \& {Kawata}}{{Mateu} et~al.}{2018}]{Mateu18}
{Mateu} C.,  {Read} J.~I.,   {Kawata} D.,  2018, \mn@doi [\mnras] {10.1093/mnras/stx2937}, \href {https://ui.adsabs.harvard.edu/abs/2018MNRAS.474.4112M} {474, 4112}

\bibitem[\protect\citeauthoryear{{Matsuno}, {Aoki}  \& {Suda}}{{Matsuno} et~al.}{2019}]{Matsuno2019}
{Matsuno} T.,  {Aoki} W.,   {Suda} T.,  2019, \mn@doi [\apjl] {10.3847/2041-8213/ab0ec0}, \href {https://ui.adsabs.harvard.edu/abs/2019ApJ...874L..35M} {874, L35}

\bibitem[\protect\citeauthoryear{{Matsuno}, {Koppelman}, {Helmi}, {Aoki}, {Ishigaki}, {Suda}, {Yuan}  \& {Hattori}}{{Matsuno} et~al.}{2022a}]{Matsuno2022}
{Matsuno} T.,  {Koppelman} H.~H.,  {Helmi} A.,  {Aoki} W.,  {Ishigaki} M.~N.,  {Suda} T.,  {Yuan} Z.,   {Hattori} K.,  2022a, \mn@doi [\aap] {10.1051/0004-6361/202142752}, \href {https://ui.adsabs.harvard.edu/abs/2022A&A...661A.103M} {661, A103}

\bibitem[\protect\citeauthoryear{{Matsuno} et~al.,}{{Matsuno} et~al.}{2022b}]{Matsuno2022b}
{Matsuno} T.,  et~al., 2022b, \mn@doi [\aap] {10.1051/0004-6361/202243609}, \href {https://ui.adsabs.harvard.edu/abs/2022A&A...665A..46M} {665, A46}

\bibitem[\protect\citeauthoryear{{Matteucci} \& {Brocato}}{{Matteucci} \& {Brocato}}{1990}]{Matteucci1990}
{Matteucci} F.,  {Brocato} E.,  1990, \mn@doi [\apj] {10.1086/169508}, \href {https://ui.adsabs.harvard.edu/abs/1990ApJ...365..539M} {365, 539}

\bibitem[\protect\citeauthoryear{{McKenzie} et~al.,}{{McKenzie} et~al.}{2022}]{McKenzie2022}
{McKenzie} M.,  et~al., 2022, \mn@doi [\mnras] {10.1093/mnras/stac2254}, \href {https://ui.adsabs.harvard.edu/abs/2022MNRAS.516.3515M} {516, 3515}

\bibitem[\protect\citeauthoryear{McKinney}{McKinney}{2010}]{pandas}
McKinney W.,  2010. pp 56--61, \mn@doi{10.25080/Majora-92bf1922-00a}

\bibitem[\protect\citeauthoryear{{McWilliam}}{{McWilliam}}{1997}]{McWilliam97}
{McWilliam} A.,  1997, \mn@doi [\araa] {10.1146/annurev.astro.35.1.503}, \href {https://ui.adsabs.harvard.edu/abs/1997ARA&A..35..503M} {35, 503}

\bibitem[\protect\citeauthoryear{{McWilliam}, {Wallerstein}  \& {Mottini}}{{McWilliam} et~al.}{2013}]{McWilliam2013}
{McWilliam} A.,  {Wallerstein} G.,   {Mottini} M.,  2013, \mn@doi [\apj] {10.1088/0004-637X/778/2/149}, \href {https://ui.adsabs.harvard.edu/abs/2013ApJ...778..149M} {778, 149}

\bibitem[\protect\citeauthoryear{{McWilliam}, {Piro}, {Badenes}  \& {Bravo}}{{McWilliam} et~al.}{2018}]{McWilliam18}
{McWilliam} A.,  {Piro} A.~L.,  {Badenes} C.,   {Bravo} E.,  2018, \mn@doi [\apj] {10.3847/1538-4357/aab772}, \href {https://ui.adsabs.harvard.edu/abs/2018ApJ...857...97M} {857, 97}

\bibitem[\protect\citeauthoryear{{Milone} et~al.,}{{Milone} et~al.}{2017}]{Milone2017}
{Milone} A.~P.,  et~al., 2017, \mn@doi [\mnras] {10.1093/mnras/stw2531}, \href {https://ui.adsabs.harvard.edu/abs/2017MNRAS.464.3636M} {464, 3636}

\bibitem[\protect\citeauthoryear{{Milone} et~al.,}{{Milone} et~al.}{2020}]{Milone2020}
{Milone} A.~P.,  et~al., 2020, \mn@doi [\mnras] {10.1093/mnras/stz2999}, \href {https://ui.adsabs.harvard.edu/abs/2020MNRAS.491..515M} {491, 515}

\bibitem[\protect\citeauthoryear{{Mucciarelli}, {Bellazzini}, {Ibata}, {Romano}, {Chapman}  \& {Monaco}}{{Mucciarelli} et~al.}{2017}]{Mucciarelli2017}
{Mucciarelli} A.,  {Bellazzini} M.,  {Ibata} R.,  {Romano} D.,  {Chapman} S.~C.,   {Monaco} L.,  2017, \mn@doi [\aap] {10.1051/0004-6361/201730707}, \href {https://ui.adsabs.harvard.edu/abs/2017A&A...605A..46M} {605, A46}

\bibitem[\protect\citeauthoryear{{Myeong}, {Vasiliev}, {Iorio}, {Evans}  \& {Belokurov}}{{Myeong} et~al.}{2019}]{Myeong19}
{Myeong} G.~C.,  {Vasiliev} E.,  {Iorio} G.,  {Evans} N.~W.,   {Belokurov} V.,  2019, \mn@doi [\mnras] {10.1093/mnras/stz1770}, \href {https://ui.adsabs.harvard.edu/abs/2019MNRAS.488.1235M} {488, 1235}

\bibitem[\protect\citeauthoryear{{Nadler} et~al.,}{{Nadler} et~al.}{2021}]{Nadler2021}
{Nadler} E.~O.,  et~al., 2021, \mn@doi [\prl] {10.1103/PhysRevLett.126.091101}, \href {https://ui.adsabs.harvard.edu/abs/2021PhRvL.126i1101N} {126, 091101}

\bibitem[\protect\citeauthoryear{{Nagasawa} et~al.,}{{Nagasawa} et~al.}{2018}]{Nagasawa2018}
{Nagasawa} D.~Q.,  et~al., 2018, \mn@doi [\apj] {10.3847/1538-4357/aaa01d}, \href {https://ui.adsabs.harvard.edu/abs/2018ApJ...852...99N} {852, 99}

\bibitem[\protect\citeauthoryear{{Naidu}, {Conroy}, {Bonaca}, {Johnson}, {Ting}, {Caldwell}, {Zaritsky}  \& {Cargile}}{{Naidu} et~al.}{2020}]{Naidu20}
{Naidu} R.~P.,  {Conroy} C.,  {Bonaca} A.,  {Johnson} B.~D.,  {Ting} Y.-S.,  {Caldwell} N.,  {Zaritsky} D.,   {Cargile} P.~A.,  2020, \mn@doi [\apj] {10.3847/1538-4357/abaef4}, \href {https://ui.adsabs.harvard.edu/abs/2020ApJ...901...48N} {901, 48}

\bibitem[\protect\citeauthoryear{{Naidu} et~al.,}{{Naidu} et~al.}{2021}]{Naidu2021}
{Naidu} R.~P.,  et~al., 2021, \mn@doi [\apj] {10.3847/1538-4357/ac2d2d}, \href {https://ui.adsabs.harvard.edu/abs/2021ApJ...923...92N} {923, 92}

\bibitem[\protect\citeauthoryear{{Newberg}, {Yanny}  \& {Willett}}{{Newberg} et~al.}{2009}]{Newberg2009}
{Newberg} H.~J.,  {Yanny} B.,   {Willett} B.~A.,  2009, \mn@doi [\apjl] {10.1088/0004-637X/700/2/L61}, \href {https://ui.adsabs.harvard.edu/abs/2009ApJ...700L..61N} {700, L61}

\bibitem[\protect\citeauthoryear{{Nidever} et~al.,}{{Nidever} et~al.}{2020}]{Nidever20APOGEE}
{Nidever} D.~L.,  et~al., 2020, \mn@doi [\apj] {10.3847/1538-4357/ab7305}, \href {https://ui.adsabs.harvard.edu/abs/2020ApJ...895...88N} {895, 88}

\bibitem[\protect\citeauthoryear{{Nissen} \& {Gustafsson}}{{Nissen} \& {Gustafsson}}{2018}]{Nissen2018}
{Nissen} P.~E.,  {Gustafsson} B.,  2018, \mn@doi [\aapr] {10.1007/s00159-018-0111-3}, \href {https://ui.adsabs.harvard.edu/abs/2018A&ARv..26....6N} {26, 6}

\bibitem[\protect\citeauthoryear{{Nordlander} \& {Lind}}{{Nordlander} \& {Lind}}{2017}]{Nordlander2017}
{Nordlander} T.,  {Lind} K.,  2017, \mn@doi [\aap] {10.1051/0004-6361/201730427}, \href {https://ui.adsabs.harvard.edu/abs/2017A&A...607A..75N} {607, A75}

\bibitem[\protect\citeauthoryear{{Norris}, {Yong}, {Gilmore}  \& {Wyse}}{{Norris} et~al.}{2010a}]{Norris2010a}
{Norris} J.~E.,  {Yong} D.,  {Gilmore} G.,   {Wyse} R. F.~G.,  2010a, \mn@doi [\apj] {10.1088/0004-637X/711/1/350}, \href {https://ui.adsabs.harvard.edu/abs/2010ApJ...711..350N} {711, 350}

\bibitem[\protect\citeauthoryear{{Norris}, {Gilmore}, {Wyse}, {Yong}  \& {Frebel}}{{Norris} et~al.}{2010b}]{Norris2010b}
{Norris} J.~E.,  {Gilmore} G.,  {Wyse} R. F.~G.,  {Yong} D.,   {Frebel} A.,  2010b, \mn@doi [\apjl] {10.1088/2041-8205/722/1/L104}, \href {https://ui.adsabs.harvard.edu/abs/2010ApJ...722L.104N} {722, L104}

\bibitem[\protect\citeauthoryear{{Norris}, {Yong}, {Venn}, {Gilmore}, {Casagrande}  \& {Dotter}}{{Norris} et~al.}{2017}]{Norris17}
{Norris} J.~E.,  {Yong} D.,  {Venn} K.~A.,  {Gilmore} G.,  {Casagrande} L.,   {Dotter} A.,  2017, \mn@doi [\apjs] {10.3847/1538-4365/aa755e}, \href {https://ui.adsabs.harvard.edu/abs/2017ApJS..230...28N} {230, 28}

\bibitem[\protect\citeauthoryear{{Ou}, {Ji}, {Frebel}, {Naidu}  \& {Limberg}}{{Ou} et~al.}{2024}]{Ou24}
{Ou} X.,  {Ji} A.~P.,  {Frebel} A.,  {Naidu} R.~P.,   {Limberg} G.,  2024, \mn@doi [\apj] {10.3847/1538-4357/ad6f9b}, \href {https://ui.adsabs.harvard.edu/abs/2024ApJ...974..232O} {974, 232}

\bibitem[\protect\citeauthoryear{{Pearson}, {Bonaca}, {Chen}  \& {Gnedin}}{{Pearson} et~al.}{2024}]{Pearson2024}
{Pearson} S.,  {Bonaca} A.,  {Chen} Y.,   {Gnedin} O.~Y.,  2024, \mn@doi [\apj] {10.3847/1538-4357/ad8348}, \href {https://ui.adsabs.harvard.edu/abs/2024ApJ...976...54P} {976, 54}

\bibitem[\protect\citeauthoryear{{Placco}, {Sneden}, {Roederer}, {Lawler}, {Den Hartog}, {Hejazi}, {Maas}  \& {Bernath}}{{Placco} et~al.}{2021}]{Placco2021}
{Placco} V.~M.,  {Sneden} C.,  {Roederer} I.~U.,  {Lawler} J.~E.,  {Den Hartog} E.~A.,  {Hejazi} N.,  {Maas} Z.,   {Bernath} P.,  2021, \mn@doi [Research Notes of the American Astronomical Society] {10.3847/2515-5172/abf651}, \href {https://ui.adsabs.harvard.edu/abs/2021RNAAS...5...92P} {5, 92}

\bibitem[\protect\citeauthoryear{{Price-Whelan} \& {Bonaca}}{{Price-Whelan} \& {Bonaca}}{2018}]{Price-Whelan2018}
{Price-Whelan} A.~M.,  {Bonaca} A.,  2018, \mn@doi [\apjl] {10.3847/2041-8213/aad7b5}, \href {https://ui.adsabs.harvard.edu/abs/2018ApJ...863L..20P} {863, L20}

\bibitem[\protect\citeauthoryear{{Price-Whelan} et~al.,}{{Price-Whelan} et~al.}{2018}]{astropy:2018}
{Price-Whelan} A.~M.,  et~al., 2018, \mn@doi [\aj] {10.3847/1538-3881/aabc4f}, \href {https://ui.adsabs.harvard.edu/#abs/2018AJ....156..123T} {156, 123}

\bibitem[\protect\citeauthoryear{{Reggiani} et~al.,}{{Reggiani} et~al.}{2019}]{Reggiani19}
{Reggiani} H.,  et~al., 2019, \mn@doi [\aap] {10.1051/0004-6361/201935156}, \href {https://ui.adsabs.harvard.edu/abs/2019A&A...627A.177R} {627, A177}

\bibitem[\protect\citeauthoryear{{Reina-Campos}, {Kruijssen}, {Pfeffer}, {Bastian}  \& {Crain}}{{Reina-Campos} et~al.}{2019}]{ReinaCampos2019}
{Reina-Campos} M.,  {Kruijssen} J.~M.~D.,  {Pfeffer} J.~L.,  {Bastian} N.,   {Crain} R.~A.,  2019, \mn@doi [\mnras] {10.1093/mnras/stz1236}, \href {https://ui.adsabs.harvard.edu/abs/2019MNRAS.486.5838R} {486, 5838}

\bibitem[\protect\citeauthoryear{{Riley} \& {Strigari}}{{Riley} \& {Strigari}}{2020}]{Riley20}
{Riley} A.~H.,  {Strigari} L.~E.,  2020, \mn@doi [\mnras] {10.1093/mnras/staa710}, \href {https://ui.adsabs.harvard.edu/abs/2020MNRAS.494..983R} {494, 983}

\bibitem[\protect\citeauthoryear{{Roederer}}{{Roederer}}{2017}]{Roederer2017}
{Roederer} I.~U.,  2017, \mn@doi [\apj] {10.3847/1538-4357/835/1/23}, \href {https://ui.adsabs.harvard.edu/abs/2017ApJ...835...23R} {835, 23}

\bibitem[\protect\citeauthoryear{{Roederer} \& {Gnedin}}{{Roederer} \& {Gnedin}}{2019}]{Roederer19}
{Roederer} I.~U.,  {Gnedin} O.~Y.,  2019, \mn@doi [\apj] {10.3847/1538-4357/ab365c}, \href {https://ui.adsabs.harvard.edu/abs/2019ApJ...883...84R} {883, 84}

\bibitem[\protect\citeauthoryear{{Roederer} \& {Kirby}}{{Roederer} \& {Kirby}}{2014}]{Roederer2014segue}
{Roederer} I.~U.,  {Kirby} E.~N.,  2014, \mn@doi [\mnras] {10.1093/mnras/stu491}, \href {https://ui.adsabs.harvard.edu/abs/2014MNRAS.440.2665R} {440, 2665}

\bibitem[\protect\citeauthoryear{{Roederer}, {Sneden}, {Thompson}, {Preston}  \& {Shectman}}{{Roederer} et~al.}{2010}]{Roederer2010}
{Roederer} I.~U.,  {Sneden} C.,  {Thompson} I.~B.,  {Preston} G.~W.,   {Shectman} S.~A.,  2010, \mn@doi [\apj] {10.1088/0004-637X/711/2/573}, \href {https://ui.adsabs.harvard.edu/abs/2010ApJ...711..573R} {711, 573}

\bibitem[\protect\citeauthoryear{{Roederer}, {Marino}  \& {Sneden}}{{Roederer} et~al.}{2011}]{Roederer11}
{Roederer} I.~U.,  {Marino} A.~F.,   {Sneden} C.,  2011, \mn@doi [\apj] {10.1088/0004-637X/742/1/37}, \href {https://ui.adsabs.harvard.edu/abs/2011ApJ...742...37R} {742, 37}

\bibitem[\protect\citeauthoryear{{Roederer}, {Preston}, {Thompson}, {Shectman}, {Sneden}, {Burley}  \& {Kelson}}{{Roederer} et~al.}{2014}]{Roederer14}
{Roederer} I.~U.,  {Preston} G.~W.,  {Thompson} I.~B.,  {Shectman} S.~A.,  {Sneden} C.,  {Burley} G.~S.,   {Kelson} D.~D.,  2014, \mn@doi [\aj] {10.1088/0004-6256/147/6/136}, \href {https://ui.adsabs.harvard.edu/abs/2014AJ....147..136R} {147, 136}

\bibitem[\protect\citeauthoryear{{Roederer} et~al.,}{{Roederer} et~al.}{2016}]{Roederer2016}
{Roederer} I.~U.,  et~al., 2016, \mn@doi [\aj] {10.3847/0004-6256/151/3/82}, \href {https://ui.adsabs.harvard.edu/abs/2016AJ....151...82R} {151, 82}

\bibitem[\protect\citeauthoryear{{Roederer}, {Hattori}  \& {Valluri}}{{Roederer} et~al.}{2018}]{Roederer18}
{Roederer} I.~U.,  {Hattori} K.,   {Valluri} M.,  2018, \mn@doi [\aj] {10.3847/1538-3881/aadd9c}, \href {https://ui.adsabs.harvard.edu/abs/2018AJ....156..179R} {156, 179}

\bibitem[\protect\citeauthoryear{{Santistevan}, {Wetzel}, {El-Badry}, {Bland-Hawthorn}, {Boylan-Kolchin}, {Bailin}, {Faucher-Gigu{\`e}re}  \& {Benincasa}}{{Santistevan} et~al.}{2020}]{Santistevan2020}
{Santistevan} I.~B.,  {Wetzel} A.,  {El-Badry} K.,  {Bland-Hawthorn} J.,  {Boylan-Kolchin} M.,  {Bailin} J.,  {Faucher-Gigu{\`e}re} C.-A.,   {Benincasa} S.,  2020, \mn@doi [\mnras] {10.1093/mnras/staa1923}, \href {https://ui.adsabs.harvard.edu/abs/2020MNRAS.497..747S} {497, 747}

\bibitem[\protect\citeauthoryear{{Sheffield} et~al.,}{{Sheffield} et~al.}{2012}]{Sheffield12}
{Sheffield} A.~A.,  et~al., 2012, \mn@doi [\apj] {10.1088/0004-637X/761/2/161}, \href {https://ui.adsabs.harvard.edu/abs/2012ApJ...761..161S} {761, 161}

\bibitem[\protect\citeauthoryear{{Shetrone} et~al.,}{{Shetrone} et~al.}{2019}]{Shetrone19}
{Shetrone} M.,  et~al., 2019, \mn@doi [\apj] {10.3847/1538-4357/aaff66}, \href {https://ui.adsabs.harvard.edu/abs/2019ApJ...872..137S} {872, 137}

\bibitem[\protect\citeauthoryear{{Shipp} et~al.,}{{Shipp} et~al.}{2018}]{Shipp18}
{Shipp} N.,  et~al., 2018, \mn@doi [\apj] {10.3847/1538-4357/aacdab}, \href {https://ui.adsabs.harvard.edu/abs/2018ApJ...862..114S} {862, 114}

\bibitem[\protect\citeauthoryear{{Simmerer}, {Sneden}, {Cowan}, {Collier}, {Woolf}  \& {Lawler}}{{Simmerer} et~al.}{2004}]{Simmerer2004}
{Simmerer} J.,  {Sneden} C.,  {Cowan} J.~J.,  {Collier} J.,  {Woolf} V.~M.,   {Lawler} J.~E.,  2004, \mn@doi [\apj] {10.1086/424504}, \href {https://ui.adsabs.harvard.edu/abs/2004ApJ...617.1091S} {617, 1091}

\bibitem[\protect\citeauthoryear{{Simon}}{{Simon}}{2019}]{Simon19}
{Simon} J.~D.,  2019, \mn@doi [\araa] {10.1146/annurev-astro-091918-104453}, \href {https://ui.adsabs.harvard.edu/abs/2019ARA&A..57..375S} {57, 375}

\bibitem[\protect\citeauthoryear{{Simon}, {Frebel}, {McWilliam}, {Kirby}  \& {Thompson}}{{Simon} et~al.}{2010}]{Simon2010}
{Simon} J.~D.,  {Frebel} A.,  {McWilliam} A.,  {Kirby} E.~N.,   {Thompson} I.~B.,  2010, \mn@doi [\apj] {10.1088/0004-637X/716/1/446}, \href {https://ui.adsabs.harvard.edu/abs/2010ApJ...716..446S} {716, 446}

\bibitem[\protect\citeauthoryear{{Simon} et~al.,}{{Simon} et~al.}{2011}]{Simon2011}
{Simon} J.~D.,  et~al., 2011, \mn@doi [\apj] {10.1088/0004-637X/733/1/46}, \href {https://ui.adsabs.harvard.edu/abs/2011ApJ...733...46S} {733, 46}

\bibitem[\protect\citeauthoryear{{Simpson} et~al.,}{{Simpson} et~al.}{2020}]{Simpson2020}
{Simpson} J.~D.,  et~al., 2020, \mn@doi [\mnras] {10.1093/mnras/stz3105}, \href {https://ui.adsabs.harvard.edu/abs/2020MNRAS.491.3374S} {491, 3374}

\bibitem[\protect\citeauthoryear{{Sitnova}, {Mashonkina}  \& {Ryabchikova}}{{Sitnova} et~al.}{2016}]{Sitnova16}
{Sitnova} T.~M.,  {Mashonkina} L.~I.,   {Ryabchikova} T.~A.,  2016, \mn@doi [\mnras] {10.1093/mnras/stw1202}, \href {https://ui.adsabs.harvard.edu/abs/2016MNRAS.461.1000S} {461, 1000}

\bibitem[\protect\citeauthoryear{{Sitnova} et~al.,}{{Sitnova} et~al.}{2024}]{Sitnova2024}
{Sitnova} T.~M.,  et~al., 2024, \mn@doi [\aap] {10.1051/0004-6361/202450981}, \href {https://ui.adsabs.harvard.edu/abs/2024A&A...690A.331S} {690, A331}

\bibitem[\protect\citeauthoryear{{Sneden}}{{Sneden}}{1973}]{Sneden73}
{Sneden} C.~A.,  1973, PhD thesis, University of Texas, Austin

\bibitem[\protect\citeauthoryear{{Sobeck} et~al.,}{{Sobeck} et~al.}{2011}]{Sobeck11}
{Sobeck} J.~S.,  et~al., 2011, \mn@doi [\aj] {10.1088/0004-6256/141/6/175}, \href {https://ui.adsabs.harvard.edu/abs/2011AJ....141..175S} {141, 175}

\bibitem[\protect\citeauthoryear{{Speagle}}{{Speagle}}{2020}]{dynesty}
{Speagle} J.~S.,  2020, \mn@doi [\mnras] {10.1093/mnras/staa278}, \href {https://ui.adsabs.harvard.edu/abs/2020MNRAS.493.3132S} {493, 3132}

\bibitem[\protect\citeauthoryear{{Spite} et~al.,}{{Spite} et~al.}{2005}]{Spite05}
{Spite} M.,  et~al., 2005, \mn@doi [\aap] {10.1051/0004-6361:20041274}, \href {https://ui.adsabs.harvard.edu/abs/2005A&A...430..655S} {430, 655}

\bibitem[\protect\citeauthoryear{{Spite} et~al.,}{{Spite} et~al.}{2012}]{Spite12}
{Spite} M.,  et~al., 2012, \mn@doi [\aap] {10.1051/0004-6361/201218773}, \href {https://ui.adsabs.harvard.edu/abs/2012A&A...541A.143S} {541, A143}

\bibitem[\protect\citeauthoryear{{Spite}, {Spite}, {Fran{\c{c}}ois}, {Bonifacio}, {Caffau}  \& {Salvadori}}{{Spite} et~al.}{2018}]{Spite2018}
{Spite} M.,  {Spite} F.,  {Fran{\c{c}}ois} P.,  {Bonifacio} P.,  {Caffau} E.,   {Salvadori} S.,  2018, \mn@doi [\aap] {10.1051/0004-6361/201833548}, \href {https://ui.adsabs.harvard.edu/abs/2018A&A...617A..56S} {617, A56}

\bibitem[\protect\citeauthoryear{{Takeda}, {Hashimoto}, {Taguchi}, {Yoshioka}, {Takada-Hidai}, {Saito}  \& {Honda}}{{Takeda} et~al.}{2005}]{Takeda05}
{Takeda} Y.,  {Hashimoto} O.,  {Taguchi} H.,  {Yoshioka} K.,  {Takada-Hidai} M.,  {Saito} Y.,   {Honda} S.,  2005, \mn@doi [\pasj] {10.1093/pasj/57.5.751}, \href {https://ui.adsabs.harvard.edu/abs/2005PASJ...57..751T} {57, 751}

\bibitem[\protect\citeauthoryear{{Theler} et~al.,}{{Theler} et~al.}{2020}]{Theler20}
{Theler} R.,  et~al., 2020, \mn@doi [\aap] {10.1051/0004-6361/201937146}, \href {https://ui.adsabs.harvard.edu/abs/2020A&A...642A.176T} {642, A176}

\bibitem[\protect\citeauthoryear{{Ting}, {Conroy}, {Rix}  \& {Cargile}}{{Ting} et~al.}{2019}]{Ting19}
{Ting} Y.-S.,  {Conroy} C.,  {Rix} H.-W.,   {Cargile} P.,  2019, \mn@doi [\apj] {10.3847/1538-4357/ab2331}, \href {https://ui.adsabs.harvard.edu/abs/2019ApJ...879...69T} {879, 69}

\bibitem[\protect\citeauthoryear{{Tinsley}}{{Tinsley}}{1979}]{Tinsley1979}
{Tinsley} B.~M.,  1979, \mn@doi [\apj] {10.1086/157039}, \href {https://ui.adsabs.harvard.edu/abs/1979ApJ...229.1046T} {229, 1046}

\bibitem[\protect\citeauthoryear{{Tolstoy}, {Hill}  \& {Tosi}}{{Tolstoy} et~al.}{2009}]{Tolstoy2009}
{Tolstoy} E.,  {Hill} V.,   {Tosi} M.,  2009, \mn@doi [\araa] {10.1146/annurev-astro-082708-101650}, \href {https://ui.adsabs.harvard.edu/abs/2009ARA&A..47..371T} {47, 371}

\bibitem[\protect\citeauthoryear{{Usman} et~al.,}{{Usman} et~al.}{2024}]{Usman24}
{Usman} S.~A.,  et~al., 2024, \mn@doi [\mnras] {10.1093/mnras/stae185}, \href {https://ui.adsabs.harvard.edu/abs/2024MNRAS.529.2413U} {529, 2413}

\bibitem[\protect\citeauthoryear{{Varghese}, {Ibata}  \& {Lewis}}{{Varghese} et~al.}{2011}]{Varghese11}
{Varghese} A.,  {Ibata} R.,   {Lewis} G.~F.,  2011, \mn@doi [\mnras] {10.1111/j.1365-2966.2011.19097.x}, \href {https://ui.adsabs.harvard.edu/abs/2011MNRAS.417..198V} {417, 198}

\bibitem[\protect\citeauthoryear{{Venn} et~al.,}{{Venn} et~al.}{2012}]{Venn12}
{Venn} K.~A.,  et~al., 2012, \mn@doi [\apj] {10.1088/0004-637X/751/2/102}, \href {https://ui.adsabs.harvard.edu/abs/2012ApJ...751..102V} {751, 102}

\bibitem[\protect\citeauthoryear{{Venn}, {Starkenburg}, {Malo}, {Martin}  \& {Laevens}}{{Venn} et~al.}{2017}]{Venn2017}
{Venn} K.~A.,  {Starkenburg} E.,  {Malo} L.,  {Martin} N.,   {Laevens} B.~P.~M.,  2017, \mn@doi [\mnras] {10.1093/mnras/stw3198}, \href {https://ui.adsabs.harvard.edu/abs/2017MNRAS.466.3741V} {466, 3741}

\bibitem[\protect\citeauthoryear{{Waller} et~al.,}{{Waller} et~al.}{2023}]{Waller2023}
{Waller} F.,  et~al., 2023, \mn@doi [\mnras] {10.1093/mnras/stac3563}, \href {https://ui.adsabs.harvard.edu/abs/2023MNRAS.519.1349W} {519, 1349}

\bibitem[\protect\citeauthoryear{{Wan} et~al.,}{{Wan} et~al.}{2020}]{Wan20}
{Wan} Z.,  et~al., 2020, \mn@doi [\nat] {10.1038/s41586-020-2483-6}, \href {https://ui.adsabs.harvard.edu/abs/2020Natur.583..768W} {583, 768}

\bibitem[\protect\citeauthoryear{Waskom et~al.,}{Waskom et~al.}{2016}]{seaborn}
Waskom M.,  et~al., 2016, seaborn: v0.7.0 (January 2016), \mn@doi{10.5281/zenodo.45133}, \url {http://dx.doi.org/10.5281/zenodo.45133}

\bibitem[\protect\citeauthoryear{{Webber} et~al.,}{{Webber} et~al.}{2023}]{Webber2023}
{Webber} K.~B.,  et~al., 2023, \mn@doi [\apj] {10.3847/1538-4357/ad0385}, \href {https://ui.adsabs.harvard.edu/abs/2023ApJ...959..141W} {959, 141}

\bibitem[\protect\citeauthoryear{{Weisz} \& {Boylan-Kolchin}}{{Weisz} \& {Boylan-Kolchin}}{2017}]{Weisz2017}
{Weisz} D.~R.,  {Boylan-Kolchin} M.,  2017, \mn@doi [\mnras] {10.1093/mnrasl/slx043}, \href {https://ui.adsabs.harvard.edu/abs/2017MNRAS.469L..83W} {469, L83}

\bibitem[\protect\citeauthoryear{{Willman} \& {Strader}}{{Willman} \& {Strader}}{2012}]{Willman12}
{Willman} B.,  {Strader} J.,  2012, \mn@doi [\aj] {10.1088/0004-6256/144/3/76}, \href {https://ui.adsabs.harvard.edu/abs/2012AJ....144...76W} {144, 76}

\bibitem[\protect\citeauthoryear{{Worley}, {Hill}, {Sobeck}  \& {Carretta}}{{Worley} et~al.}{2013}]{Worley13}
{Worley} C.~C.,  {Hill} V.,  {Sobeck} J.,   {Carretta} E.,  2013, \memsai, \href {https://ui.adsabs.harvard.edu/abs/2013MmSAI..84..269W} {84, 269}

\bibitem[\protect\citeauthoryear{York et~al.,}{York et~al.}{2000}]{York00}
York D.~G.,  et~al., 2000, \mn@doi [The Astronomical Journal] {10.1086/301513}, 120, 1579

\bibitem[\protect\citeauthoryear{{Yuan}, {Chang}, {Beers}  \& {Huang}}{{Yuan} et~al.}{2020}]{Yuan2020}
{Yuan} Z.,  {Chang} J.,  {Beers} T.~C.,   {Huang} Y.,  2020, \mn@doi [\apjl] {10.3847/2041-8213/aba49f}, \href {https://ui.adsabs.harvard.edu/abs/2020ApJ...898L..37Y} {898, L37}

\bibitem[\protect\citeauthoryear{{Yuan} et~al.,}{{Yuan} et~al.}{2022a}]{Yuan2022}
{Yuan} Z.,  et~al., 2022a, \mn@doi [\mnras] {10.1093/mnras/stac1399}, \href {https://ui.adsabs.harvard.edu/abs/2022MNRAS.514.1664Y} {514, 1664}

\bibitem[\protect\citeauthoryear{{Yuan} et~al.,}{{Yuan} et~al.}{2022b}]{Yuan2022cetus}
{Yuan} Z.,  et~al., 2022b, \mn@doi [\apj] {10.3847/1538-4357/ac616f}, \href {https://ui.adsabs.harvard.edu/abs/2022ApJ...930..103Y} {930, 103}

\bibitem[\protect\citeauthoryear{{Zhang}, {Gehren}  \& {Zhao}}{{Zhang} et~al.}{2014}]{Zhang14}
{Zhang} H.~W.,  {Gehren} T.,   {Zhao} G.,  2014, in {Feltzing} S.,  {Zhao} G.,  {Walton} N.~A.,   {Whitelock} P.,  eds,  IAU Symposium Vol. 298, Setting the scene for Gaia and LAMOST. pp 453--453, \mn@doi{10.1017/S1743921313007187}

\bibitem[\protect\citeauthoryear{{van~der~Walt}, Colbert  \& Varoquaux}{{van~der~Walt} et~al.}{2011}]{numpy}
{van~der~Walt} S.,  Colbert S.~C.,   Varoquaux G.,  2011, \mn@doi [Computing in Science \& Engineering] {http://dx.doi.org/10.1109/MCSE.2011.37}, 13, 22

\makeatother
\end{thebibliography}
\bibliographystyle{mnras}

\end{document}